\documentclass[11pt]{article} 
\pdfoutput=1 

\usepackage{jheppub}
\usepackage{graphicx} 
\usepackage[T1]{fontenc}
\usepackage[utf8]{inputenc}
\usepackage{amsmath,amsfonts,amssymb,amsthm}
\usepackage{float}
\usepackage{mathrsfs}
\usepackage{bm}
\usepackage{booktabs}
\usepackage{caption}
\usepackage{subcaption}
\usepackage{makecell}
\usepackage{cancel}
\usepackage[formats]{listings}
\usepackage[font=small]{caption}
\usepackage{array}
\usepackage{slashed}
\DeclareMathOperator{\sgn}{sgn}
\usepackage[dvipsnames]{xcolor}
\usepackage{subcaption}
\usepackage{multirow}

\usepackage{stackengine}
\stackMath
\usepackage{rotating}
\usepackage{enumerate}
\usepackage{physics}
\usepackage{booktabs}
\setcellgapes{3.5pt}
\makegapedcells   
\DeclareUnicodeCharacter{2212}{-}

\begin{document} 

\title{
Higgs decays to four leptons to $\mathcal{O}(1/\Lambda^4)$ in SMEFT}
\author[]{Mario Flores-Hernandez \&}
\emailAdd{mfloresh@nd.edu}
\author[]{Adam Martin}
\emailAdd{amarti41@nd.edu}
\affiliation[]{Department of Physics and Astronomy, University of Notre Dame, South Bend, IN, 46556 USA}

\abstract{
We study the decays $h \to \ell \bar{\ell} \left(Z \to \ell' \bar{\ell'}\right)$ and $h\to\ell\bar{\nu}_\ell\nu_{\ell'}\bar{\ell'}$ within the SMEFT framework and including effects up to $\mathcal O(1/\Lambda^4)$, where $\Lambda$ is the new physics scale suppressing higher dimensional operators. To work to this order, we must include the square of dimension-six operators and the interference of dimension-eight operators with the Standard Model. We study angular asymmetries and other differential decay observables and determine which are most sensitive to $\mathcal O(1/\Lambda^4)$ effects. While new kinematic structures arising in higher dimensional operators have the potential to induce novel angular dependency, we find this does not occur for  $h\to\ell\bar{\ell}\left(Z\xrightarrow{}\ell'\bar{\ell'}\right)$. For $h \to \ell \bar{\nu}_\ell \nu_{\ell'} \bar{\ell'}$, new angular dependencies do arise at $\mathcal O(1/\Lambda^4)$, though they require a fully reconstructible (meaning we can go to the Higgs rest frame) final state. For non-reconstructible final states such as  $\ell \bar{\nu}_\ell \nu_{\ell'} \bar{\ell'}$, we must study Higgs production and decay together with the appropriate observables, which we find obscures the new angular effects.
}

\maketitle
\setcounter{page}{2}
\clearpage

\section{Introduction}\label{Section 1}

As we head into the precision phase of the LHC era, clean, high-statistics observables are an invaluable laboratory to test the Standard Model (SM). The Higgs decay $h \to ZZ^* \to \ell \bar{\ell} \ell' \bar{\ell'}$ and its charged current counterpart $h \to WW^* \to \ell \bar{\nu}_\ell \nu_{\ell'} \bar{\ell'} $ are two such observables. The fully visible 'golden' four lepton channel will yield $\mathcal O(10^4)$ events with the full $3$ ab$^{-1}$ HL-LHC luminosity, and the two lepton plus two neutrino channel has a larger rate, $\mathcal O(10^6)$ events. Both decays are also prime targets for a future lepton collider such as the FCC-ee.

To interpret potential deviations in these channels in a systematic and model-independent way, it is convenient to work within the Standard Model Effective Field Theory (SMEFT), where nonstandard interaction strengths or kinematic structures are captured by higher-dimensional operators formed from the SM fields alone \cite{Buchmuller:1985jz, Grzadkowski:2010es, Brivio:2017vri}. The widths and differential distributions for $h \to \ell \bar{\ell} \ell' \bar{\ell'}, h \to \ell \bar{\nu}_\ell \nu_{\ell'} \bar{\ell'}$ have been studied in the SMEFT framework extensively at $\mathcal{O}(1/\Lambda^2)$ in the SMEFT expansion \cite{Buchalla:2013mpa, Beneke:2014sba, Boselli:2017pef, Brivio:2019myy, He:2019kgh, Banerjee:2020vtm, Dawson:2024pft, Bellafronte:2026mhp}, which comes from the interference between the SM amplitude and effects induced by dimension six operators. To isolate or enhance the visibility of SMEFT effects, several observables have proven useful, including dilepton invariant-mass distributions, angular distributions \cite{Stolarski:2012ps,Chen:2012jy,Chen:2013ejz}, and various asymmetries \cite{Buchalla:2013mpa,Beneke:2014sba}.

 Recently, interest has grown in extending SMEFT analyses to include  $\mathcal{O}(1/\Lambda^4)$ contributions \cite{Hays:2018zze, Hays:2020scx, Corbett:2021jox, Corbett:2023qtg, Boughezal:2021tih, Corbett:2021cil, Alioli:2020kez, Boughezal:2022nof,Kim:2022amu, Allwicher:2022gkm, Dawson:2021xei, Degrande:2023iob, Corbett:2021eux, Martin:2023tvi, Gillies:2024mqp}.  One important motivation is the "inverse problem", where the imprints of different UV compositions are degenerate at dimension six. This degeneracy is reduced at dimension eight, enabling the distinction between UV scenarios \cite{Zhang:2021eeo}. Additionally, dimension eight operators can help to quantify the "truncation uncertainty", that is, the theoretical error introduced when only dimension six operators are considered \cite{Trott:2021vqa, Martin:2021cvs}.  Another motivation stems from energetics, as dimension eight contributions of the form $\hat{s}^2/\Lambda^4$  become "energy enhanced" in the regime where $\sqrt{\hat{s}}\gg v$, making dimension-eight effects comparable with dimension six operators \cite{Assi:2024zap, Assi:2025zmp}.  Perhaps most relevant to this work is the emergence of novel kinematics at dimension eight, potentially leading to very distinct phenomena that would be entirely missed by considering only dimension-six operators \cite{Corbett:2021iob, Corbett:2023yhk}.

Motivated by the high precision of measurements on Higgs decay to four leptons,  the potential novel kinematics of dimension-eight operators, and the key role the Higgs width plays in testing the SM~\cite{Brivio:2019myy}, in this work we carry out the first analysis of $h \to \ell \bar{\ell} \left(Z \to \ell' \bar{\ell'}\right)$,
$h \to \ell \bar{\nu}_\ell \nu_{\ell'} \bar{\ell'}$ to $\mathcal{O}(1/\Lambda^4)$. At this order, we must include (at the squared amplitude level) contributions that are linear in dimension eight operators, and contributions that are quadratic in dimension six operators. We divide our analysis into two parts: the four charged lepton case which is fully reconstructible, and the two charged lepton case which is not. In both cases, we are particularly interested in the $\mathcal{O}(1/\Lambda^4)$  SMEFT effects to the angular distribution of the final state leptons.  

Because the four charged lepton final state is fully reconstructible, we perform the analysis completely analytically.  We use the narrow-width approximation to calculate the lowest order (in SM couplings) differential decay rate for  $h \to \ell \bar{\ell} \left(Z \to \ell' \bar{\ell'}\right)$, neglecting loop corrections and in the massless lepton limit. For this purpose, we utilize the geometric formulation of SMEFT (geoSMEFT) and the massless spinor-helicity formalism, including new structures from higher dimensional operators. Squaring the amplitude and summing over final state helicity combinations, we obtain an analytic expression for the differential decay rate to $\mathcal O(1/\Lambda^4)$. To study the angular distributions of the final state leptons, we perform a similar analysis as the one presented in \cite{Buchalla:2013mpa, Beneke:2014sba}. Specifically, we decompose the differential decay rate in terms of a linear combination of angular functions of three different angles, each of which is accompanied by a coefficient. These coefficients are functions of SM parameters and SMEFT Wilson coefficients. By considering various differential observables (dilepton masses, angular asymmetries), it is possible to tease out different coefficient functions. 

For $h \to \ell \bar{\nu}_\ell \nu_{\ell'} \bar{\ell'}$, we obtain the analogous analytic expression for the differential decay rate. However, as the final state is not fully reconstructible, we cannot consider Higgs decay in isolation and we must work with the full production and decay process, $pp \to h \to \ell \bar{\nu}_\ell \nu_{\ell'} \bar{\ell'}$.~\footnote{This is true at a hadron collider. At a lepton collider, we can determine the Higgs rest frame by measuring the momenta of all other (non-Higgs) final states; however neutrinos will still be invisible.} Including production and decay, the fully differential results are analytically intractable, so we will turn to Monte Carlo. In addition, the presence of the neutrinos means we are limited in what observables we can use.

 This paper is organized as follows. In Section \ref{section 2}, we present the list of dimension six and dimension eight operators involved in the analysis and study their different roles. In Section \ref{section 3} we present the results for $h \to \ell \bar{\ell} \left(Z \to \ell' \bar{\ell'}\right)$. We detail the calculation to obtain the analytic expression of the differential decay rate, then introduce the observables sensitive to SMEFT effects. In Sec. \ref{section 4} we present the results for $h \to \ell \bar{\nu}_\ell \nu_{\ell'} \bar{\ell'}$. The calculation details are presented analogously to those in Section \ref{section 3}, however, different observables are introduced for this section. Finally, in Sec.~\ref{section 5} we present our conclusions. 
 
\section{Contributing SMEFT Operators}\label{section 2}
The SMEFT assumes the existence of physics beyond the Standard Model (SM) at an UV energy scale $\Lambda$, which is systematically captured by extending the SM with a series of higher dimensional operators of the form:

\begin{equation}
    \mathcal{L}_{\text{SMEFT}}=\mathcal{L}_{\text{SM}}+\sum_{d>4,i}\frac{C^{(d)}_{i}}{\Lambda^{(d-4)}}\mathcal{O}_{i}^{(d)}.
\end{equation}
The operators $\mathcal{O}_{i}^{(d)}$ are formed from SM fields and are invariant under $\text{U(1)}_Y\times\text{SU(2)}_L\times\text{SU(3)}_C$. The operators are organized by their mass dimension $d$ and are labeled by $i$, which runs over the set of independent operators at a given dimension. The Wilson coefficients, $C^{(d)}_i$, get an imprint of the UV physics that has been integrated out.  Rather than assuming a particular UV competition, we take a bottom-up perspective, meaning that we treat the Wilson coefficients as free parameters with no relations between them. We assume lepton and baryon number conservation, as constrains on these type of operators push $\Lambda$ close to the GUT scale . This assumption forbids all odd dimension operators (and also some even dimensional operators). We further reduce the number of operators by assuming CP conservation, since these CP interactions are tightly constrained by low-energy experiments \cite{DAmbrosio:2002vsn}.

In a decay process like the one in consideration, SMEFT operators contribute to the amplitude through their associated Wilson coefficients. To remain consistent with the SMEFT expansion, at  $\mathcal{O}(1/\Lambda^4)$  the square amplitude takes the form:
\begin{equation}\label{square amplitude}
    |\mathcal{A}|^2=|A_\text{SM}|^2 +2\operatorname{Re}(A_\text{SM}^*A_6)+|A_6|^2 +2\operatorname{Re}(A_\text{SM}^*A_8),
\end{equation}
where $A_{SM}$ is the SM amplitude and $A_6$ and $A_8$ are functions of the Wilson coefficients of the dimensions six and eight operators, respectively.  Clearly, dimension eight operators only contribute as interference terms with the SM, while dimension six operators contribute both via interference with the Standard Model and through squared terms, $|A_6|^2$. Note that $|A_6|^2$ also includes interference between different dimension-six operators.

\begin{figure}[h!]
  \begin{subfigure}[b]{0.325\textwidth}
    \includegraphics[width=\textwidth]{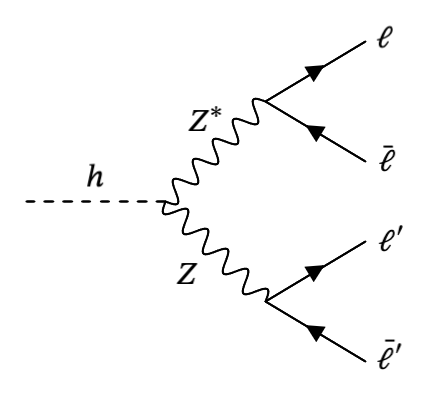}
    \subcaption*{(a)}
  \end{subfigure}
  \begin{subfigure}[b]{0.325\textwidth}
    \includegraphics[width=\textwidth]{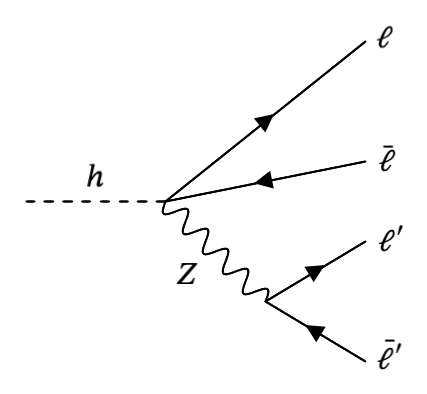}
    \subcaption*{(b)}
  \end{subfigure}
  \begin{subfigure}[b]{0.325\textwidth}
    \includegraphics[width=\textwidth]{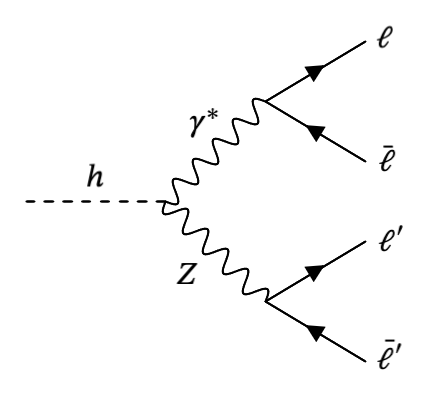}
    \subcaption*{(c)}
  \end{subfigure}  
  \caption{Topologies for $h\to\ell\bar{\ell}\left(Z\to\ell'\bar{\ell'}\right)$ }
  \label{htoZlldiagrams}
\end{figure}

\begin{figure}[h!]
  \begin{subfigure}[b]{0.325\textwidth}
    \includegraphics[width=\textwidth]{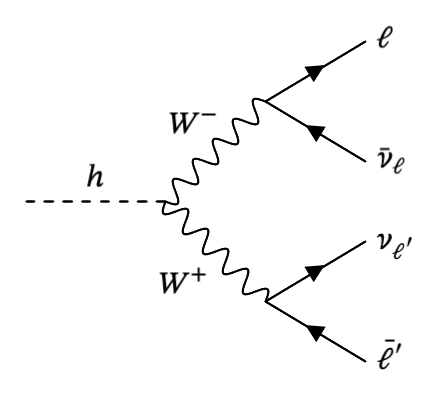}
    \subcaption*{(d)}
  \end{subfigure}
  \begin{subfigure}[b]{0.325\textwidth}
    \includegraphics[width=\textwidth]{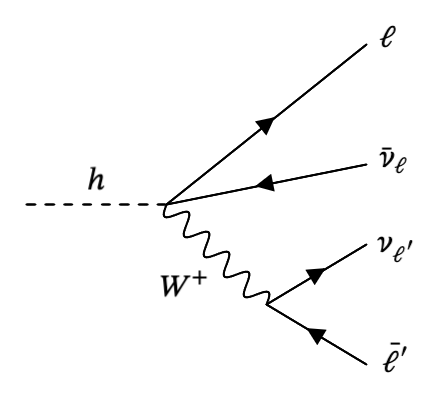}
    \subcaption*{(e)}
  \end{subfigure}
  \begin{subfigure}[b]{0.325\textwidth}
    \includegraphics[width=\textwidth]{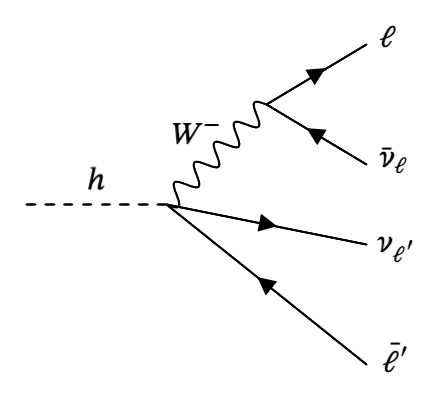}
    \subcaption*{(f)}
  \end{subfigure}  
  \caption{Topologies for $h \to \ell \bar{\nu}_\ell \nu_{\ell'} \bar{\ell'}$ }
  \label{htoWW diagrams}
\end{figure}

To depict the possible ways in which SMEFT effects can enter Eq. (\ref{square amplitude}), let us first present the tree-level topologies for $h\xrightarrow{}Z(\xrightarrow{}\ell'\bar{\ell'})\ell\bar{\ell}$ and $h\xrightarrow{}\ell\bar{\nu}_\ell\nu_{\ell'}\bar{\ell'}$, in Fig.~\ref{htoZlldiagrams} and Fig.~\ref{htoWW diagrams}. There could be another diagram where  $h\xrightarrow{}\ell'\bar{\ell'}\ell\bar{\ell}$ directly, meaning no intermediate, on-shell $Z$, but this will be strongly suppressed by kinematic cuts (at the detector level, which we will assume throughout) requiring $m_{\ell\ell} \sim m_Z$ for at least one pair of same-flavor leptons. The requirement $m_{\ell\ell} \sim m_Z$ also suppresses variations of topology (b) where the intermediate $Z$ is off-shell, $Z \to Z^*$. For $h\xrightarrow{}\ell\bar{\nu}_\ell\nu_{\ell'}\bar{\ell'}$, we cannot reconstruct the $W$, so we allow topologies where either boson ($W^\pm$ or $W^{*\pm}$) is involved in the four-point vertex with the Higgs (topology (e) or (f) in Fig.~\ref{htoWW diagrams}). We also include the five-point vertex
\(h \to \ell \bar{\nu}_\ell \nu_{\ell'} \bar{\ell'}\),
which, unlike the
\(h \to Z(\to \ell' \bar{\ell'})\,\ell \bar{\ell}\)
contribution, cannot be removed by analysis cuts.
Its effect, however, is negligible: at
\(\mathcal{O}(1/\Lambda^4)\)
it enters only through interference with the SM amplitude,
and the two exhibit markedly different kinematic structures; see Sec.~\ref{sec:hzzasymm}. We therefore neglect this contribution in the remainder of our analysis.


From these topologies, we see the different ways higher-dimensional operators can enter:
\begin{enumerate}[i)]
    \item They can alter the vertices $hVV$ and $Vf\bar{f'}$ ($V=W^\pm,Z,A$,  $f,f'=e,l,\nu$) which exist in the SM. This can be through the strength of the vertex or by inducing new Lorentz structures.
    \item They can introduce vertices with no SM analog. Specifically, the 3-point vertex $hZA$, and the 4-point contact vertices $f\bar{f'} V h $. 
    \item The third way is less obvious from the diagrams. Higher dimensional operators can affect the relationship between Lagrangian parameters and experimental inputs like $\alpha_{em}, M_Z, G_F$.
\end{enumerate}

The geometric formulation of the SMEFT (geoSMEFT) greatly facilitates calculations such as $h\to\ell\bar{\ell}\left(Z\to\ell'\bar{\ell'}\right)$  and $h \to \ell \bar\nu_{\ell}\nu_{\ell'}\ell'$, as it neatly separates SMEFT effects that can affect EW inputs and modify the strength of SM vertices from SMEFT effects that generate new vertices and new Lorentz structures \cite{Helset:2020yio}. The geoSMEFT reorganizes a subset of SMEFT operators in terms of field-space dependent -- but derivative independent -- quantities, referred to as 'connections'. The geoSMEFT connections relevant for this work are:
\begin{align}
\label{eq:Lgeo}
 \nonumber
    \mathcal{L}_{\mbox{geoSMEFT}}\supset& \,h_{IJ}(\phi)(D_\mu\phi)^I(D^\mu\phi)^J+g_{AB}(\phi)\mathcal{W}^A_{\mu\nu}\mathcal{W}^{B,\mu\nu}+k^A_{IJ}(\phi)(D_\mu\phi)^I(D^\mu\phi)^J\mathcal{W}_A^{\mu\nu}\\
    & +L_{J,A}^{\psi,pr}(\phi) (D_\mu\phi)(\bar{\psi}_p\gamma^\mu\tilde{\sigma}^A\psi_r)+(d_A^{e,pr}(\phi)\bar{l}_p\sigma^{\mu\nu}e_r\mathcal{W}_A^{\mu\nu}+h.c.).
\end{align}
Here $\mathcal{W}^A=\{W^1,W^2,W^3,B\}$ are the electroweak gauge fields and $\phi^I=\{\phi^1,\phi^2,\phi^3,\phi^4\}$ is a re-expression of the Higgs complex doublet $H$ into four real scalar fields~\footnote{The connections are shown to be functions of $\phi$, which is a shorthand for $\phi^I\phi_I=H^\dagger H$. }. The fermionic fields relevant for $h\to\ell\bar{\ell}\left(Z\to\ell'\bar{\ell'}\right)$  and $h \to \ell \bar\nu_{\ell}\nu_{\ell'}\ell'$ are $\psi=l$ for left-handed fermions, $\psi=e$ for right-handed fermions; $p$ and $r$ are flavor indices. In Table \ref{list of operators} we list the operators, up to dimension eight, contributing to these connections. 

\begin{table}[t]
\centering
\resizebox{0.95\textwidth}{!}{
\begin{tabular}{|l|l|}
\hline
\multicolumn{2}{|c|}{\textbf{geoSMEFT operators}}  \\ \hline
\multicolumn{1}{|c|}{$h_{IJ}$ Connection} & \multicolumn{1}{c|}{$g_{AB}$ Connection} \\ \hline
\begin{tabular}[c|]{@{}l@{}}$Q_{H\Box}=(H^\dag H)\Box(H^\dag H)$ \\ $Q^{(6)}_{HD}=(H^\dag D_\mu H)^*(H^\dag D^\mu H) $ \\ $Q^{(8)}_{HD}=(H^\dag H)^2(D_\mu H)^\dag(D^\mu H)$ \\ $Q^{(8)}_{HD,2}=(H^\dag H)(H^\dag \sigma^I H)(D_\mu H)^\dag\sigma^I(D^\mu H)$ \end{tabular}     
& \begin{tabular}[c]{@{}l@{}}$Q^{(6+2n)}_{HB}=(H^\dag H)^{n+1} B_{\mu \nu}B^{\mu \nu}$ \\ $Q^{(6+2n)}_{HW}=(H^\dag H)^{n+1} W^I_{\mu \nu}W^{I,\mu \nu}$ \\ $Q^{(6+2n)}_{HBW}=(H^\dag H)^{n}(H^\dag \sigma^I H) W^I_{\mu \nu}B^{\mu \nu}$ \\ $Q^{(8)}_{HW,2}=(H^\dag \sigma^I H)(H^\dag \sigma^J H)W^I_{\mu \nu}W^{J,\mu \nu}$ \end{tabular}     \\ \hline
\multicolumn{1}{|c|}{$k^A_{IJ}$ Connection} & \multicolumn{1}{c|}{$L^\psi_{I,A}$ Connection $(\psi=l,e)$} \\ \hline
\begin{tabular}[c|]{@{}l@{}}$Q^{(8)}_{HDHB}=i(H^\dag H) (D^\mu H)^\dag(D^\nu H)B_{\mu \nu}$ \\ $Q^{(8)}_{HDHW}=i(H^\dag H) (D^\mu H)^\dag\sigma^I(D^\nu H)W^I_{\mu \nu}$ \\ $Q^{(8)}_{HDHW,2}=i\epsilon_{IJK}(H^\dag \sigma^I H)(D^\mu H)^\dag\sigma^J(D^\nu H)W^K_{\mu \nu}$ \end{tabular} 
& \begin{tabular}[c]{@{}l@{}}$Q^{1,(6+2n)}_{H\psi}=i(H^\dag H)^n(H^\dag \overleftrightarrow{D}_\mu H)\bar{\psi}_p \gamma^\mu \psi_r$ \\ $Q^{3,(6+2n)}_{Hl}=i(H^\dag H)^n(H^\dag \overleftrightarrow{D}_\mu^I H)\bar{l}_p \gamma^\mu \sigma^I l_r $ \\ $Q^{2,(8)}_{Hl}=i(H^\dag H)(H^\dag \sigma^I H)(H^\dag \overleftrightarrow{D}_\mu H)\bar{l}_p \gamma^\mu \sigma^Il_r$ \\ $Q^{\epsilon,(8)}_{Hl}=i\epsilon_{IJK}(H^\dag \sigma^I H)(H^\dag \overleftrightarrow{D}^J_\mu H)\bar{l}_p \gamma^\mu \sigma^Kl_r$ \end{tabular} \\ \hline
\multicolumn{2}{|c|}{$d_{A}$ connection} \\ \hline
\multicolumn{2}{|c|}{
\begin{tabular}{r l} $Q_{eB}=(\bar{l}_p\sigma_{\mu\nu} e_r) H B^{\mu\nu}$,& $Q_{eW}= (\bar{l}_p\sigma_{\mu\nu}\sigma^I e_r) H W^{I,\mu\nu}$ \end{tabular}} \\ \hline
\end{tabular}
}
\caption{GeoSMEFT operators contributing to Higgs decay to four leptons. The superscript $2n+1$ indicates that the operator can be dimension six, for $n=0$, or dimension eight, for $n=1$. The subindex $\psi=e,l$, stands for the relevant fields for the process in consideration.  We have suppress the flavor indices in the operator name to simplify the notation.}
\label{list of operators}
\end{table}

The bilinear connections $h_{IJ}$ and $g_{AB}$ impact the normalization of electroweak gauge bosons and the Higgs, which affects how electroweak observables are related to Lagrangian parameters (point iii.) above). In addition, $h_{IJ}$ modifies the coupling strength of SM-like $hVV$ vertex, which enters in topologies (a) and (d) from Fig.~\ref{htoZlldiagrams}, \ref{htoWW diagrams}, while $g_{AB}$ and $k^A_{IJ}$ introduce $hVV$ vertices with novel Lorentz structure and generate the non-SM vertex $hZA$ in topology (c). The fermion connection $L_{J,A}^{\psi,pr}$ shifts the fermion gauge vertices $\bar{\ell}\ell Z$, $\ell\nu W$ that appear in all topologies from their SM values without changing the structure. Furthermore, the fermion connection also generates a subset of the contributions to the contact vertices $ff' Vh$ present in diagrams (b), (e), and (f). 

Finally, the dipole connection $d_A^{e,pr}$ contributes to the $ff' V$ and $ff'Vh$ vertices. Usually, dipole operators are neglected because these operators have a different chirality structure than the SM and therefore do not interfere ($\mathcal O(1/\Lambda^2)$) if the leptons are massless. However, once $\mathcal{O}(1/\Lambda^4)$ effects are considered, dimension-six dipole operators can contribute as squared terms -- either multiplying themselves or other dimension-six SMEFT effects. One can still neglect dipole operators by assuming a $U(3)^5=U(3)_Q\times U(3)_u\times U(3)_d\times U(3)_L \times U(3)_e$ flavor symmetry, but in this work we don't make such an assumption. Instead, we assume a $U(3)_Q\times U(3)_u\times U(3)_d\times [U(1)_{L+e}]^3$, i.e. lepton-flavor diagonal, which allows dipoles to enter. In the following, we will only consider dimension-six dipole operators, since to dimension-eight dipole operators do not interfere with the SM (for the same reasons dimension-six do not) and therefore cannot appear in the square amplitude at $\mathcal O(1/\Lambda^4)$. 

The four-point (and higher) vertices generated by Eq.~\eqref{eq:Lgeo} come from expanding out the connections about the Higgs vev. For example, using $\langle \rangle$ to indicates the vacuum expectation value and expanding $L_{J,A}(\phi) = \langle L_{J,A}(\phi )\rangle + \langle \frac{\partial L_{J,A}(\phi)}{\partial \phi_K} \rangle \phi_K + \cdots$, the $\phi_K$ combines with $(D_\mu\phi)(\bar{\psi}_p\gamma^\mu\tilde{\sigma}^A\psi_r)$ to give a four particle interaction. This means geoSMEFT only includes four-point vertices from operators that {\em also} generate a two or three-point vertex (the $\langle L_{J,A}(\phi )\rangle$ term in the example above. To get the full set of four-point vertices, we need to add the contributions from operators that do not fit into one of the geoSMEFT connections (and therefore, by definition, cannot contribute to two or three-point vertices at tree level). The complete set of dimension-eight operators can be found in \cite{Li:2020gnx,Murphy:2020rsh}, however neither of these sets fits exactly into geoSMEFT -- meaning that they have not utilized integration by parts and equation of motion redundancies to their full extent to minimize the number of operators contributing to two and three-point vertices. A geoSMEFT-compatible basis for dimension eight was established in Ref. \cite{Corbett:2023yhk}, along with the set of dimension eight operators generating four point vertices for $\bar f f \to h V(\bar f' f')$. As this process is related to our process by crossing, the operator set, shown in Table \ref{non-geoSMEFT operators}, is the same~\footnote{We do not include dimension eight operators of the class $H^2\psi^4$ as these only contribute via five-point contact terms involving the Higgs and four fermions. As explained earlier, this type of contribution is negligible, either as a consequence of analysis cuts on the dilepton invariant or because the contact terms have a very different kinematic structure than the SM pieces they interfere with. There are no dimension six operators that generate five-point vertices.}.

\begin{table}[t]\centering
\begin{tabular}{|l|l|}
\hline
\multicolumn{2}{|c|}{\textbf{non-geoSMEFT operators}} \\ \hline
\multicolumn{2}{|c|}{Class $\psi^2XH^2D$ $(\psi=l,e)$} \\ \hline
\multicolumn{2}{|c|}{\begin{tabular}{l l} $Q^{(1)}_{\psi^2BH^2D}=(\bar{\psi_p}\gamma^\nu\psi_r)D^\mu(H^\dag H)B_{\mu\nu}$ & $Q^{(3)}_{l^2BH^2D}=(\bar{l_p}\gamma^\nu\sigma^Il_r)D^\mu(H^\dag \sigma^I H)B_{\mu\nu}$ \\
$Q^{(1)}_{\psi^2WH^2D}=(\bar{\psi_p}\gamma^\nu\psi_r)D^\mu(H^\dag \sigma^IH)W^I_{\mu\nu}$ & $Q^{(3)}_{l^2WH^2D}=(\bar{l_p}\gamma^\nu \sigma^Il_r)D^\mu(H^\dag H)W^I_{\mu\nu}$ \\
 $Q^{(4)}_{l^2WH^2D}=i(\bar{l_p}\gamma^\nu \sigma^I l_r)(H^\dag \overleftrightarrow{D}^{\mu} H)W^I_{\mu\nu}$, & $Q^{(5)}_{\psi^2WH^2D}=\epsilon_{IJK}(\bar{l}_p\gamma^\nu\sigma^Il_r)D^\mu(H^\dag\sigma^J H)W^K_{\mu\nu}$ \\ 
\end{tabular}} \\ \hline
\multicolumn{2}{|c|}{Class $\psi^2H^2D^3$ $(\psi=l,e)$} \\ \hline
\multicolumn{2}{|c|}{\begin{tabular}{l l} 
\multicolumn{2}{l}{$Q^{(1)}_{\psi^2H^2D^3}=i(\bar{\psi_p}\gamma^\mu\psi_r)\left[ (D_\nu H)^\dag (D^2_{(\mu,\nu)} H)-(D^2_{(\mu,\nu)} H)^\dag (D_\nu H) \right ]$} \\
\multicolumn{2}{l}{$Q^{(3)}_{l^2H^2D^3}=i(\bar{l}_p\gamma^\mu\sigma^Il_r)\left[(D_\nu H)^\dag \sigma^I(D^2_{(\mu,\nu)} H)-(D^2_{(\mu,\nu)} H)^\dag \sigma^I(D_\nu H) \right ]$}\\
\multicolumn{2}{l}{$Q^{(4)}_{l^2H^2D^3}=i(\bar{l_p}\gamma^\mu\sigma^I\overleftrightarrow{D}^{\nu}l_r)\left[ (D_\mu H)^\dag \sigma^I(D_\nu H)+(D_\nu H)^\dag \sigma^I(D_\mu H)\right ]$}
\end{tabular}} \\ \hline
\end{tabular}
\caption{Dimension Eight Operators contributing to Higgs decay to four leptons, not included in geoSMEFT. The subindex $\psi=e,l$, stands for the relevant fields for the process in consideration Flavor indices in the operator name to simplify the notation.}
\label{non-geoSMEFT operators}
\end{table}

\section{Results for $h\to\ell\bar{\ell}\left(Z\to\ell'\bar{\ell'}\right)$ }\label{section 3}

\subsection{Amplitude}\label{3.1}

In this section, we present the spinor-helicity amplitude for  $h(p)\xrightarrow{}\ell(p_1)\bar{\ell}(p_2)\left(Z\xrightarrow{}\ell'(p_3)\bar{\ell'}(p_4)\right)$. We classify the contributions to the amplitude into two types, based on the helicities of the final-state fermions. The first type includes contributions with helicity configurations that match those of the Standard Model (SM) in the massless lepton limit, and can therefore interfere with the SM amplitude. We refer to these as \textit{interfering helicity} contributions. The second type, \textit{non-interfering helicity} contributions, consists of amplitudes with helicity configurations that differ from those in the SM. For operators with dimension $\le$ eight, only the dipole operators contribute to the non-interfering helicity type -- all other contributions fall into the interfering category

\subsubsection{Interfering contribution}

After introducing the classification of the amplitude terms, we now show the full interfering contribution to the amplitude. We take all fermions to be massless and all momenta outgoing. With these convention, the allowed helicity configurations for the dilepton pairs are the usual right-right and left-left chiralities. We decompose the amplitude into contributions from each topology, referred to as subamplitudes. In this case, there are three subamplitudes:  $\mathcal{A}_{hZZ}$, $\mathcal{A}_{hZ\psi}$, and $\mathcal{A}_{hZA}$, corresponding to topologies (a), (b), and (c) in Fig.~\ref{htoZlldiagrams}, named after the Higgs vertex involved in each topology. To better organize the SMEFT effects, we choose to present each subamplitude in the following form:
\begin{equation}                       
\mathcal{A}_I(p,p_1^{h_1},p_2^{h_2},p_3^{h_3},p_4^{h_4})=\sum_i G_I^{(i)}(\{C^{(d)}_j\}) A^{(i)}_I(p,p_1^{h_1},p_2^{h_2},p_3^{h_3},p_4^{h_4}),
\label{amplitude decomposition}
\end{equation}
where $\mathcal{A}_I$ is the subamplitude for a particular topology $I$, with a fixed helicity combination $(h_1,h_2,h_3,h_4)$. All the kinematic dependence is contained in the kinematic structures $A^{(i)}_I$, while the coupling factors $G_I^{(i)}$ encapsulate the Wilson coefficients dependence. This form is convenient as it makes it easy to spot the order in the SMEFT expansion where different effects arise.

\begin{table}[t!]
\centering
\renewcommand{\arraystretch}{1.5}
\begin{tabular}{|l|l|l|l|l|}
\hline
\multicolumn{5}{|c|}{\textbf{$\mathcal{A}(h\xrightarrow{}\ell\bar{\ell}(Z\xrightarrow{}\ell'\bar{\ell'})$})} \\ \hline
Subamplitude & \makecell{Kinematic \\  Structure}          & Coupling Factor                        & SMEFT order               & Generating Operators\\ \hline
\multirow{3}{*}{$\mathcal{A}_{hZZ}$} & $A^{(1)}_{hZZ}$ & $g^{Z\psi}g^{Z\psi}c^{(1)}_{hZZ}$ &$\mathcal{O}(1)$           & \makecell{SM , $h_{IJ}$}\\ \cline{2-5}
 & $A^{(2)}_{hZZ}$    &$g^{Z\psi}g^{Z\psi} c^{(2)}_{hZZ}$ &$\mathcal{O}(1/\Lambda^2)$ & \makecell{$g_{AB}$}  \\ \cline{2-5}
 & $A^{(3)}_{hZZ}$    &$g^{Z\psi}g^{Z\psi}c^{(3)}_{hZZ}$ &$\mathcal{O}(1/\Lambda^4)$ & \makecell{$k^A_{IJ}$} \\ \hline
\multirow{4}{*}{$\mathcal{A}_{hZ\psi}$} & $A^{(1)}_{hZ\psi}$ &$g^{Z\psi} g^{(1)}_{hZ\psi}$            &$\mathcal{O}(1/\Lambda^2)$ & \makecell{$Q^{1,(6+2n)}_{H\psi}$, $Q^{2,(8)}_{Hl}$, \\ $Q^{3,(6+2n)}_{Hl}$}\\ \cline{2-5}
 & $A^{(2)}_{hZ\psi}$ &$g^{Z\psi} g^{(2)}_{hZ\psi}$            &$\mathcal{O}(1/\Lambda^4)$ & \makecell{$Q^{(1)}_{\psi^2VH^2D}$, $Q^{(3)}_{l^2VH^2D}$\\$Q^{(1)}_{\psi^2H^2D^3}$, $Q^{(3)}_{l^2H^2D^3}$}\\ \cline{2-5}
 & $A^{(3)}_{hZ\psi}$ &$g^{Z\psi} g^{(3)}_{hZ\psi}$            &$\mathcal{O}(1/\Lambda^4)$ & \makecell{$Q^{(1)}_{\psi^2VH^2D}$, $Q^{(3)}_{l^2VH^2D}$\\$Q^{(1)}_{\psi^2H^2D^3}$, $Q^{(3)}_{l^2H^2D^3}$}\\ \cline{2-5}
 &$A^{(4)}_{hZ\psi}$ &$g^{Z\psi} g^{(4)}_{hZ\psi}$            &$\mathcal{O}(1/\Lambda^4)$ & \makecell{$Q^{(1)}_{\psi^2H^2D^3}$, $Q^{(3)}_{l^2H^2D^3}$}\\ \hline
\multirow{2}{*}{$\mathcal{A}_{hZA}$} &$A^{(1)}_{hZA}$    &$eg^{Z\psi} c^{(1)}_{hZA}$              &$\mathcal{O}(1/\Lambda^2)$ & \makecell{$g_{AB}$}   \\ \cline{2-5}
 &$A^{(2)}_{hZA}$    &$eg^{Z\psi} c^{(2)}_{hZA}$              &$\mathcal{O}(1/\Lambda^4)$ & \makecell{$k^A_{IJ}$} \\ \hline
\end{tabular}
\caption{Subamplitudes contributing to $\mathcal{A}(h\xrightarrow{}\ell\bar{\ell}(Z\xrightarrow{}\ell'\bar{\ell'})$ . To the right of each subamplitude we find the contributing  kinematic structures next to it's respective coupling factor. Third column specifies the arising SMEFT order, meaning the lowest SMEFT order the kinematic structure appears. Fourth column indicates the operators that generates the kinematic structure, meaning that at least one of this operators is necessary to generate the corresponding kinematic structures. When referred to a particular geoSMEFT connection, it is understood that any of the operators contributing to that connection can generate the respective kinematic structure. By SM we mean that no SMEFT operators are neccessary to generate this kinematic structure, however it can get SMEFT corrections. Above $V=B,W$, and $\psi=e,l$.  }
\label{coupling factors table h->ZZ}
\end{table}

Table \ref{coupling factors table h->ZZ} summarizes the amplitude decomposition for$h(p)\xrightarrow{}\ell(p_1)\bar{\ell}(p_2)\left(Z\xrightarrow{}\ell'(p_3)\bar{\ell'}(p_4)\right)$. The coupling factors accompanying the kinematic structures are given by the products of the various effective couplings appearing in the subamplitude. The effective couplings are defined as the coefficients multiplying the different Lorentz structures. The expansion of these effective couplings in terms of Wilson coefficients has been worked out in \cite{Corbett:2023yhk, Hays:2020scx}. For completeness, we include these expressions in Appendix \ref{effective couplings appendix}. Here we present the effective couplings in terms of the Feynman Rules, extracted for the three-point vertices from \cite{Helset:2020yio}, and for contact terms by implementing the operators in \textit{FeynRules} \cite{Christensen:2008py,Alloul:2013bka}. These results were cross-checked in the dimension-six limit against \cite{Dedes:2017zog}. Table \ref{coupling factors table h->ZZ} contains the following effective couplings:

\begin{itemize}
    \item $g^{Z\psi}$ is the coupling of $Z$ with left-handed leptons for $\psi=l$, and right-handed leptons for $\psi=e$. This vertex has the same Lorentz structure as the SM, but the coupling strength gets corrections from SMEFT operators.
    \item $c^{(i)}_{hZZ}$ are the couplings of the different Lorentz structures involved in the $hZZ$ vertex. Explicitly, these Lorentz structures are: 
    \begin{align}
    \label{hZZ vertex}
            h(p)Z^\mu(k_1) Z^\nu(k_2): \ \ \ & ic^{(1)}_{hZZ}g^{\mu\nu}+ic^{(2)}_{hZZ}4(k_1^\nu k_2^\mu -k_1\cdot k_2 g^{\mu\nu})
             \\ 
            \nonumber &+ic^{(3)}_{hZZ}\left[(p\cdot k_1+p\cdot k_2)g^{\mu\nu}-p^\mu k_1^\nu -p^\nu k_2^\mu\right].
    \end{align}
    \item $g^{(i)}_{hZ\psi}$ are the couplings of the Lorentz structures for the 4-point contact vertex $hZ\psi \psi$.
    \begin{align}
    \label{hZpsipsi vertex}
            h(p)Z^\mu(k_2)\psi(p_1)\bar\psi(p_2): \ \ \ & ig^{(1)}_{hZ\psi}\gamma^\mu-ig^{(2)}_{hZ\psi}p \cdot k_2\gamma^\mu-ig^{(3)}_{hZ\psi}p^\mu \slashed{k_2}-ig^{(4)}_{hZ\psi}p^\mu \slashed{p}.
    \end{align}
    Similarly to $g^{Z\psi}$, $\psi=l,e$ denotes the coupling with left-handed and right-handed leptons. Notice (from Appendix \ref{effective couplings appendix}) the effective couplings $g^{(2)}_{hZ\psi}$, $g^{(3)}_{hZ\psi}$, and $g^{(4)}_{hZ\psi}$,  are the only effective couplings generated by non-geoSMEFT operators.
    \item $c^{(i)}_{hZA}$ are the couplings of the Lorentz structures for the SMEFT vertex $hZA$
    \begin{align}
    \label{hZA vertex}
            h(p)A^\mu(k_1) Z^\nu(k_2): \ \ \ & ic^{(1)}_{hZA}2(k_1^\nu k_2^\mu -k_1\cdot k_2 g^{\mu\nu})+ic^{(3)}_{hZA}(p\cdot k_1 g^{\mu\nu}-p^\mu k_1^\nu).
    \end{align}
    \item $e$ naturally is the electromagnetic coupling for $A\psi\psi$, this vertex remains unchanged by gauge invariance.
\end{itemize}
To illustrate the subamplitude decomposition (\ref{amplitude decomposition}) and the use of Table \ref{coupling factors table h->ZZ}, consider the subamplitude for the topology (a) in Figure \ref{htoZlldiagrams} ($\mathcal{A}_{hZZ}$) for all leptons left-handed (helicity combination $(h_1,h_2,h_3,h_4)=(-,+,-,+)$). According to Table \ref{coupling factors table h->ZZ}, this subamplitude contains three kinematic pieces, corresponding to the three different Lorentz structures in Eq.~\eqref{hZZ vertex}. Dressing each of the three kinematic structures with its corresponding coupling factors in Table \ref{coupling factors table h->ZZ}, the explicit subamplitude decomposition for $\mathcal{A}_{hZZ}$ is
\begin{align}
    \mathcal{A}_{hZZ}=&\left(\left[g^{Zl}\right]^2c^{(1)}_{hZZ}\right)A^{(1)}_{hZZ}+\left(\left[g^{Zl}\right]^2c^{(2)}_{hZZ}\right)A^{(2)}_{hZZ}+\left(\left[g^{Zl}\right]^2c^{(1)}_{hZZ}\right)A^{(3)}_{hZZ},
    \label{example}
\end{align}
where we have suppressed the momentum dependence. Note that the left-handed coupling appears squared, $\left[g^{Zl}\right]^2$, for this helicity configuration. One factor comes from the left-handed leptons with $(h_1, h_2)=(-,+)$, and the other from the leptons with $(h_3, h_4)=(-,+)$.

All the Wilson coefficient dependence is contained in the coupling factors, so the SMEFT order of the coupling factors indicates the order at which the respective kinematic structure arises. The minimal SMEFT order of the different coupling factors is summarized in the fourth column of Table \ref{coupling factors table h->ZZ}. We see that $A^{(1)}_{hZZ}$ arises at  SM level, $A^{(2)}_{hZZ}$ at $\mathcal{O}(1/\Lambda^2)$, and $A^{(3)}_{hZZ}$ at $\mathcal{O}(1/\Lambda^4)$. Therefore, when squaring $\mathcal{A}_{hZZ}$ and expanding to $\mathcal{O}(1/\Lambda^4)$ we need to keep: i.) $(A^{(1)}_{hZZ})^2$ and $(A^{(2)}_{hZZ})^2$, ii.) the interference between $A^{(1)}_{hZZ}$ and $A^{(2)}_{hZZ}$, and iii.) the interference between $A^{(1)}_{hZZ}$ and $A^{(3)}_{hZZ}$. Of course, in the full calculation we add all subamplitudes before squaring -- so we must also consider potential interference between $\mathcal A_{hZZ}$ and $\mathcal A_{hZ\psi}, \mathcal A_{hZA}$.

Now that we have explained the organization of the amplitude and the content of Table \ref{coupling factors table h->ZZ}, let's go into greater detail for an example helicity configuration. We chose $(h_2,h_2,h_3,h_4)=(-,+,-,+)$, this is, all leptons are taken as left-handed. In the following, we will express things in terms of spinor-helicity variables, using the conventions of \cite{Elvang:2013cua, Henn:2014yza} --  angle brackets $\langle  \ , \  \rangle$ for positive helicity fermions, and square brackets $[ \ , \ ]$ for negative helicity fermions.

We start with $\mathcal{A}_{hZZ}$ for topology (a), involving the $hZZ$ vertex already present in the SM. As we saw in the Eq.~\eqref{example}), once we include SMEFT effects there are three different kinematic pieces:
\begin{align}
\label{A1hZZ}
    A^{(1)}_{hZZ}(p,p_1^{-},p_2^{+},p_3^{-},p_4^{+})=& \ 2\frac{\hat{p}^{Z}_{12}}{s_{12}}\frac{\hat{p}^{Z}_{34}}{s_{34}}\langle 3 \gamma_\mu 4] \langle 1 \gamma^\mu2], \\
    \label{A2hZZ}
    A^{(2)}_{hZZ}(p,p_1^{-},p_2^{+},p_3^{-},p_4^{+})=& \ 4\frac{\hat{p}^{Z}_{12}}{s_{12}}\frac{\hat{p}^{Z}_{34}}{s_{34}}\Bigl( \langle 3 \slashed{k_1}4] \langle 1 \slashed{k_2}2]-(k_1\cdot k_2)\langle 3 \gamma_\mu 4] \langle 1 \gamma^\mu2] \Bigr), \\ 
    A^{(3)}_{hZZ}(p,p_1^{-},p_2^{+},p_3^{-},p_4^{+})=& \ \frac{\hat{p}^{Z}_{12}}{s_{12}}\frac{\hat{p}^{Z}_{34}}{s_{34}}\Bigl(2 \langle 3 \slashed{k_1}4] \langle 1 \slashed{k_2}2]-m_h^2\langle 3 \gamma_\mu 4] \langle 1 \gamma^\mu2]\Bigr).
\end{align}
where $k_1=p_1+p_2$ and $k_1=p_3+p_4$, and we have used the notation
\begin{equation}
    \nonumber
    s_{ij}=(p_i+p_j)^2, \ \ \  \ \ \ \ \hat{p}^{V}_{ij}=\frac{s_{ij}}{s_{ij}-m^2_V+i\Gamma_Vm_V}.
\end{equation}
$A^{(1)}_{hZZ}$ arises at $\mathcal{O}(1)$ and contains the SM part of $\mathcal{A}_{hZZ}$. It has the SM kinematic structure, so SMEFT effects only enter in the coupling factor. The second term, $A^{(2)}_{hZZ}$ arises at $\mathcal{O}(1/\Lambda^2)$ and contains non-SM kinematics coming from the $g_{AB}$ connection. More specifically, the the di-lepton momentum $k_1$ and $k_2$ appear inside spinor chains $\langle 3 \slashed{k_1}4]$ and $\langle 1 \slashed{k_2}2]$ along with the dot product $(k_1\cdot k_2)$. As $g_{AB}$ contains dimension-eight operators, the coupling factor also gets $\mathcal{O}(1/\Lambda^4)$ corrections. This can be compared with $A^{(3)}_{hZZ}$, which also has non-SM kinematics but  only contributes at $\mathcal{O}(1/\Lambda^4)$ (and can be traced to the $k^A_{IJ}$ connection);  $A^{(3)}_{hZZ}$ also features the dilepton momentum (in spinor chains $\langle 3 \slashed{k_1}4]$ and $\langle 1 \slashed{k_2}2]$) and $m_h^2$, which comes from momentum conservation and the fact that the Higgs is on-shell.

Moving on to $\mathcal{A}_{hZ\psi}$ to topology (b), all terms contain a $hZ\psi\psi$ contact vertex. There are four kinematic structures, corresponding to the four Lorentz structures in (\ref{hZpsipsi vertex}). These are given by
\begin{align}
    \label{A1hZpsi}
    A^{(1)}_{hZ\psi}(p,p_1^{-},p_2^{+},p_3^{-},p_4^{+})=& \ \frac{\hat{p}^Z_{34}}{s_{34}}\langle 3 \gamma_\mu 4] \langle 1 \gamma^\mu2], \\ \label{A2hZpsi}
    A^{(2)}_{hZ\psi}(p,p_1^{-},p_2^{+},p_3^{-},p_4^{+})=& \ -\frac{\hat{p}^Z_{34}}{s_{34}}(p\cdot k_2)\langle 3 \gamma_\mu 4] \langle 1 \gamma^\mu2],  \\ \label{A3hZpsi}
    A^{(3)}_{hZ\psi}(p,p_1^{-},p_2^{+},p_3^{-},p_4^{+})=& \ \frac{\hat{p}^Z_{34}}{s_{34}}\langle 3 \slashed{k_1}4] \langle 1 \slashed{k_2}2],\\ \label{A4hZpsi}
    A^{(4)}_{hZ\psi}(p,p_1^{-},p_2^{+},p_3^{-},p_4^{+})=& \ -\frac{\hat{p}^Z_{34}}{s_{34}}\langle 3 \slashed{k_1}4] \langle 1 \slashed{k_2}2]. 
\end{align}
Where $p=p_1+p_2+p_3+p_4 \equiv k_1+k_2$. With the same logic as above, we can track the operators that generate these $A^{(i)}_{hZ\psi}$ and the SMEFT order they arise using Tables \ref{coupling factors table h->ZZ}.  Notice the absence of the propagator $p^Z_{12}$ and the appearance of the the Higgs momenta in  $A^{(2)}_{hZ\psi} \propto p\cdot k_2$. Using momentum conservation, this can be rewritten as  $p\cdot k_2=k_2^2+k_1\cdot k_2$, where $k_2^2$ is the di-lepton invariant mass. Interestingly, $A^{(3)}_{hZ\psi}$ and $A^{(4)}_{hZ\psi}$ have the same kinematics, up to a minus sign. In principle, these amplitudes could have been different as their Lorentz structures differ. However, they become identical (up to a sign) for this process after imposing momentum conservation.

Finally, we have topology (c) ($\mathcal A_{hZA}$), which contains an $hZ\gamma$ effective vertex. This topology also gets a SM loop contributions, which we neglect here. There are two different Lorentz structures in (\ref{hZA vertex}) leading to two distinct kinematic structures:
\begin{align}
    A^{(1)}_{hZA}(p,p_1^{-},p_2^{+},p_3^{-},p_4^{+})=& \ 2\frac{1}{s_{12}}\frac{\hat{p}^Z_{34}}{s_{34}}\bigl( \langle 3 \slashed{k_1}4] \langle 1 \slashed{k_2}2]-(k_1\cdot k_2) \langle 3 \gamma_\mu 4] \langle 1 \gamma^\mu2]\bigr),\label{A1hZA} \\ 
    A^{(2)}_{hZA}(p,p_1^{-},p_2^{+},p_3^{-},p_4^{+})=& \ \frac{1}{s_{12}}\frac{\hat{p}^Z_{34}}{s_{34}}\bigl( \langle 3 \slashed{k_1}4] \langle 1 \slashed{k_2}2]-(k_1\cdot p) \langle 3 \gamma_\mu 4] \langle 1 \gamma^\mu2]\bigr). \label{A2hZA}
\end{align}
Note the presence of the photon propagator $1/s_{12}$. The rest of the non-zero helicity configurations in the interfering category are $(h_1,h_2,h_3,h_4)=(-,+,+,-), (+,-,-,+)$ for two right-handed and two left-handed leptons, and $(h_1,h_2,h_3,h_4)=(+,-,+,-)$ for all right-handed leptons. The subamplitudes for these configurations can be obtained from equations (\ref{A1hZZ})-(\ref{A2hZA}) with the appropriate angle and square bracket replacements. Particularly, to make a left-handed spinor chain into a right-handed chain, the replacement is  
\begin{equation}
    \label{replacement 1}
    \langle i \gamma^\mu j] \xrightarrow{} [i \gamma^\mu j \rangle.
\end{equation}

\subsubsection{Non-Interfering contribution}

Now we turn to the non-interfering contribution, which arises solely from operators in the dipole $d_A^e$ connection. 

Each of  $\mathcal{A}_{hZZ}$, $\mathcal{A}_{hZ\psi}$, and $\mathcal{A}_{hZA}$ receive a dipole contribution, denoted as $A^{dipole}_{hZZ}$, $A^{dipole}_{hZ\psi\psi}$, and $A^{dipole}_{hZA}$, respectively. We display the possible helicity combinations for these dipole kinematic structures in Table \ref{dipole helicities hZZ} below, along with their coupling factors. We have only included contributions at the amplitude of $\mathcal{O}(1/\Lambda^2)$ or lower, as $\mathcal{O}(1/\Lambda^4)$ dipole contribution won't interfere with the SM in the squared amplitude and therefore can't contribute to the Higgs width at $\mathcal O(1/\Lambda^4)$. 

\begin{table}[t!]
\centering
\renewcommand{\arraystretch}{1.5}
\begin{tabular}{|l|l|l|}
\hline
\multicolumn{3}{|c|}{\textbf{Dipole contribution to $h\xrightarrow{}\ell\bar{\ell}Z(\xrightarrow{}\ell'\bar{\ell'})$}}          \\ \hline  
Helicity    & $A^{dipole}_{hZZ}$        &$A^{dipole}_{hZ\psi\psi}$   \\ \hline
$(+,+,-,+)$ &$c^{(1)}_{hZZ}g^{Zl}d^*_Z$ &$g^{ZL}d^*_{hZ}$            \\ \hline
$(+,+,+,-)$ &$c^{(1)}_{hZZ}g^{Ze}d^*_Z$ &$g^{Ze}d^*_{hZ}$            \\ \hline
$(-,-,-,+)$ &$c^{(1)}_{hZZ}g^{Zl}d_Z$   &$g^{ZL}d_{hZ}$              \\ \hline
$(-,-,+,-)$ &$c^{(1)}_{hZZ}g^{Ze}d_Z$   &$g^{Ze}d_{hZ}$              \\ \hline
$(-,+,+,+)$ &$c^{(1)}_{hZZ}g^{Zl}d^*_Z$ &$\mathcal{O}(1/\Lambda^4)$  \\ \hline
$(+,-,+,+)$ &$c^{(1)}_{hZZ}g^{Ze}d^*_Z$ &$\mathcal{O}(1/\Lambda^4)$  \\ \hline
$(-,+,-,-)$ &$c^{(1)}_{hZZ}g^{Zl}d_Z$   &$\mathcal{O}(1/\Lambda^4)$  \\ \hline
$(+,-,-,-)$ &$c^{(1)}_{hZZ}g^{Ze}d_Z$   &$\mathcal{O}(1/\Lambda^4)$  \\ \hline
\end{tabular}
\caption{Dipole kinematic structures with their respective coupling factors for a particular helicity combination. The coupling factor stated as $\mathcal{O}(1/\Lambda^4)$ means that this particular helicity combination doesn't contribute to the square amplitude to the order we are working. }
\label{dipole helicities hZZ}
\end{table}
In Table \ref{dipole helicities hZZ} we introduced the following couplings, whose expansion in terms of Wilson coefficients can be found in Appendix \ref{effective couplings appendix}:
\begin{itemize}
    \item $d_{Z}$ is the dipole coupling of the $Z\psi\psi$ vertex
    \begin{align}\label{Zpsipsi dipole vertex}
    Z^\mu(k_2)\ell(p_3)\bar{\ell}(p_4) \  :  \ \  -d_Z^*k_{2,\nu}\sigma^{\mu\nu}P_L-d_Zk_{2,\nu}\sigma^{\mu\nu} P_R .
    \end{align}
    Notice that $d_z$ couples right-handed with left-handed leptons. This allows for left-right or right-left helicity combinations for the dilepton pairs.
    \item $d_{hZ}$ is the dipole coupling for the contact term $hZ\psi\psi$
    \begin{align}\label{hZpsipsi dipole vertex}
    h(p)Z^\mu(k_2)\ell(p_3)\bar{\ell}(p_4) \  :  \ \  -d_{hZ}^*k_{2,\nu}\sigma^{\mu\nu}P_L-d_{hZ}k_{2,\nu}\sigma^{\mu\nu} P_R .
    \end{align}
    Similarly, $d_{hZ}$ couples right-handed with left-handed leptons. At dimension six, $d_Z$ and $d_{hZ}$ are related by $d_{hZ}=vd_Z$, which further simplifies the analytic expression of the differential decay rate.
\end{itemize}

We have omitted $A^{dipole}_{hZA}$ in Table \ref{dipole helicities hZZ} because all helicity combinations involve at least one dipole vertex {\em and} an $hZA$ effective vertex. Thus, $A^{dipole}_{hZA}$ contributes at least to $\mathcal{O}(1/\Lambda^4)$ in the amplitude. The same logic applies to the last four combinations of $A^{dipole}_{hZ\psi}$ -- they require a dipole vertex $Z\ell\bar{\ell}$ and a $hZ\ell\bar{\ell}$ contact vertex, each of which are $\mathcal O(1/\Lambda^2)$. In comparison, the first four helicity combinations for $A^{dipole}_{hZ\psi}$ only involve the dipole contribution to the contact vertex $hZ\ell\bar{\ell}$, hence are $\mathcal{O}(1/\Lambda^2)$.\footnote{Also note that helicity combinations of the form $(h_1,h_2,h_3,h_4)=\pm(1,1,1,1)$ were not included. These combinations involve two dipole vertices, which contribute to $\mathcal{O}(1/\Lambda^4)$ to the amplitude level.}

Let us work out the non-zero helicity configurations in more detail, starting with \linebreak $(h_1,h_2,h_3,h_4)=(+,+,-,+)$. Keeping only $\mathcal{O}(1/\Lambda^2)$ terms, the dipole kinematic structures are:
\begin{align}
    \label{AhZZdipole}
    A^{dipole}_{hZZ}(p,p_1^+,p_2^+,p_3^-,p_4^+)&=\hat{p}^Z_{12}\hat{p}^Z_{34}\big([1\gamma_{\mu}\slashed{k}_1 2]-[1\slashed{k}_1\gamma_{\mu} 2]\big)\langle3\gamma^{\mu}4],\\ 
    \label{AhZpsidipole}
    A^{dipole}_{hZ\psi}(p,p_1^+,p_2^+,p_3^-,p_4^+)&=-\frac{1}{2}\hat{p}^Z_{34}\big([1\gamma_{\mu}\slashed{k}_2 2]-[1\slashed{k}_2\gamma_{\mu} 2]\big)\langle3\gamma^{\mu}4] .
\end{align}
The amplitude for $(h_1,h_2,h_3,h_4)=(+,+,+,-),(-,-,-,+),(-,-,+,-)$ can be obtained from Eq. \eqref{AhZZdipole} and \eqref{AhZpsidipole} using \eqref{replacement 1} and
\begin{equation}
    \label{replacement 2}
    [i\gamma_{\mu}\gamma_{\nu}j] \leftrightarrow \langle i\gamma_{\mu}\gamma_{\nu}j\rangle  
\end{equation}
to exchange a right-left spinor chain into a left-right spinor chain.

For helicity $(h_1,h_2,h_3,h_4)=(-,+,+,+)$, only diagram (a) gets a $\mathcal{O}(1/\Lambda^2)$ contribution:
\begin{align}
    \label{AhZZdipole2}
    A^{dipole}_{hZZ}(p,p_1^-,p_2^+,p_3^+,p_4^+)&=\hat{p}^Z_{12}\hat{p}^Z_{34}\big([3\gamma_{\mu}\slashed{k}_2 4]-[3\slashed{k}_2\gamma_{\mu} 4]\big)\langle1\gamma^{\mu}2],
\end{align}
The amplitudes for $(h_1,h_2,h_3,h_4)=(+,-,+,+),(-,+,-,-),(+,-,-,+)$  can be obtained from (\ref{AhZZdipole2}) with replacements (\ref{replacement 1}) and (\ref{replacement 2}). For these helicities, $A^{dipole}_{hZZ}$ won't interfere with  $A^{dipole}_{hZ\psi}$ to $\mathcal{O}(1/\Lambda^4)$. 
\subsection{Observables}\label{section 3.2}

Now that we have the full SMEFT amplitude for $h\to\ell\bar{\ell}\left(Z\xrightarrow{}\ell'\bar{\ell'}\right)$, we can compute observables and look for $\mathcal{O}(1/\Lambda^4)$ SMEFT  effects in the differential decay rate.

\subsubsection{Angular Distribution}\label{Section 3.2.1}

In the Higgs rest frame, in massless leptons limit, and the narrow-width approximation, the differential decay rate for $h(p)\xrightarrow{}\ell(p_1)\bar{\ell}(p_2)\left(Z(k_2)\xrightarrow{}\ell'(p_3)\bar{\ell'}(p_4)\right)$ is given by:
\begin{equation}
    \frac{d\Gamma}{ds d\cos \theta d\cos{\Delta}d\phi}=\frac{\lambda}{(2\pi)^5 2^{10}\sqrt{r}\Gamma_Z} \left|h\to\ell\bar{\ell}\left(Z\to\ell'\bar{\ell'}\right) \right|^2,
    \label{Angular Distribution h->ZZ}
\end{equation}
where 
\begin{equation}
    s=\frac{k_1^2}{m_h^2}, \quad r=\frac{m_Z^2}{m_h^2}\approx.53, \quad \lambda=\sqrt{1+s^2+r^2-2s-2r-2rs},
\end{equation}
and $\left|\mathcal{A}(h\to\ell\bar{\ell}\left(Z\to\ell'\bar{\ell'}\right))\right|^2$ is the full spin-summed, squared amplitude of the process \footnote{We used the input parameter scheme $\{\hat{G}_F,\hat{m}_Z,\hat{\alpha}\}$ with  $\hat{G}_F=1.1663787\times10^{-5}\text{ GeV}^{-2}$, $\hat{m}_Z=91.1976\text{ GeV}$ and $\hat{\alpha}(m_Z)=1/127.944$.}. Note that $s$ is associated with the momentum of the off-shell $Z$ boson, so it's constrained to the range $0\leq s \lesssim 0.075$. The details of the kinematics and the definitions of the angular variables $(\theta,\Delta,\phi)$ are discussed in Appendix \ref{kinematics appendix}. To compute the spin summed squared amplitude, each kinematic structure is dressed with its corresponding coupling factor, retaining all terms up to $\mathcal{O} (1/\Lambda^4)$ after squaring. To calculate and manipulate spinor products and Lorentz invariants, we used SpinorHelicity4D~\cite{AccettulliHuber:2023ldr} and FeynCalc~\cite{Shtabovenko:2023idz}. 

In this frame, the spin-summed, squared amplitude decomposes into the following linearly independent combination angular functions:\footnote{Comparing our expression (\ref{Angular Distribution h->ZZ}) with that in \cite{Buchalla:2013mpa}, at $\mathcal{O}(1/\Lambda^2)$ one finds $J_1^Z=J_2^Z$. At $\mathcal{O}(1/\Lambda^4)$, however, this degeneracy is broken by the dipole operators. The same happens with $J_3^Z$ and $J_4^Z$ at $\mathcal{O}(1/\Lambda^2)$, which allows one to trade $\cos^2{\theta}$ and $\cos^2{\Delta}$ in favor of $\sin^2{\theta}\sin^2{\Delta}$ by means of the identity $\cos^2{\theta}+\cos^2{\Delta}=\cos^2{\theta}\cos^2{\Delta}-\sin^2{\theta}\sin^2{\Delta}+1$. We have deliberately chosen to work in a basis without $\sin^2{\theta}\sin^2{\Delta}$ to highlight that  $J_1^Z\neq J_2^Z$ and $J_3^Z\neq J_4^Z$, at $\mathcal{O}(1/\Lambda^4)$. On the other hand, in \cite{Beneke:2014sba}, no trigonometric functions are traded in favor of the other one. Instead, the $J$-functions are massaged in such a way that $J_1^Z=J_2^Z=J_3^Z=J_4^Z$ and $\sin^2{\theta}\sin^2{\Delta}$ appears with a different $J$-Function. This is only possible due to the degeneracy $J_1^Z=J_2^Z$ and $J_3^Z=J_4^Z$ at $\mathcal{O}(1/\Lambda^2)$ but this is not possible at $\mathcal{O}(1/\Lambda^4)$ due to dipole operators.}

\begin{align}
     \label{J^Z functions}
    \left|\mathcal{A}(h\to\ell\bar{\ell}\left(Z\to\ell'\bar{\ell'}\right))\right|^2=
    &J^Z_1+J^Z_2 \cos^2{\theta}\cos^2{\Delta}\\ \nonumber 
    &+J^Z_3 \cos^2{\theta}+J^Z_4 \cos^2{\Delta}+J^Z_5\cos{\theta}\cos{\Delta} \\ \nonumber 
    &+(J^Z_6 \sin{\theta}\sin{\Delta}+J^Z_7 \sin{(2\theta)}\sin{(2\Delta)})\cos{\phi}\\  \nonumber
    &+J^Z_8\sin^2{\theta}\sin^2{\Delta}\cos{(2\phi)} 
\end{align}
In the above, $J^Z_i=J^Z_i(r,s)$, so that all the angular dependence has been factorized from these functions. We will refer to these as $J$-functions.  The explicit expressions for the $J$-functions are rather complex. Instead of displaying them here, we have summarized which operators contribute to each J-function in Table \ref{J^Z operators}. Their full expressions are given in Appendix \ref{J^Z explicit expressions}. 

\begin{table}[t!]
\centering
\begin{tabular}{|l|l|l|l|l|}
\hline
& \multirow{2}{*}{\centering{\textbf{$J$-Function}}}& \multicolumn{3}{c|}{\textbf{Contribution SMEFT Operators}} \\  \cline{3-5}
                   &  & \makecell{Dimension 6} & Dipole & \makecell{Dimension 8} \\ \hline
\multirow{7}{*}{\begin{sideways}SM-like Distribution\end{sideways}}  
& \makecell{$J^Z_1$, $J^Z_2$, $J^Z_3$, \\  $J^Z_4$, $J^Z_7$, $J^Z_8$} & \makecell{$Q_{H\Box}^{(6)}$,$Q_{HD}^{(6)}$,  \\$Q_{HV}^{(6)}$, $Q_{HWB}^{(6)}$,\\$Q_{H\psi}^{1,(6)}$, $Q_{Hl}^{3,(6)}$} & \makecell{$Q_{eB}^{(6)}$, $Q_{eW}^{(6)}$} & \makecell{$Q_{HD}^{(8)}$, $Q_{HD,2}^{(8)}$, \\$Q_{HV}^{(8)}$ $Q_{HWB}^{(8)}$,  $Q_{HW,2}^{(8)}$, \\ $Q_{HDHV}^{(8)}$, $Q_{HDHW,2}^{(8)}$ \\ $Q_{H\psi}^{1,(8)}$, $Q_{H}^{2,(8)}$, $Q_{Hl}^{\epsilon,(8)}$ \\ $Q^{(1)}_{\psi^2VH^2D}$, $Q^{(3)}_{l^2VH^2D}$\\$Q^{(1)}_{\psi^2H^2D^3}$, $Q^{(3)}_{l^2H^2D^3}$} \\ \cline{2-5}
&   \makecell{$J^Z_5$, $J^Z_6$} & \makecell{$Q_{H\Box}^{(6)}$,$Q_{HD}^{(6)}$,  \\$Q_{HV}^{(6)}$, $Q_{HWB}^{(6)}$,\\$Q_{H\psi}^{1,(6)}$, $Q_{Hl}^{3,(6)}$} &   & \makecell{$Q_{HD}^{(8)}$, $Q_{HD,2}^{(8)}$, \\$Q_{HV}^{(8)}$ $Q_{HWB}^{(8)}$,  $Q_{HW,2}^{(8)}$, \\ $Q_{HDHV}^{(8)}$, $Q_{HDHW,2}^{(8)}$ \\ $Q_{H\psi}^{1,(8)}$, $Q_{H}^{2,(8)}$, $Q_{Hl}^{\epsilon,(8)}$ \\ $Q^{(1)}_{\psi^2VH^2D}$, $Q^{(3)}_{l^2VH^2D}$\\$Q^{(1)}_{\psi^2H^2D^3}$, $Q^{(3)}_{l^2H^2D^3}$} \\ \hline
\end{tabular}
\caption{SMEFT operators contribution to J-functions. Being this a \textit{SM-like angular distribution}, all the $J$-functions are present without the need of SMEFT operators. Above, $V=W,B$, and $\psi=e,l$. }
\label{J^Z operators}
\end{table}

We are particularly interested in how the $\mathcal{O}(1/\Lambda^4)$ SMEFT effects modify the angular distribution (\ref{Angular Distribution h->ZZ}). This can manifest in two ways: 1.) by the appearance of trigonometric functions (with a respective $J$-function prefactor) that are not present in (\ref{J^Z functions}), for example, $\cos{(2\Delta)\cos{\theta}\cos{\phi}}$, or 2.) by modifying the relative size of the existing $J$-functions.  This first case corresponds to a \textit{non SM-like angular distribution} where new $J$-functions arise purely from SMEFT. This leads to a distribution with a different shape compared to the SM (one can think of it as introducing additional "valleys" and "hills" to the profile of the angular distribution). This situation, in principle, is more encouraging as SMEFT effects are easier to identify.  However, this is  not  what  we find for this decay. Instead, we find the second case, a \textit{SM-like angular distribution}, where the existing $J$-functions have a leading SM contribution but receive subleading SMEFT corrections. In this case, SMEFT operators don't change the overall shape of the distribution, only modify it's amplitude (think of this as the same number of "hills" and "valleys", but with different sizes). Despite these changes being more subtle, SMEFT effects on the $J$-functions can still be significant.

From Table \ref{J^Z operators} alone we can't directly determine how the operators modify the size of the $J$-functions. However, we can observe that the dipole operators $Q_{eB}^{(6)}$ and $Q_{eW}^{(6)}$ don't contribute to the functions $J_5^Z$ and $J_6^Z$, while all non-dipole operators contribute to every $J$-functions. Examining the explicit expressions for the $J$-Functions reveals that each depends differently on the various operators, this means that certain $J$-Functions are potentially more sensitive to specific operators. Therefore, identifying observables sensitive to individual $J$-functions will help us isolate the SMEFT effects and analyze their $\mathcal{O}(1/\Lambda^4)$ contributions.

\subsubsection{Asymmetries and Di-lepton Invariant Mass}\label{sec:hzzasymm}

In this subsection, we identify observables sensitive to the $J$-functions. In the literature, two types of observables have been used to extract the $J$-functions from (\ref{J^Z functions}): the dilepton invariant mass distribution and angular asymmetries \cite{Buchalla:2013mpa,Beneke:2014sba}. We explore both of those here, following Ref.~\cite{Beneke:2014sba} for definitions and naming convention. 

The dilepton invariant mass distribution is obtained by integrating the angular variables in (\ref{J^Z functions}). The result is:
\begin{equation}
    \frac{d\Gamma}{ds}=\frac{\lambda}{(2\pi)^5 2^{10}\sqrt{r}\Gamma_Z}\frac{8\pi}{9}(9 J_1^Z + J_2^Z + 3 J_3^Z+ 3 J_4^Z).
\end{equation}
Note that we use the dimensionless variable $s=k_1^2/m_h^2$ instead of $k_1^2$. We see that after integration, this observable allows us to study a reduced set of the $J$-functions.

Two of the asymmetries explored in \cite{Beneke:2014sba} ($\mathcal{A}^{(1)}_{\phi}$ and $\mathcal{A}^{(2)}_{\phi}$) are CP-odd and therefore vanish under our assumptions. The remaining asymmetries are: 
\begin{align}
  \mathcal{A}_{\theta,\Delta}= & \left(\frac{d\Gamma}{ds}\right)^{-1} \int_{-1}^{1} \sgn{(\cos\theta)}\left(\int_{-1}^{1} \sgn{(\cos{\Delta})}\frac{d \Gamma}{dsd\cos{\Delta}d\cos{\Delta}}d\cos{\Delta}\right)d\cos{\theta}\\  \nonumber
  =&\frac{9}{4}\frac{J_5^Z}{9 J_1^Z + J_2^Z + 3 J_3^Z+ 3 J_4^Z} \\
  \mathcal{A}^{(2)}_{\theta}= & \left(\frac{d\Gamma}{ds}\right)^{-1} \int_{-1}^{1} \sgn{(\cos{2\theta})}\frac{d\Gamma}{dsd\cos{\theta}} \, d\cos{\theta} \\ \nonumber
  = & 1 - \sqrt{2}+\frac{J_2^Z + 3 J_3^Z}{\sqrt{2}(9 J_1^Z + J_2^Z + 3 J_3^Z+ 3 J_4^Z)}\\
  \mathcal{A}^{(2)}_{\Delta}= & \left(\frac{d\Gamma}{ds}\right)^{-1} \int_{-1}^{1} \sgn{(\cos{2\Delta})}\frac{d\Gamma}{dsd\cos{\Delta}} \, d\cos{\Delta} \\ \nonumber
  = & 1 - \sqrt{2}+\frac{J_2^Z + 3 J_4^Z}{\sqrt{2}(9 J_1^Z + J_2^Z + 3 J_3^Z+ 3 J_4^Z)}
\end{align}    
\begin{align}
  \mathcal{A}^{(3)}_{\phi}= & \left(\frac{d\Gamma}{ds}\right)^{-1} \int_{0}^{2\pi} \sgn{(\cos{\phi})}\frac{d \Gamma}{dsd\phi} \, d\phi=\frac{9\pi}{8}\frac{J_6^Z}{9 J_1^Z + J_2^Z + 3 J_3^Z+ 3 J_4^Z}\\
  \mathcal{A}^{(4)}_{\phi}= & \left(\frac{d\Gamma}{ds}\right)^{-1} \int_{0}^{2\pi} \sgn{(\cos{(2\phi)})}\frac{d \Gamma}{dsd\phi} \, d\phi=\frac{8}{\pi}\frac{J_8^Z}{9 J_1^Z + J_2^Z + 3 J_3^Z+ 3 J_4^Z}
\end{align} 
Here, the sign function is defined as $\sgn(\pm |x|)=\pm1$. 

From this list, we focus on $\mathcal{A}^{(4)}_{\phi}$, $\mathcal{A}^{(2)}_{\theta}$ and $\mathcal{A}^{(2)}_{\Delta}$. We choose these because they are more sensitive to dipole and dimension eight effects (i.e. they don't depend on $J_5^Z$ and $J_6^Z$, which are insensitive to the dipoles). The observable $d\Gamma/ds$ was extensively explored in \cite{Buchalla:2013mpa} to $\mathcal{O}(1/\Lambda^2)$, while $\mathcal{A}_{\theta,\Delta}$ and $\mathcal{A}^{(3)}_{\phi}$ were studied to $\mathcal{O}(1/\Lambda^2)$ in \cite{Beneke:2014sba}. Although both studied the $\mathcal{O}(1/\Lambda^2)$ SMEFT effects, these were characterized by effective couplings rather than by specific Wilson coefficients. We want a more direct connection to the SMEFT operators, so we work with the Wilson coefficients instead. 

To explore the $\mathcal{O}(1/\Lambda^4)$ effects these operators induce on these asymmetries and $d\Gamma/ds$, we turn on dimension six and dimension eight operators one at a time, equivalent to zeroing all but one Wilson coefficient. To illustrate the range of possible SMEFT effects, we select one representative operator for each effective coupling. For the dipole contribution, we choose the operator $Q^{(6)}_{eW}$. The coupling $c^{(1)}_{hZZ}$ is represented by $Q^{(6)}_{HD}$, while $c^{(2)}_{hZZ}$ and $c^{(1)}_{hZA}$ correspond to $Q^{(6)}_{HW}$ and $Q^{(8)}_{HW}$, respectively. The couplings $c^{(3)}_{hZZ}$ and $c^{(2)}_{hZA}$ receive contributions only at dimension eight; for these we use $Q^{(8)}_{HDHW}$. We include the operator $Q^{(6)}_{He}$, which contributes to both $g^{Ze}$ and $g^{(1)}_{hZe}$, as well as the analogous $Q^{1,(6)}_{H\ell}$ for $g^{Z\ell}$ and $g^{(1)}_{hZ\ell}$. Finally, we choose  $\mathcal{Q}^{(1)}_{\psi^2H^2D^3}$ ($\psi=e,\ell$) for the effective couplings $g^{(2)}_{hZ\psi}$, $g^{(3)}_{hZ\psi}$, and $g^{(4)}_{hZ\psi}$.\footnote{For the coupling factor $c^{(1)}_{hZA}$ we have chosen two operators instead of one. The reason for this is that, even though they contribute with the same kinematics at the amplitude level, at the squared amplitude level the dimension six operator $Q^{(6)}_{HW}$ gets a radical enhancement, compared to the dimension eight operator $Q^{(8)}_{HW}$. This become will evident latter and will be explained deeper.} For those couplings getting both dimension six and dimension eight contribution, we have chosen a representative operator of dimension six (except for $c^{(2)}_{hZZ}$ and  $c^{(1)}_{hZA}$) as the dimension eight operator will contribute with the same kinematics but will be  suppressed by an additional factor of $v^2/\Lambda^2$ relative to the dimension six operator.

\begin{figure}[t]
  \begin{subfigure}[b]{0.5\textwidth}
    \includegraphics[width=\textwidth]{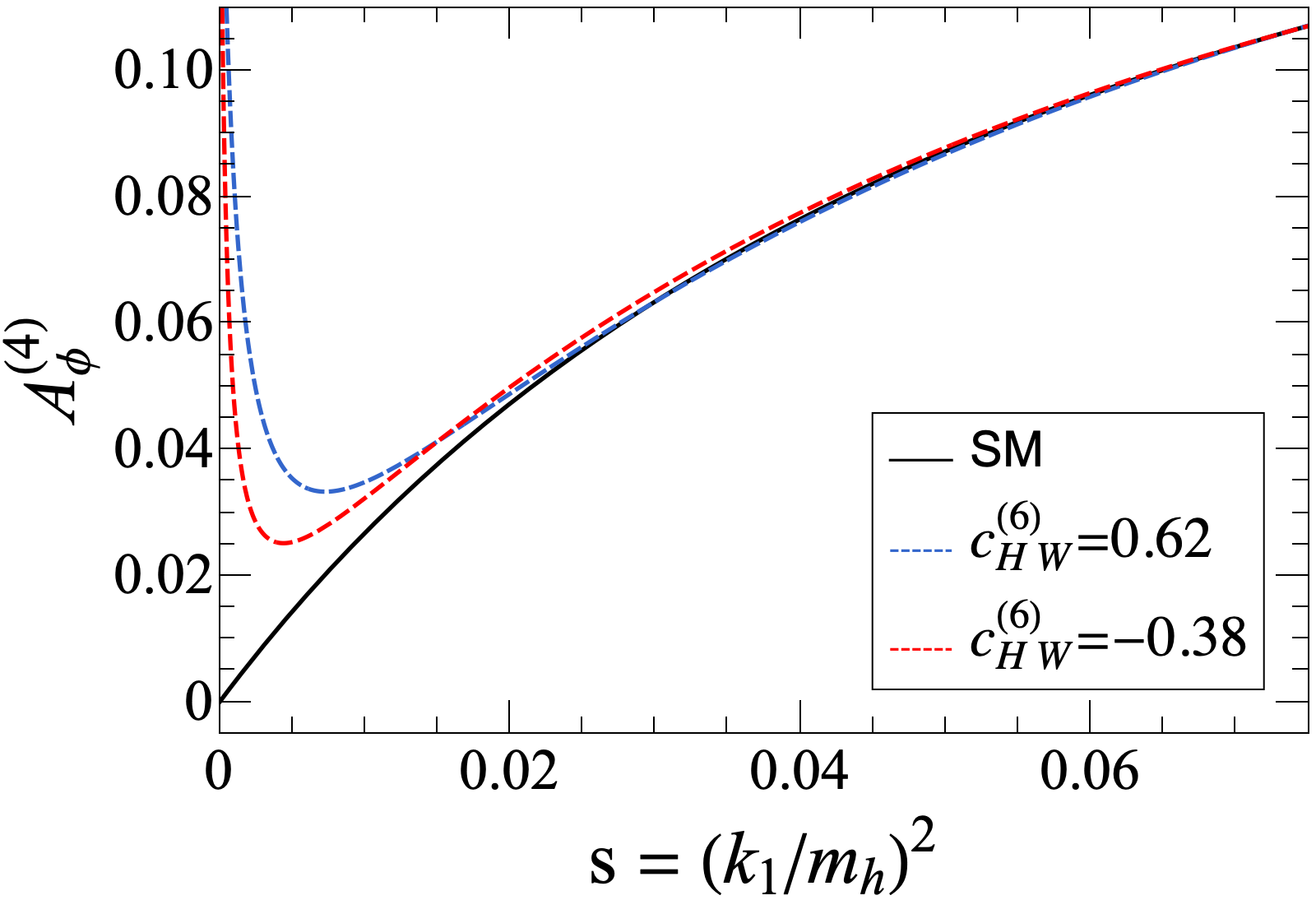}
    \subcaption{}
    \label{A4phi dim6:a}
  \end{subfigure}
    \begin{subfigure}[b]{0.5\textwidth}    \includegraphics[width=\textwidth]{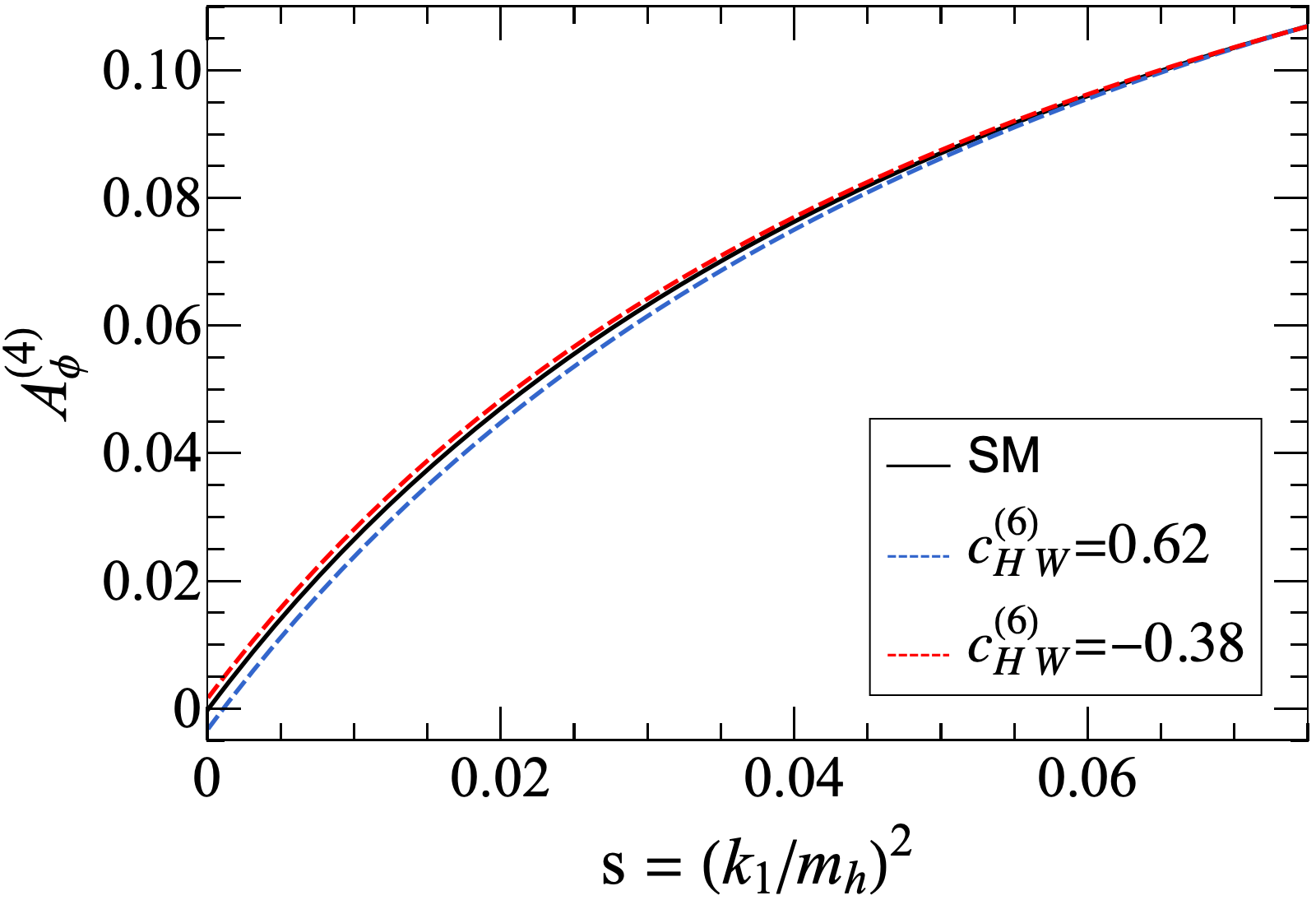}
    \subcaption{}
    \label{A4phi dim6:b}
  \end{subfigure}
  \begin{subfigure}[b]{0.5\textwidth}
    \centering
    \includegraphics[width=\textwidth]{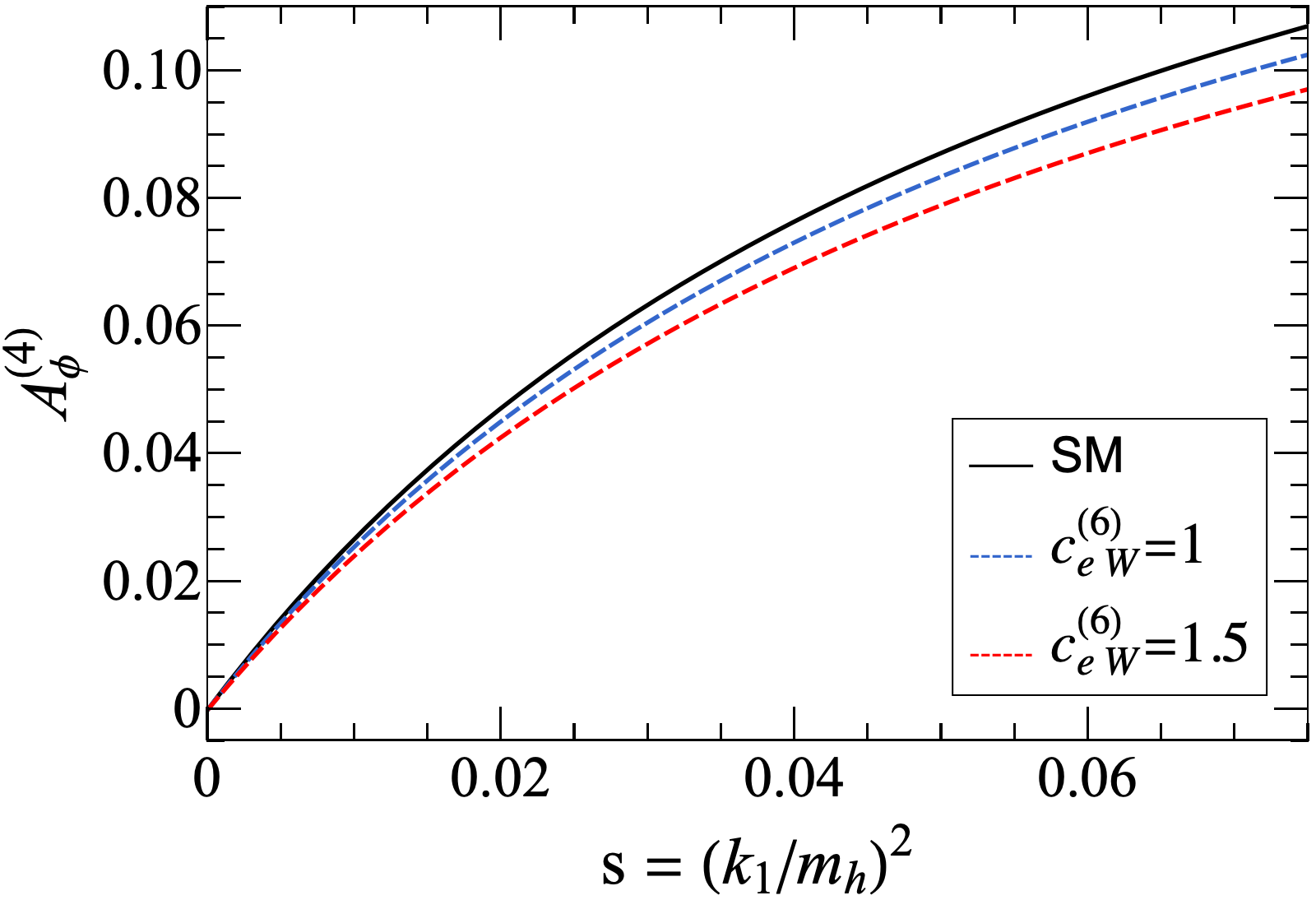}
    \subcaption{}
    \label{A4phi dim6:c}
  \end{subfigure}
   \begin{subfigure}[b]{0.5\textwidth}            
    \includegraphics[width=\textwidth]{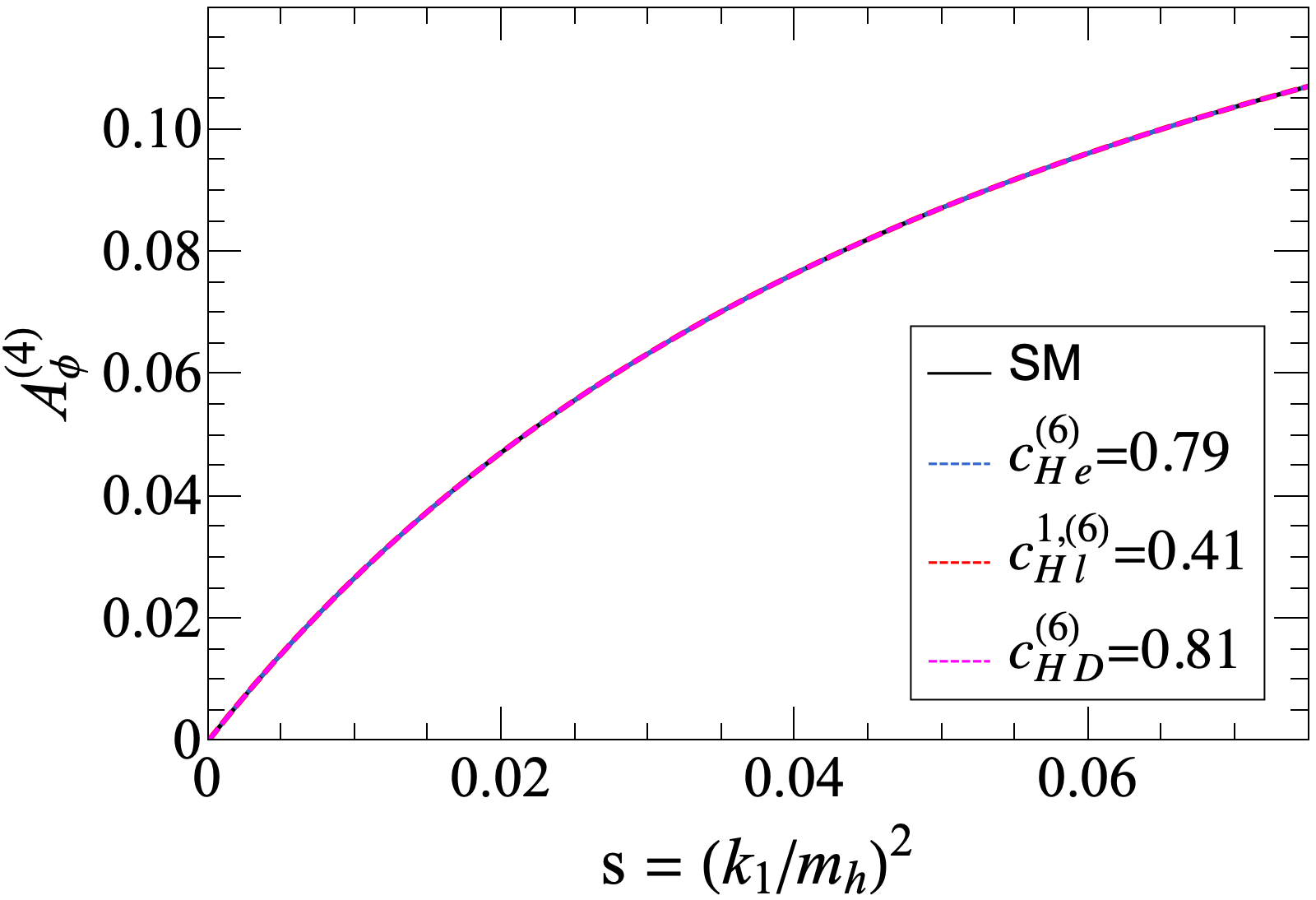}
    \subcaption{}
    \label{A4phi dim6:d}
  \end{subfigure}
  \caption{Dimension 6 SMEFT operators' contribution to $\mathcal{A}_\phi^{(4)}$. The solid black lines represent the SM with all Wilson coefficients set to zero. The other curves represent the operator's contribution by setting it's respective Wilson coefficient to the indicated value, taking $\Lambda=1\, \text{TeV}$,  and all other Wilson coefficients are set to zero.  In figure (a) we considered interference terms and squared terms of $Q_{HW}^{(6)}$, while in figure (b) we have only considered the interference  of $Q_{HW}^{(6)}$.}
  \label{A4phi dim6}
\end{figure}

\begin{figure}[!t]
\centering
  \begin{subfigure}[b]{0.48\textwidth}
    \includegraphics[width=\textwidth]{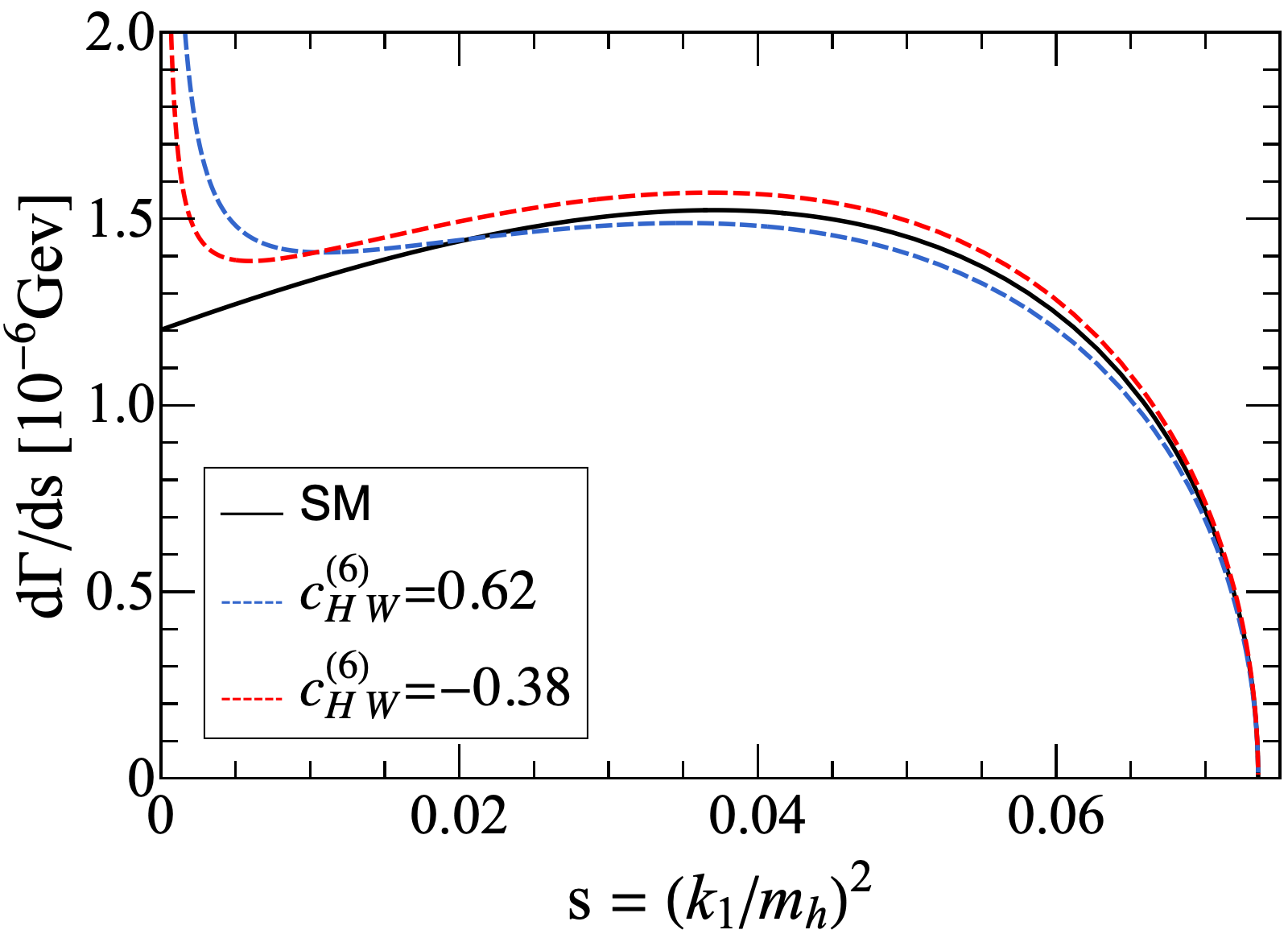}
    \subcaption{}
    \label{invmass cHW}
  \end{subfigure}
  \begin{subfigure}{0.48\textwidth}
    \centering
    \includegraphics[width=\textwidth]{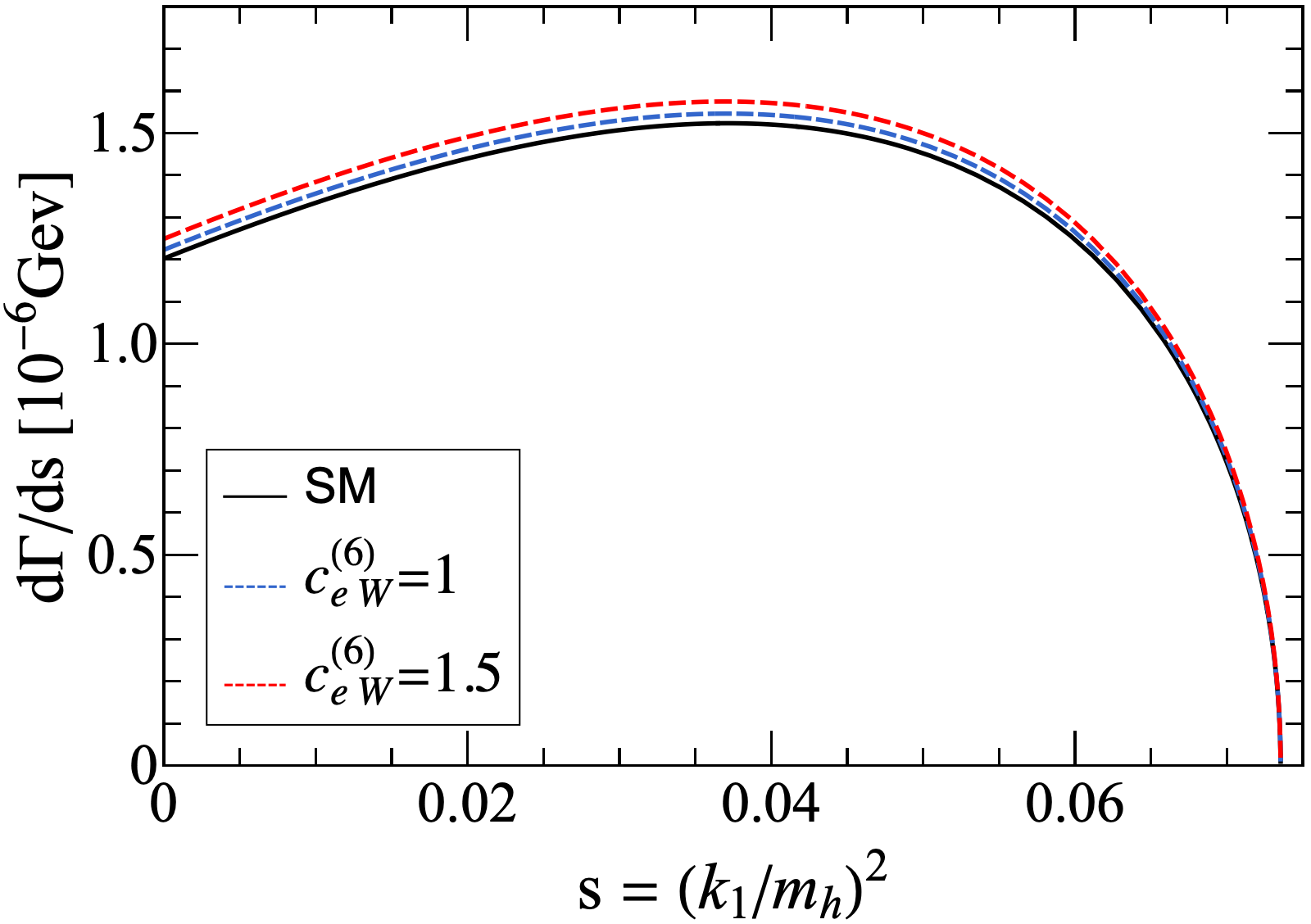}
    \subcaption{}
    \label{invmass dipole}
  \end{subfigure}
  \begin{subfigure}[b]{0.48\textwidth}          
    \includegraphics[width=\textwidth]{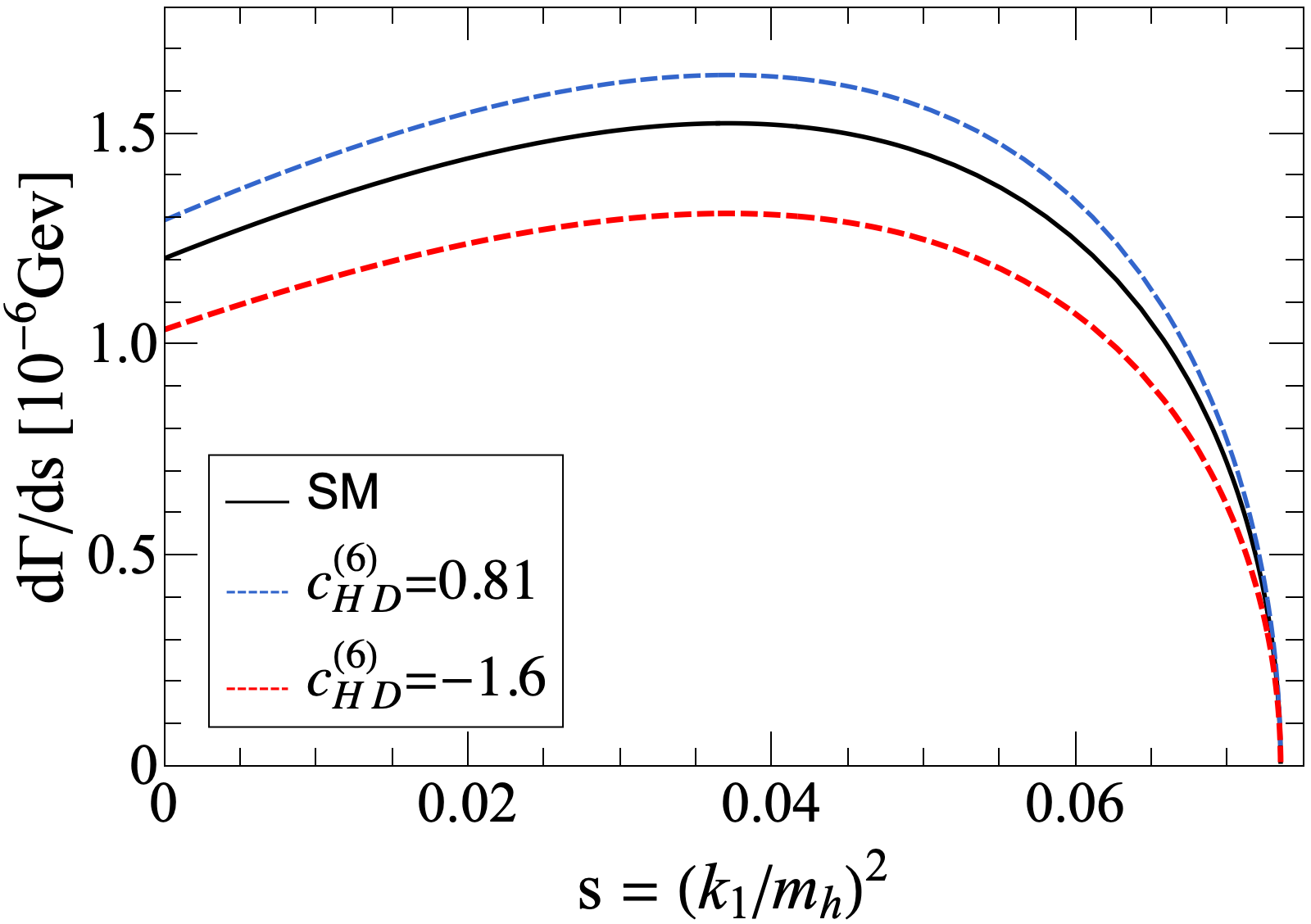}
    \subcaption{}
    \label{invmass cHD}
  \end{subfigure}
  \begin{subfigure}[b]{0.48\textwidth} 
    \centering
    \includegraphics[width=\textwidth]{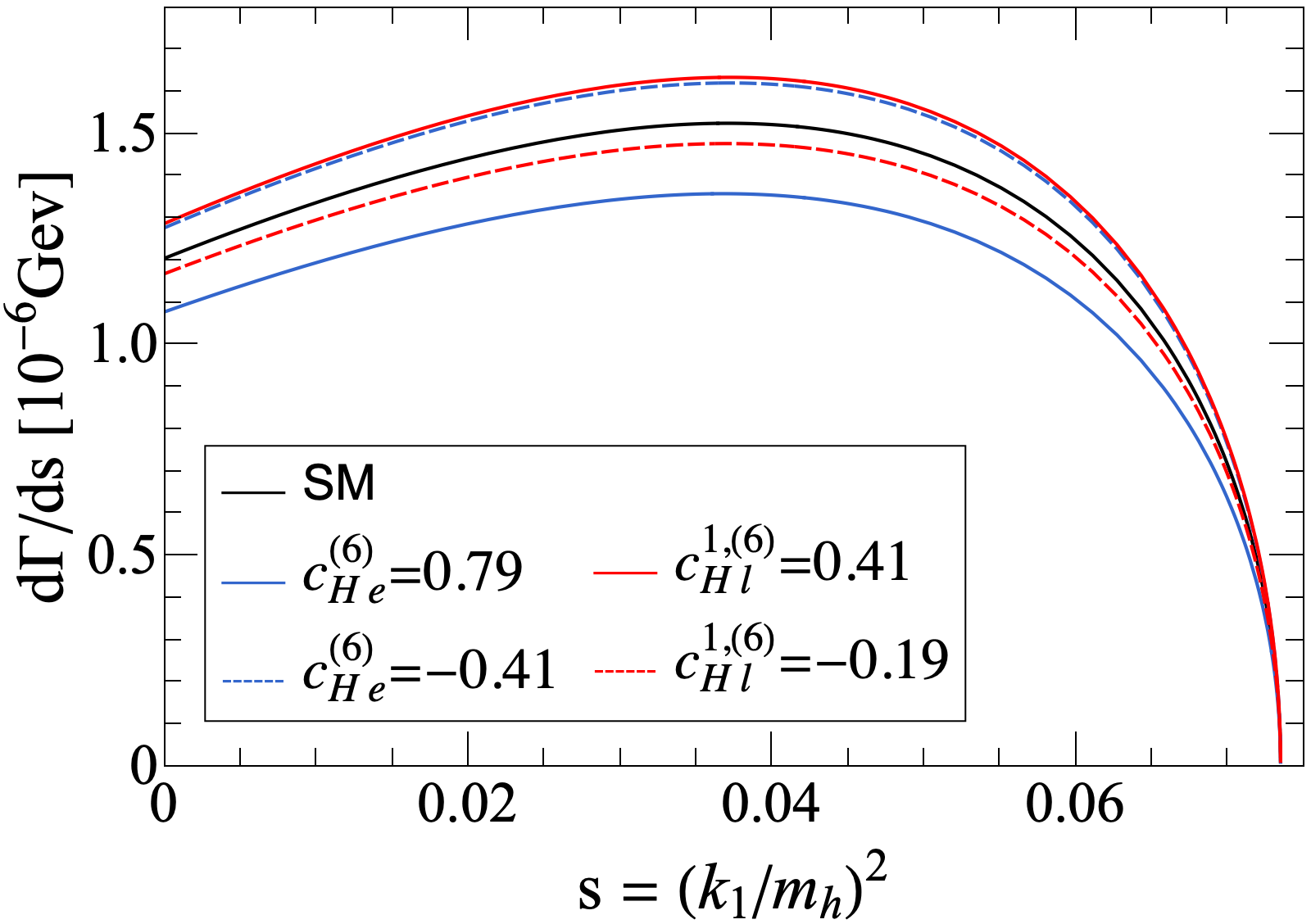}
    \subcaption{}
    \label{invmass cHpsi}
  \end{subfigure}
  \caption{Dimension 6 SMEFT operators' contribution to $d\Gamma/ds$.  The solid black lines represent the SM with all Wilson coefficients set to zero. The other curves represent the operator's contribution by setting it's respective Wilson coefficient to the indicated value, taking $\Lambda=1\, \text{TeV}$, and all other Wilson coefficients are set to zero. Notice in figure (d), solid lines represent positive values of the Wilson coefficients, and dashed lines represent negative values.}
  \label{invmass dim6}
\end{figure}

In Figures \ref{A4phi dim6}-\ref{invmass dim6}, we show the effect of dimension six operators by plotting $\mathcal{A}^{(4)}_{\phi}$, $\mathcal{A}^{(2)}_{\theta}$ and $\mathcal{A}^{(2)}_{\Delta}$ as a function of $s=k_1^2/m_h^2$, setting the respective Wilson coefficient to a non-zero value, while keeping the rest as zero. Non-dipole operators are constrained by electroweak precision observables, top, Higgs and diboson measurements in \cite{Ellis:2020unq}. Dipole operators are significantly constrained by electric and magnetic dipole measurements in \cite{Kley:2021yhn,Aebischer:2021uvt}. All of these constraints ignore dimension eight effects, and the possibility of cancellations among SMEFT effects. We take as guidelines the bounds for non-dipole operators, and set the Wilson coefficients to the upper and lower 95\% CL values according to \cite{Ellis:2020unq}. On the other hand, the bounds from \cite{Aebischer:2021uvt} on $Q_{eW}^{(6)}$ are so stringent that there is no effect in Higgs decay. To show the possible effect dipole operators have, we therefore relax these bounds. Of course, to be consistent with \cite{Aebischer:2021uvt}, this relaxed scenario implicitly assumes a cancellation is taking place in dipole observables or that the light fermion ($e/u/d$) Yukawa couplings are significantly different than their SM values~\cite{Brod:2013cka}.

\begin{figure}[t]
\centering
  \begin{subfigure}[b]{0.48\textwidth}
    \includegraphics[width=\textwidth]{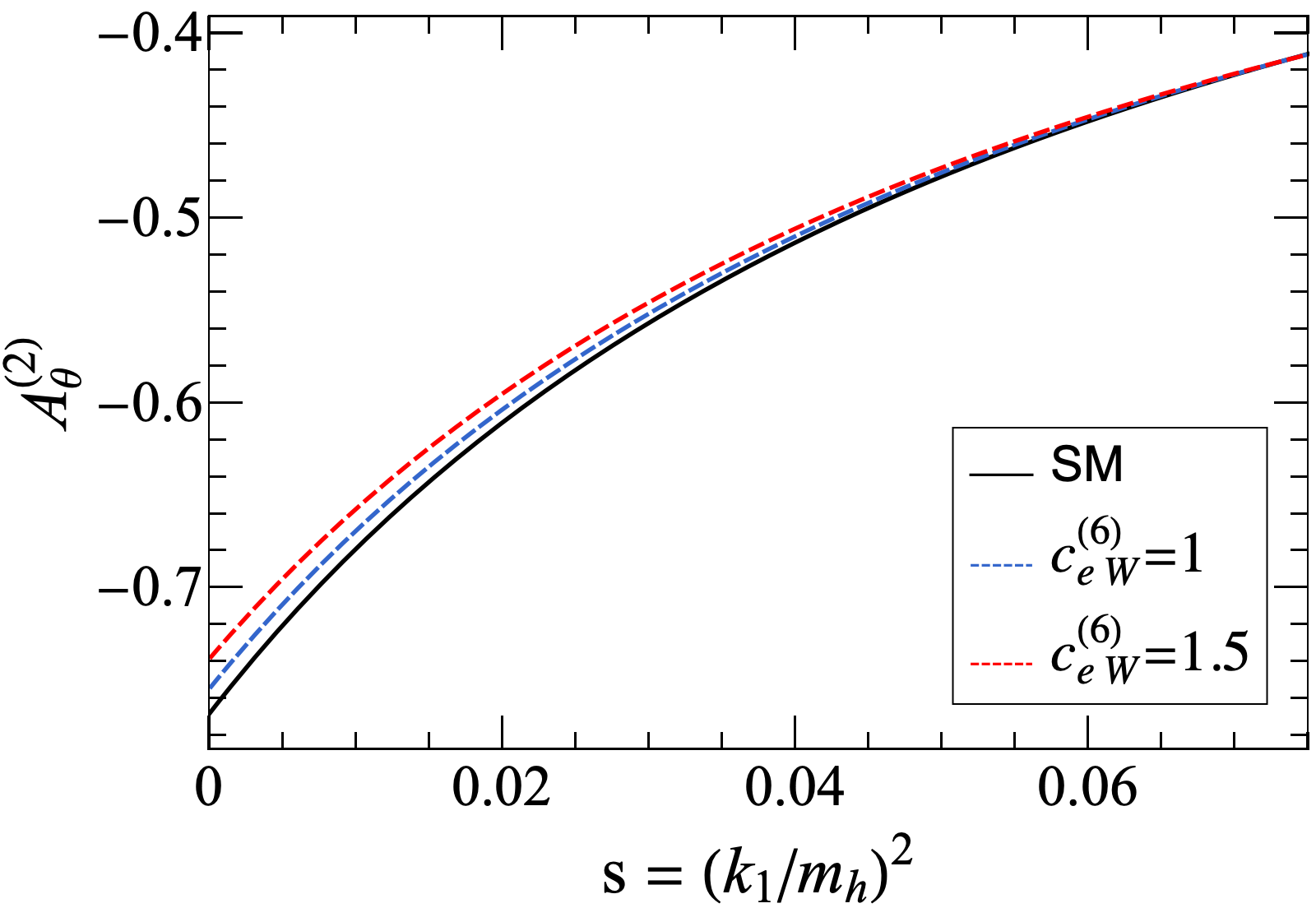}
    \subcaption{}
    \label{Atheta_dipole}
  \end{subfigure}
    \begin{subfigure}[b]{0.48\textwidth}    \includegraphics[width=\textwidth]{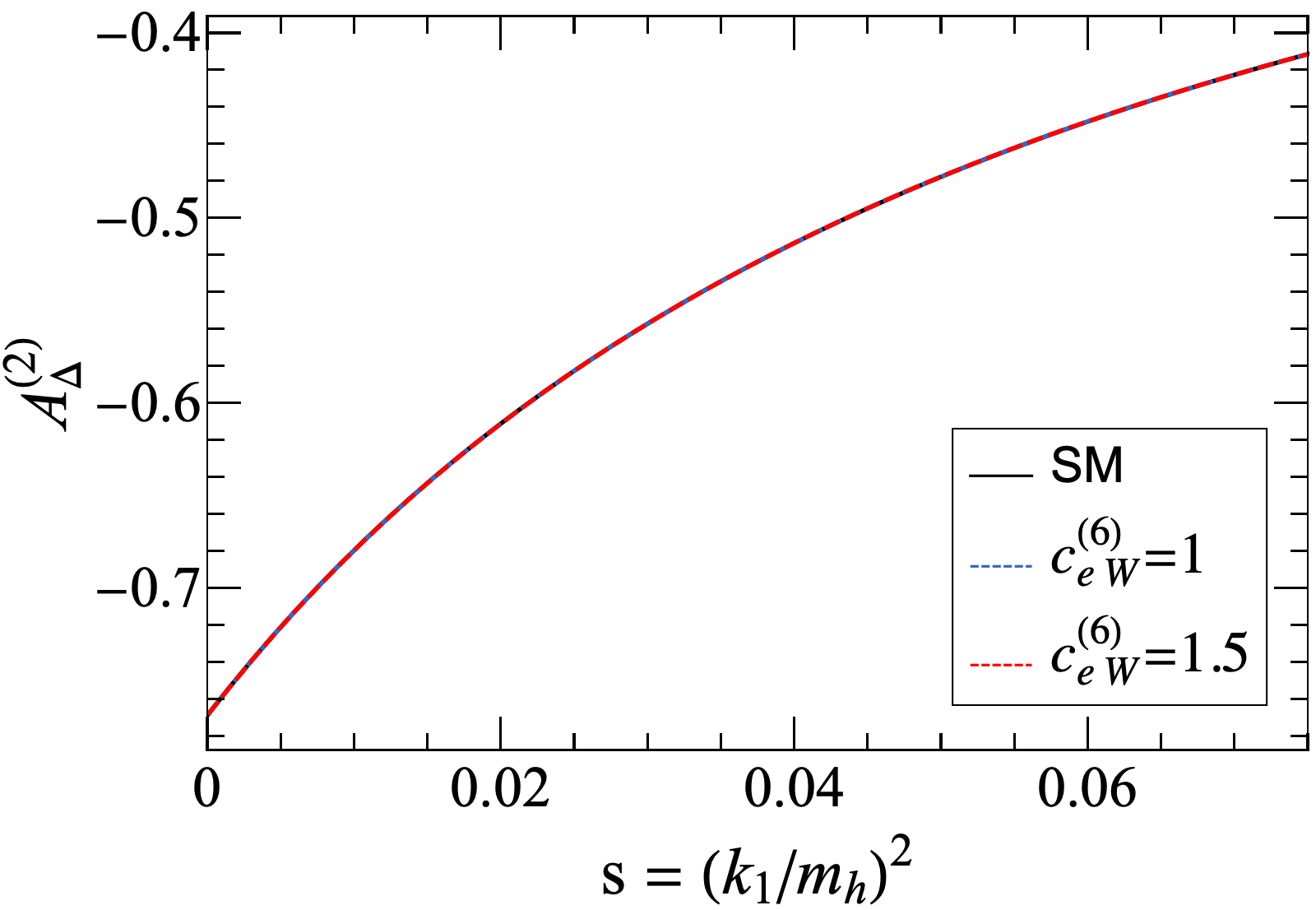}
    \subcaption{}
    \label{Adelta_dipole}
  \end{subfigure}
  \begin{subfigure}{0.48\textwidth}
    \centering
    \includegraphics[width=\textwidth]{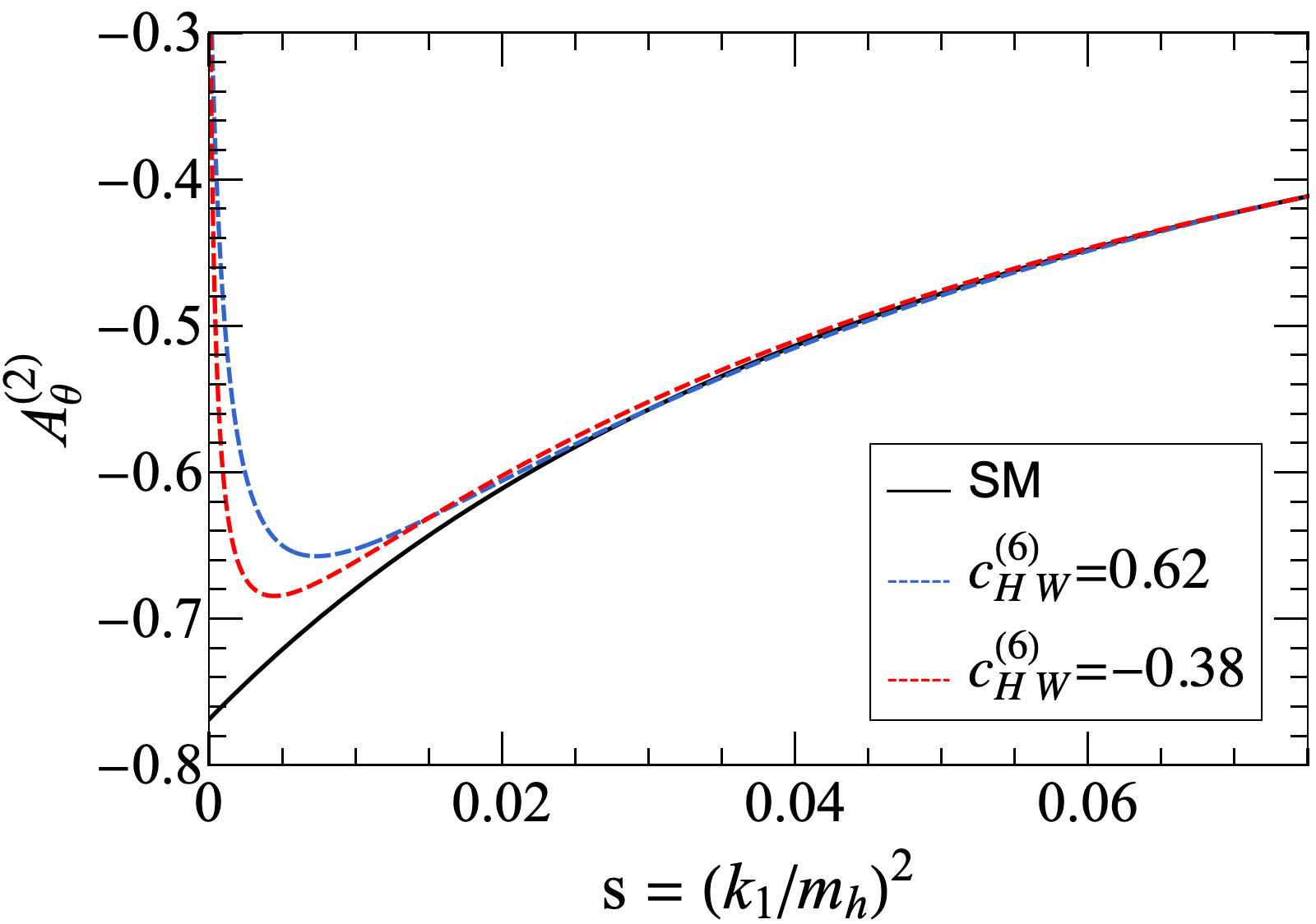}
    \subcaption{}
    \label{Atheta_cHW}
  \end{subfigure}
  \caption{Dimension 6 SMEFT operators' contribution to $A_\theta^{(2)}$ and $A_\Delta^{(2)}$.  The solid black lines represent the SM with all Wilson coefficients set to zero. The other curves represent the operator's contribution by setting it's respective Wilson coefficient to the indicated value, taking $\Lambda=1\, \text{TeV}$,  and all other Wilson coefficients are set to zero. In figures (a) and (b) we compare the dipole contribution to  $A_\theta^{(2)}$ and $A_\Delta^{(2)}$, respectively. Since for non-dipole operators $A_\Delta^{(2)}=A_\theta^{(2)}$, in figure (c) we only show the operators' contribution to $A_\theta^{(2)}$ . }
  \label{other asymmetries}
\end{figure}

Let's first look at the dimension six operators' contribution to $\mathcal{A}^{(4)}_\phi$, shown in Figure \ref{A4phi dim6}. We can see from Fig.~\ref{A4phi dim6}\subref{A4phi dim6:a} the first sign of a significant $\mathcal{O}(1/\Lambda^4)$ effect, coming from $Q^{(6)}_{HW}$. The curve representing this operator presents a radically different shape from the rest, showing a growing behavior around $s\sim0$. This is due to a photon pole enhancement, coming from the kinematic structure $A_{hZA}^{(1)}$ in Eq. (\ref{A1hZA}) were we can see the photon propagator $1/s_{12}\sim1/s$. After squaring, we get linear contributions of the form $c^{(6)}_{HW}/s$ and quadratic contributions $(c^{(6)}_{HW})^2/s^2$ (this can be explicitly seen in Appendix \ref{J^Z explicit expressions}). The quadratic contributions are responsible for the $(1/\Lambda^4)$ enhancement in the $s\sim0$ region. To emphasize that this is a purely $\mathcal{O}(1/\Lambda^4)$ effect, we have plotted the $\mathcal{O}(1/\Lambda^2)$ contribution from $Q^{(6)}_{HW}$ in Figure \ref{A4phi dim6}\subref{A4phi dim6:b}. In this plot we have only considered the SM contribution and the interference with $Q^{(6)}_{HW}$. 

The other noticeable $\mathcal{O}(1/\Lambda^4)$ comes from the dipole operator $Q^{(6)}_{eW}$ in Figure \ref{A4phi dim6}\subref{A4phi dim6:c}. Although not as radical as $Q^{(6)}_{HW}$, the curve representing $Q^{(6)}_{eW}$ has a very distinct shape from the one for $Q^{(6)}_{HW}$. The reason its contribution so significant is that in addition to restricting the form of the higher dimensional operators that can appear, interference with the SM also introduces factors of small SM couplings. Since dipole operators don't interfere with the SM, they only contribute as squared terms. While this means they are inherently $\mathcal O(1/\Lambda^4)$, they aren't suppressed by additional SM couplings.

In comparison, we see that there is no effect whatsoever on $\mathcal{A}^{(4)}_\phi$ from $Q^{(6)}_{HD}$, $Q^{(6)}_{He}$, and $Q^{1,(6)}_{H\ell}$ (Fig. \ref{A4phi dim6}\subref{A4phi dim6:d}). This can be explained from a geoSMEFT perspective. Recall that $Q^{(6)}_{HD}$ belongs to the field space connection $h_{IJ}$. The role of this connection is to change the normalization of the fields and to shift the SM-like $hZZ$ vertex, which is encapsulated by $c^{(1)}_{hZZ}$. As this coupling is present in practically every $J$-Function, much of this shift cancels when taking ratios as in $\mathcal{A}^{(4)}_\phi$. Something similar happens with $Q^{(6)}_{He}$ and $Q^{1,(6)}_{H\ell}$ shifting the $Z\psi\psi$ vertex. Additionally, both operators generate a 4-point contact vertex $hZ\bar{\psi}\psi$, which leads to an $s$-dependent contribution in the denominator and numerator of $\mathcal{A}^{(4)}_\phi$. However, this $s$-dependence is the same as the SM (see Appendix \ref{J^Z explicit expressions}), so these SMEFT contributions just lead to a change in normalization that cancels once we take the ratio.\footnote{This can be understood from the kinematic structures $A_{hZZ}^{(1)}$ in Eq.~(\ref{A1hZZ}), corresponding to the SM contribution, and $A_{hZ\psi}^{(1)}$ in eq. (\ref{A1hWpsipsi}), the dimension six contact term contribution. Note these are equal, up to a propagator, which translates in the square amplitude as the same kinematic dependence.}

Moving on to $d\Gamma/ds$, in Fig.~\ref{invmass dim6} we plot the differential distribution in the SM and compare to the distribution including the SMEFT operators motivated earlier. In Fig.~\ref{invmass dim6}\subref{invmass cHW}, we see the same radical enhancement coming from the photon pole. Again, this is an $\mathcal{O}(1/\Lambda^4)$ effect coming from the quadratic contributions of $Q_{HW}^{(6)}$. This behavior was already noted in \cite{Beneke:2014sba}. However, as mentioned in that work, a calculation considering the full $\mathcal{O}(1/\Lambda^4)$ contribution was needed to obtain a consistent EFT result. In Figure \ref{invmass dim6}\subref{invmass cHD} and \ref{invmass dim6}\subref{invmass cHpsi} we show the effects of operators $Q^{(6)}_{HD}$, $Q^{(6)}_{He}$ and $Q^{1,(6)}_{Hl}$. These operators had negligible effects on $\mathcal{A}^{(4)}_{\phi}$, because it's defined as a ratio of $J$-functions. In contrast, $d\Gamma/ds$ is not a ratio and we see a much larger effect. Interestingly, $Q^{1,(6)}_{Hl}$ show constructive interference --depending on the sign of the Wilson coefficient-- while $Q^{(6)}_{He}$ shows destructive interference. A similar effect was observed by \cite{Buchalla:2013mpa} and~\cite{Beneke:2014sba} at $\mathcal{O}(1/\Lambda^2)$, but using axial and vector couplings. $Q^{(6)}_{HD}$ also shows a large constructive interference contribution. Meanwhile, the dipole operator $Q_{eW}^{(6)}$ has a small effect compared to the other dimension six operators.

We briefly discuss the asymmetries $\mathcal{A}^{(2)}_{\theta}$ and $\mathcal{A}^{(2)}_{\Delta}$, shown in Figure \ref{other asymmetries}, as the same comments for apply here $\mathcal{A}^{(4)}_\phi$. The most interesting feature of these asymmetries is that they are sensitive to the degeneracy broken by dipole operators. As it can be explicitly seen in Appendix \ref{J^Z explicit expressions}, the $J$-functions $J^Z_3$ and $J^Z_4$ only differ by their dipole contributions. Hence, when dipole operators are not considered, $J^Z_3=J^Z_4$, and therefore $\mathcal{A}^{(2)}_{\theta}$ and $\mathcal{A}^{(2)}_{\Delta}$ are degenerate. This is due to an `imbalance' in the possible helicity configurations, by which we mean that the last four helicity configurations for $A_{hZ\psi}^{dipole}$ in Table \ref{dipole helicities hZZ} don't contribute to $\mathcal{O}(1/\Lambda^4)$ in the square amplitude. As a result, $\mathcal{A}^{(2)}_{\theta}$ in Fig.~\ref{other asymmetries}\subref{Atheta_dipole} shows a bigger sensitivity to the dipole operator than $\mathcal{A}^{(2)}_{\Delta}$ in Fig.~\ref{other asymmetries}\subref{Adelta_dipole}. Regarding the other dimension six operators, we also have a photon pole enhancement due to $Q_{HW}^{(6)}$ (see Figure \ref{other asymmetries}\subref{Atheta_cHW}), and because we are not considering a dipole operator, $\mathcal{A}^{(2)}_{\theta}$ and $\mathcal{A}^{(2)}_{\Delta}$ show the same enhancement. Like in $\mathcal{A}^{(4)}_\phi$, the operators $Q^{(6)}_{HD}$, $Q^{(6)}_{He}$, and $Q^{1,(6)}_{H\ell}$ have no impact on these asymmetries, which is why we don't show a plot for these operators' contribution. 

\begin{figure}[t]
  \begin{subfigure}[b]{0.49\textwidth}    
    \includegraphics[width=\textwidth]{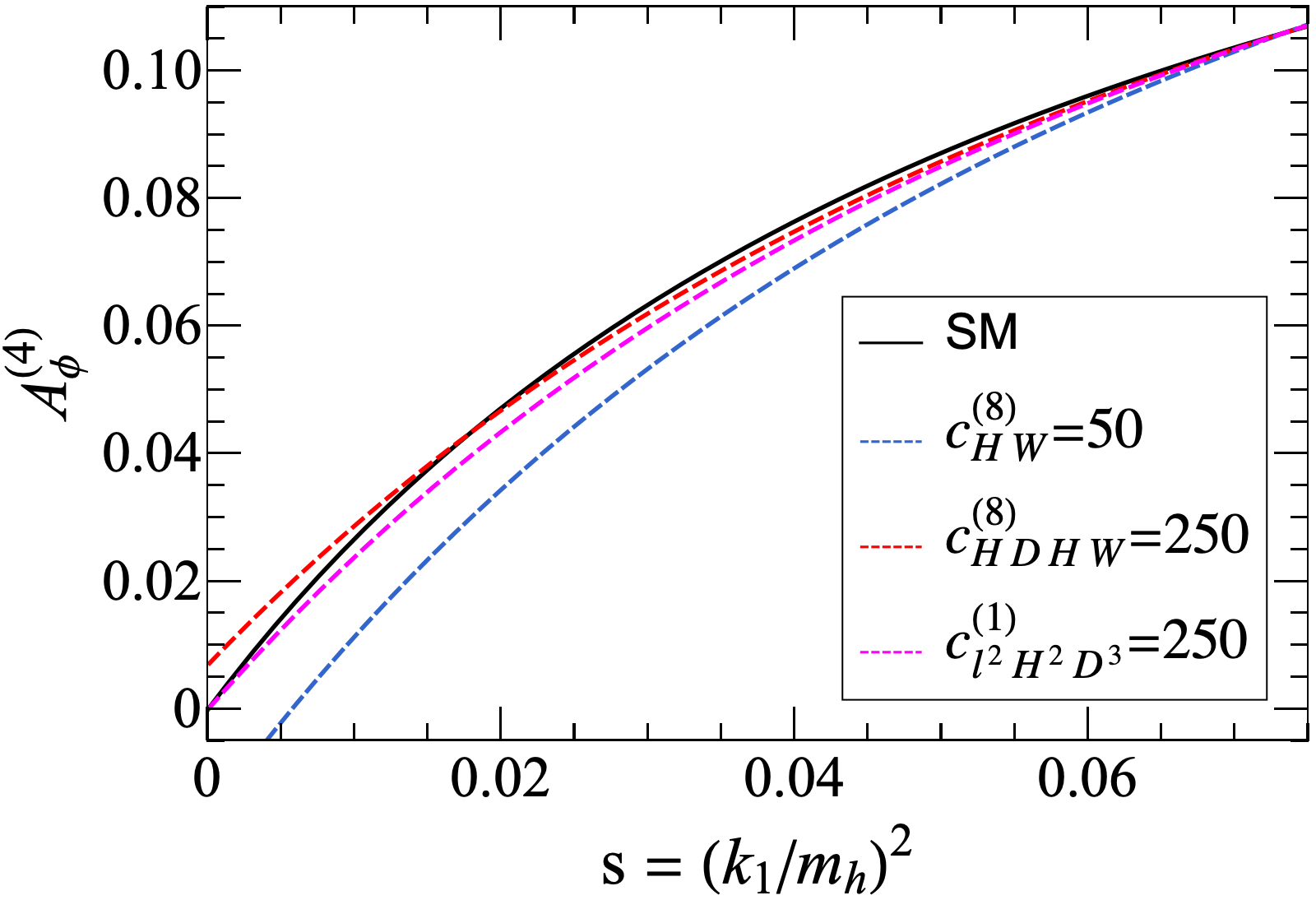}
    \subcaption{}
    \label{dim8 plots:a}
  \end{subfigure}
  \begin{subfigure}[b]{0.49\textwidth}
    \includegraphics[width=\textwidth]{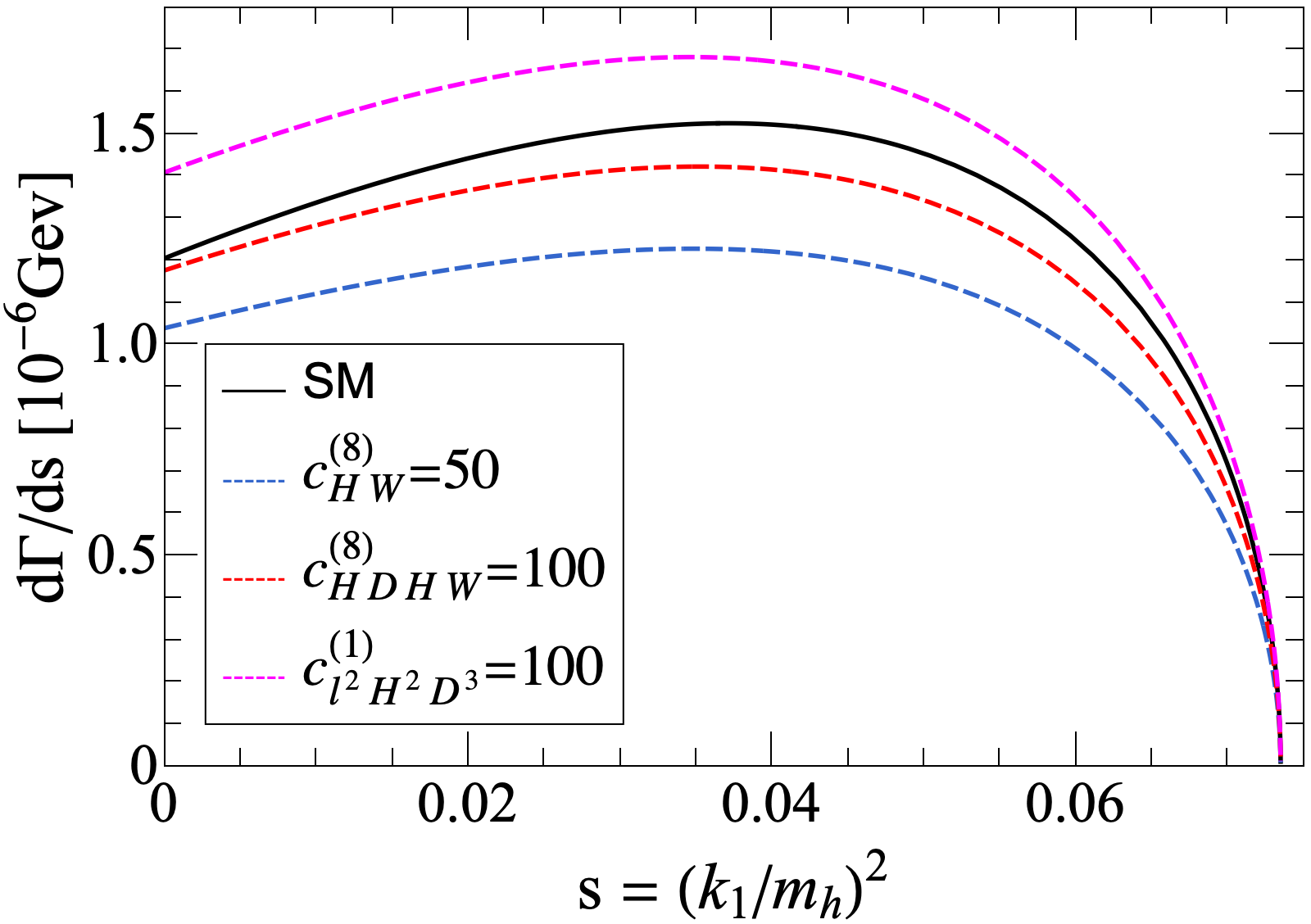}
    \subcaption{}
    \label{dim8 plots:b}
  \end{subfigure}
  \centering
  \begin{subfigure}[b]{0.49\textwidth}  
    \includegraphics[width=\textwidth]{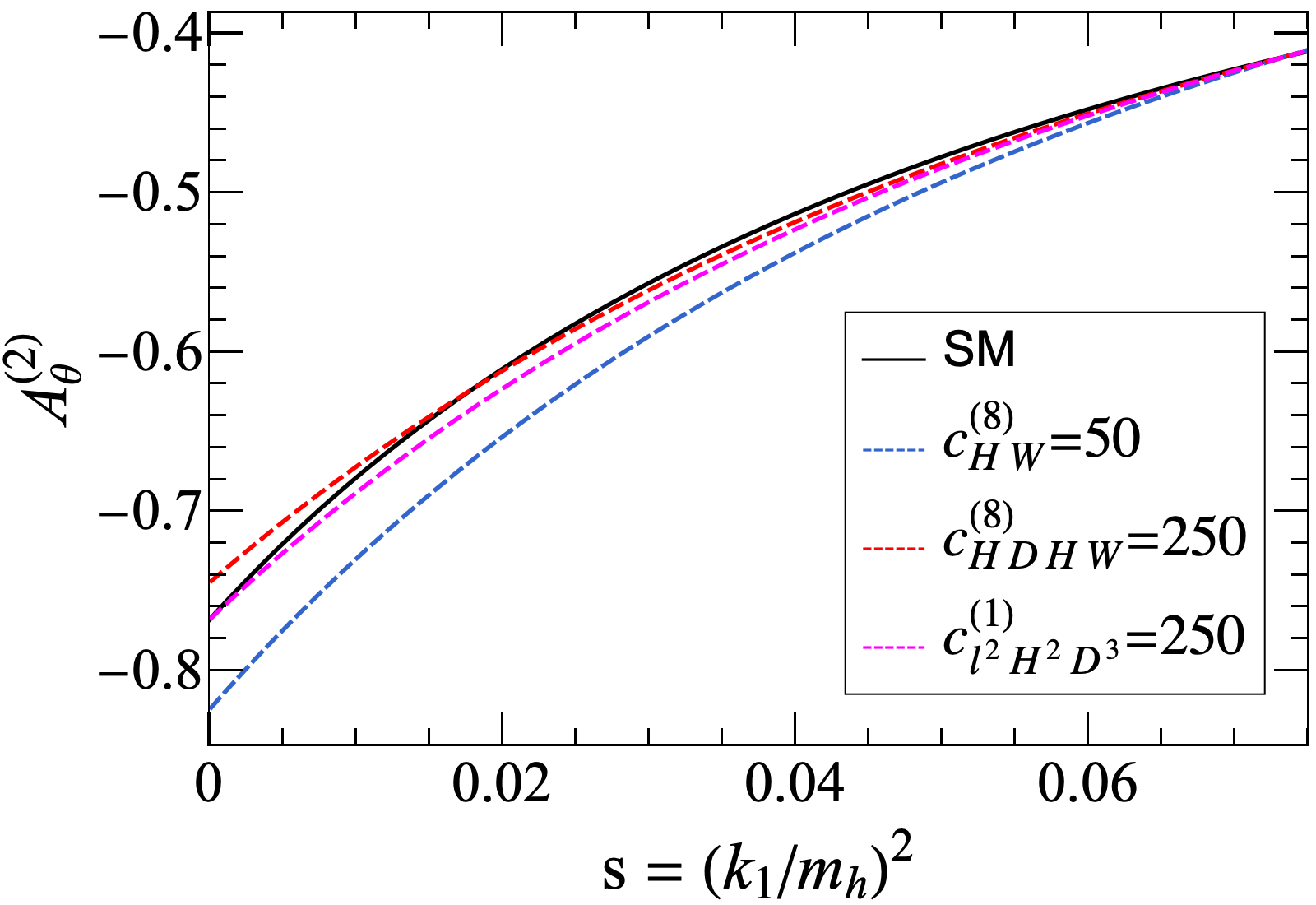}
    \subcaption{}
  \end{subfigure}
  \caption{Dimension Eight operators' contribution to $\mathcal{A}_\phi^{(4)}$, Figure (a), and $d\Gamma/ds$, Figure (b).  The solid black lines represent the SM with all Wilson coefficients set to zero. The solid black lines represent the SM with all Wilson coefficients set to zero. The other curves represent the operator's contribution by setting it's respective Wilson coefficient to the indicated value, taking $\Lambda=1\, \text{TeV}$,  and all other Wilson coefficients are set to zero.}
  \label{dim8 plots}
\end{figure}

Continuing with the dimension eight operators, in Figure \ref{dim8 plots} we plot as a function of $s=k_1^2/m_h^2$ the effects that dimension eight operators have on $\mathcal{A}^{(4)}_{\phi}$, $\mathcal{A}^{(2)}_{\theta}$ and $d\Gamma/ds$ ~\footnote{We don't show $Q_{e^2H^2D^3}^{(1)}$ since it's contribution it's the same as $Q_{l^2H^2D^3}^{(1)}$. This is because it's contribution to the effective couplings is equal if only $Q_{e^2H^2D^3}^{(1)}$ or $Q_{l^2H^2D^3}^{(1)}$ are considered (see Appendix \ref{effective couplings appendix}).}. We don't show the contribution to $\mathcal{A}^{(2)}_{\Delta}$ because it is degenerate to $\mathcal{A}^{(2)}_{\theta}$ when only dimension eight operators are considered. While some bounds for some of the dimension eight Wilson coefficients exist \cite{Boughezal:2021tih, Boughezal:2022nof, Dawson:2021xei, Corbett:2023qtg}, most of them are currently not bounded~\footnote{A broader set of operators can be bounded (either alone or in certain combinations) to have a definite sign via analyticity arguments, see~\cite{Adams:2006sv, Bellazzini:2014waa, Li:2022rag}}. A natural choice is to take the size of the Wilson coefficients as $\mathcal{O}(1)$. However, if we stick with coefficients of $\mathcal O(1)$ and a scale $\Lambda=1\, \text{TeV}$, we observe that effects of dimension eight operators are minimal compared to dimension six operators.

There are a number of reasons that conspire to suppress the dimension eight operators' contribution. The first and most obvious is the $1/\Lambda^4$ suppression. As explained in \cite{Assi:2025zmp}, dimension eight contributions appear accompanied by factors of either $s^2/\Lambda^4$, $v^2s/\Lambda^4$, or $v^4/\Lambda^4$. If $s \gg v$, the former is much less of a suppression than the latter and can lead to `enhanced' dimension eight effects. However, as we are studying on-shell Higgs decay, $\sqrt{s}=m_h\sim v$, so no such energy enhancement is possible and all the dimension eight contributions become suppressed. While this logic explains why dimension eight effects are small compared to linear dimension six terms, it does not explain why we find dimension eight is small compared to effects from dimension six squared, formally the same order in $ v/\Lambda$.

The second source of suppression are the small SM couplings. As dimension eight operators enter observables at $\mathcal O(1/\Lambda^4)$ by interfering with the SM, their amplitude contributions contain factors of the SM couplings.  This should be contrasted with `self-square contributions' such as we find from the dipole operators. Despite being suppressed by $1/\Lambda^4$, due to its non-interfering nature, these terms aren't accompanied by SM model couplings.

The last reason is counterintuitive, so allow us to elaborate. In the squared amplitude, there are four types of contributions: 1.) terms involving the photon propagator, $1/s$, 2) terms with one $Z$ boson propagator ~\footnote{Recall that we used the narrow-width approximation, so the $Z$ boson with the propagator $1/(s_{34}-m_Z^2)$ is on shell.}, $1/(s_{12}-m_Z^2)\sim1/(s-r)$ , 3) contributions with two $Z$ boson propagator, $1/(s-r)^2$,  and 4) terms with no propagator.  We have already mentioned the terms with photon propagator, which are associated with topology (c) in Fig.~\ref{htoWW diagrams}. The terms with $1/(s-r)^2$ come from squared terms of topology (a),  the terms with $1/(s-r)$ are due to 4-point contact terms and come from interference between topology (a) and (b), while terms with no propagators are 4-point contact term squared from topology (b). The key point is that, rather than the two $Z$ propagators being suppressed, as one finds in the high energy limit $s \gg r$, these terms become enhanced. For the kinematics here, $s$ is constrained and $s<r$, such that resulting in $1/(r-s)\sim2$. Therefore, terms with two $Z$ boson propagators are actually enhanced by at least a factor of 2 over 4-point contact terms. Moreover, this contributes to some dimension six squared effects being larger than some dimension eight. In particular, dimension six contributions from topology (a) are enhanced over 4-point dimension eight contributions.~\footnote{This propagator effect is exactly what suppresses contact $h \to $ four fermion contributions.}

While neither of these effects in isolation is very large, their combined effect is the significant suppression of dimension eight terms compared to dimension six squared that we observe. To explore  what size dimension eight couplings the asymmetries are sensitive to, in Fig.~\ref{dim8 plots} we enhance the Wilson coefficients beyond their estimated $\mathcal O(1)$ size. The necessary couplings are $\mathcal O(50-250)$; the lower limit of this matches what one would expect is needed to overcome $\mathcal O(v^2/\Lambda^2)$ suppression compared to dimension six effects. While there are few current bounds on dimension eight operators, taking these large coefficients seriously implies huge hierarchies between coefficients. Hierarchies are possible, either coming from cancellation or if operators arise at different orders when matched to a UV theory. Nevertheless, we will regard Fig.~\ref{dim8 plots} (and others like it) as purely illustrative.

\begin{figure}
\centering
\end{figure}

Of the dimension eight operators we consider, $Q^{(8)}_{HW}$ shows the largest effect in all three observables. One may have expected that this operator also gets a photon propagator enhancement when $s\sim0$ like  its dimension six counterpart $Q^{(6)}_{HW}$, however this is not the case. Unlike $Q^{(6)}_{HW}$, $Q^{(8)}_{HW}$ contributes to the square amplitude through an interference term of the form $c^{(8)}_{HW}/s$, but we still need to take the spinor products into account. After simplifying the spinor products, this $1/s$ factor gets canceled, as can be seen from the explicit expression for the $J$-functions in Appendix \ref{J^Z explicit expressions}. 

The operator $Q^{(1)}_{l^2H^2D^3}$ has the smallest effect on the asymmetries. Similarly to the 4-point contact term generated by $Q^{3,(6)}_{Hl}$, this is because the asymmetries are defined as ratios of $J$-functions and effects cancel between the numerator and denominator.  The cancellations for $Q^{(1)}_{l^2H^2D^3}$ are not as exact as for $Q^{3,(6)}_{Hl}$, resulting in a small but noticeable effect. For $d\Gamma/ds$, the impact from $Q^{(1)}_{l^2H^2D^3}$ is larger (for a given Wilson coefficient value) as $d\Gamma/ds$ is not a ratio. The effect of $Q^{(1)}_{l^2H^2D^3}$ on $d\Gamma/ds$ is smaller than from $Q^{(8)}_{HW}$ because the latter (not a contact term) is enhanced by propagator factors $\sim 1/(r-s)$.

Finally, we find $Q_{HDHW}^{(8)}$ has a very small effect, requiring Wilson coefficients to $\mathcal O(200)$ to be noticeable. Contributions from this operator are enhanced by the propagator, but this is overwhelmed by a suppression from additional SM couplings (compared to e.g. $Q^{(8)}_{HW}$). The additional SM couplings come from the definition of the effective couplings in terms of Wilson coefficients, as can be seen in Appendix \ref{effective couplings appendix}. To show this, we take the ratio of effective couplings generated by $Q_{HW}^{(8)}$ and those generated by $Q_{HDHW}^{(8)}$, we find $\left(c_{hZZ}^{(2)}/c_{hZZ}^{(3)}\right)\sim4\left(c_{HW}^{(8)}/_{HDHW}^{(8)}\right)$.

To conclude this section, we point out that asymmetries, being defined as ratios, generally suppress some operators' contributions. This cancellation can be exact, as in the case of $Q^{3,(6)}_{Hl}$, or approximate as for $Q^{(1)}_{l^2H^2D^3}$. Conversely, $d\Gamma/ds$ is not a ratio is generically more sensitive to SMEFT effects (except, interestingly, for dipole operators). Although SMEFT effects on asymmetries are less pronounced than those on $d\Gamma/ds$, asymmetries provide complementary information to distinguish the operators responsible for the effects on $d\Gamma/ds$. Another advantage of asymmetries is the possibility to extract non-SM $J$-functions in processes with a \textit{non SM-like angular distribution}. While $h\to\ell\bar{\ell}\left(Z\to\ell'\bar{\ell'}\right)$ exhibits a \textit{SM-like angular distribution}, in the next section we will find a \textit{non SM-like angular distribution} for which asymmetries, with some caveats, are a good observable to extract non-SM $J$-functions and thereby study SMEFT effects.

\section{Results for $h \to \ell \bar{\nu}_\ell \nu_{\ell'} \bar{\ell'}$}\label{section 4}

\subsection{Amplitude}\label{section 4.1}

Now we turn to the second part of our analysis. In this section, we present the spinor-helicity amplitude for $h(p)\xrightarrow{}\ell(p_1)\bar{\nu}_\ell(p_2)\nu_\ell(p_3)\bar{\ell}(p_4)$. Analytically, this process shows several similarities to $h\to\ell\bar{\ell}\left(Z\xrightarrow{}\ell'\bar{\ell'}\right)$, Nevertheless, it also shows some important differences, which can be summarized as follows:

\begin{enumerate}
    \item Absence of photon propagator. For this process,  there is no analogous topology to diagram (c) in Figure \ref{htoZlldiagrams}, involving the propagating photon. As a result, the photon propagator enhancement observed in the previous analysis does not occur here.
    \item Additional topology (f) in Figure \ref{htoZlldiagrams}. Since the final state is not reconstructible, diagram (f) can't be discarded by imposing kinematic cuts on the dilepton invariant mass.
    \item Complex effective couplings. In the SM, the coupling for the  $W\ell\nu$ vertex is real. However, SMEFT corrections can generate an imaginary part (see Appendix \ref{effective couplings appendix}). Something similar happens with the $hWW$ vertex and $hW\ell\nu$ contact term.
    \item More kinematic structures. The kinematic structures $A^{(i)}_{hZZ}$ and $A^{(i)}_{hZ\psi}$ have an analogous counterpart in this process, but we also find new kinematic structures with no \linebreak $h\to\ell\bar{\ell}\left(Z\to\ell'\bar{\ell'}\right)$ analog.
    \item Reduced number of helicity configurations. In the SM level, only left-handed leptons couple to $W^\pm$. Except for dipole operators, SMEFT preserves this coupling structure because only left-handed neutrinos are considered by the SMEFT. Dipole operators, on the other hand, allow $W^\pm$ to couple with neutrinos and right-handed charged leptons, introducing two other helicity configurations.
\end{enumerate}
With the exception of the first point, all of these differences lead to \textit{non SM-like angular distributions}. To study $h(p)\xrightarrow{}\ell(p_1)\bar{\nu}_\ell(p_2)\nu_\ell(p_3)\bar{\ell}(p_4)$, we adopt the same conventions as in Section \ref{3.1} and follow a similar organization style, separating the \textit{interfering contributions} from the \textit{non-interfering contributions} and decomposing the subamplitudes as in Eq.~(\ref{amplitude decomposition}). As the procedure to construct the amplitude is the same as the one presented in Section \ref{3.1}, the details and explicit expressions are relegated to Appendix \ref{example Appendix}. Here we focus on the main differences with the four charged lepton case. 

\begin{table}[t!]
\centering
\renewcommand{\arraystretch}{1.5}
\begin{tabular}{|c|c|c|c|c|}
\hline
\multicolumn{5}{|c|}{\textbf{$\mathcal{A}(h\xrightarrow{}\ell\bar{\nu}_\ell\nu_\ell\bar{\ell})$}}   \\ \hline 
Amplitude & \makecell{Kinematic \\ Structure} & Coupling Factor & SMEFT order & Generating Operator \\ \hline
\multirow{4}{*}{$\mathcal{A}_{hWW}$} 
& $A^{(1)}_{HWW}$          & $\left|g^{W}\right|^2c^{(1)}_{HWW}$ & $\mathcal{O}(1)$           & SM, $h_{IJ}$                  \\ \cline{2-5}
& $A^{(2)}_{hWW}$          & $\left|g^{W}\right|^2c^{(2)}_{hWW}$    & $\mathcal{O}(1/\Lambda^2)$ & $Q_{HW}^{(6)}$, $Q_{HW}^{(8)}$                        \\ \cline{2-5}
& $A^{(3)}_{hWW}$          & $\left|g^{W}\right|^2c^{(3)}_{hWW}$    & $\mathcal{O}(1/\Lambda^4)$ & \multirow{2}{*}{$Q_{HDHW}^{(8)}$, $Q_{HDHW,2}^{(8)}$} \\ \cline{2-4}
& $A^{(4)}_{hWW}$          & $\left|g^{W}\right|^2c^{(4)}_{hWW}$    & $\mathcal{O}(1/\Lambda^4)$ &                                                       \\ \hline
\multirow{8}{*}{$\mathcal{A}_{hW^{\pm}\psi}$}
& $A^{(1)}_{hW^{\pm}\psi}$ & $g^{W^\pm} g^{(1)}_{hW^\mp l}$        & $\mathcal{O}(1/\Lambda^2)$ & $Q_{Hl}^{3,(6+2n)}$, $Q_{Hl}^{\epsilon,(8)}$           \\ \cline{2-5}
& $A^{(2)}_{hW^{\pm}\psi}$ & $g^{W^\pm} g^{(2)}_{hW^\mp l}$        & $\mathcal{O}(1/\Lambda^4)$ & \multirow{2}{*}{$Q_{l^2H^2D^3}^{(4)}$}                 \\ \cline{2-4}
& $A^{(3)}_{hW^{\pm}\psi}$ & $g^{W^\pm} g^{(3)}_{hW^\mp l}$        & $\mathcal{O}(1/\Lambda^4)$ &                                                        \\ \cline{2-5}
& $A^{(4)}_{hW^{\pm}\psi}$ & $g^{W^\pm} g^{(4)}_{hW^\mp l}$        & $\mathcal{O}(1/\Lambda^4)$ & $Q_{l^2WH^2D}^{(3)}$, $Q_{l^2WH^2D}^{(5)}$             \\ \cline{2-4}
& $A^{(5)}_{hW^{\pm}\psi}$ & $g^{W^\pm} g^{(5)}_{hW^\mp l}$        & $\mathcal{O}(1/\Lambda^4)$ & $Q_{l^2H^2D^3}^{(3)}$                                  \\ \cline{2-5}
& $A^{(6)}_{hW^{\pm}\psi}$ & $g^{W^\pm} g^{(6)}_{hW^\mp l}$        & $\mathcal{O}(1/\Lambda^4)$ & $Q_{l^2H^2D^3}^{(4)}$                \\ \cline{2-5}
& $A^{(7)}_{hW^{\pm}\psi}$ & $g^{W^\pm} g^{(7)}_{hW^\mp l}$        & $\mathcal{O}(1/\Lambda^4)$ & $Q_{l^2H^2D^3}^{(3)}$                \\ \cline{2-5}
& $A^{(8)}_{hW^{\pm}\psi}$ & $g^{W^\pm} g^{(8)}_{hW^\mp l}$        & $\mathcal{O}(1/\Lambda^4)$ & $Q_{l^2H^2D^3}^{(4)}$                                                      \\ \hline
\end{tabular}
\caption{Subamplitudes contributing to $\mathcal{A}(h\xrightarrow{}\ell\bar{\nu}_\ell\nu_\ell\bar{\ell})$. To the right of each subamplitude we find the contributing  kinematic structures next to its respective coupling factor. Third column specifies the arising SMEFT order and the fourth column the generating operators. Above $(g^{W^-})^*=g^{W^+}$ and $\left|g^{W}\right|^2=g^{W^-}g^{W^+}$}
\label{coupling factors hww}
\end{table}

Starting with the \textit{interfering contribution}, the only interfering helicity configuration is $(h_1,h_2,h_3,h_4)=(-,+,-,+)$. The subamplitude decomposition (\ref{amplitude decomposition}) for this process has been organized in Table \ref{coupling factors hww}. Here $\mathcal{A}_{hWW}$, $\mathcal{A}_{hW^{+}\psi}$, where $\mathcal{A}_{hW^-\psi}$ are the corresponding subamplitudes for diagrams (d), (e) and (f) in Figure \ref{htoWW diagrams} and effective couplings are defined in Appendix \ref{example Appendix} and expressed in terms of Wilson coefficients in Appendix \ref{effective couplings appendix}.

From Table \ref{coupling factors hww}, we see that there are more kinematic structures in this process than in $h\to\ell\bar{\ell}Z(\to\ell'\bar{\ell'})$. This occurs because $ A^{(2)}_{hW^\pm\psi}$, $A^{(3)}_{hW^\pm\psi}$, $ A^{(6)}_{hW^\pm\psi}$, and $ A^{(8)}_{hW^\pm\psi}$ depend explicitly on the individual lepton momentum (as opposed to di-lepton momenta), either residing within spinor chains or as dot products. The appearance of the individual lepton momenta can be traced back to the operator $Q_{l^2H^2D^3}^{(4)}$, which contains a covariant derivative acting on one of the fermions.\footnote{This operator does not contain any three or four point vertices which can contribute to $h\to\ell\bar{\ell}\left(Z\to\ell'\bar{\ell'}\right)$, hence it did not appear in the last section}. As we will see, these new kinematic structures involving the individual lepton momenta lead to novel angular dependence.

\begin{table}[t!]
\centering
\renewcommand{\arraystretch}{1.5}
\begin{tabular}{|l|l|l|l|}
\hline
\multicolumn{4}{|c|}{\textbf{Dipole Contribution to $h\xrightarrow{}\ell\bar{\nu}_\ell\nu_\ell\bar{\ell}$}} \\ \hline  
Helicity    & $A^{dipole}_{hWW}$            &$A^{dipole}_{hW^+\psi}$    & $A^{dipole}_{hW^-\psi}$   \\ \hline
$(+,+,-,+)$ &$c^{(1)}_{hWW}g^{W^+}d_{W^-}$ &$d_{hW^-}g^{W^+}$             & $\mathcal{O}(1/\Lambda^4)$    \\ \hline
$(-,+,+,+)$ &$c^{(1)}_{hWW}g^{W^-}d_{W^+}$ &$\mathcal{O}(1/\Lambda^4)$     & $d_{hW^+}g^{W^-}$            \\ \hline
\end{tabular}
\caption{Dipole kinematic structures with their respective coupling factors for a particular helicity combination. The coupling factors stated as $\mathcal{O}(1/\Lambda^4)$ means that this particular helicity combination doesn't contribute to the square amplitude.} 
\label{helicity combinations hww}
\end{table}

For the \textit{non-interfering contributions}, only two helicity configurations are possible: \linebreak $(h_1,h_2,h_3,h_4)$ $=(+,+,-,+)$ and $(h_1,h_2,h_3,h_4)=(-,+,+,+)$. The dipole kinematic structures that contribute to $\mathcal{A}_{hWW}$, $\mathcal{A}_{hW^{+}\psi}$, and $\mathcal{A}_{hW^-\psi}$ are all generated by the same operator, $Q^{(6)}_{eW}$. In Table \ref{helicity combinations hww} we show the various subamplitudes where $Q^{(6)}_{eW}$ can enter (for each non-zero helicity configuration), along with the corresponding coupling\footnote{$A^{dipole}_{hW^\pm\psi}$, we get terms at $\mathcal O(1/\Lambda^4)$ from the combination of one dipole $W\nu \ell$ vertex and one non-dipole $hW \nu \ell$ contact vertex. Configurations with a dipole operator contribution to the $hW \nu \ell$ contact vertex are $\mathcal{O}(1/\Lambda^6)$ or higher.}. The definitions of the couplings and their expansion in terms of Wilson coefficients can be found in Appendix \ref{example Appendix} and \ref{effective couplings appendix} respectively, while the explicit expressions for the kinematic structures can be found in Appendix \ref{example Appendix}. The main difference between these expressions and their counterparts in the $h\xrightarrow{}\ell\bar{\ell}Z(\xrightarrow{}\ell'\bar{\ell'})$ case is that  $A^{dipole}_{hW^-\psi}$, can interfere with $A^{dipole}_{hWW}$.

\subsection{Observables}
 In this section, first we present the differential decay rate analogously as in Section \ref{section 3.2}. Then, since the final state is not fully reconstructible, we introduce the appropriate observables for this process and present the SMEFT contribution to such observables.

\subsubsection{Angular Distribution}\label{Section 4.2.1}
In the Higgs rest frame, the differential decay rate for $h(p)\xrightarrow{}l(p_1)\bar{\nu}_l(p_2)\nu_l(p_3)\bar{l}(p_4)$ is given by:
\begin{equation}
\label{htoWWwidth}
    \frac{d\Gamma}{ds dr d\cos \theta d\cos{\Delta}d\phi}=\frac{m_h\lambda}{(2\pi)^62^9}\left|\mathcal{A}(h\xrightarrow{}l\bar{\nu}_l\nu_l\bar{l} \ )\right|^2,
\end{equation}
with
\begin{equation}
    s=\frac{k_1^2}{m_h^2}, \quad r=\frac{k_2^2}{m_h^2}.
\end{equation}
and $k_1=p_1+p_2,\,k_2=p_3+p_4$ being the momenta carried by the $W$ bosons. Notice we don't use the narrow-width approximation here, since a narrow mass peak can't be reconstructed due to the neutrinos. The remaining notation and conventions follow those of Section \ref{Section 3.2.1} . The spin-summed, squared amplitude can be written as:
\begin{align}
    \label{J^W functions}
    \left|\mathcal{A}(h\xrightarrow{}l\bar{\nu}_l\nu_l\bar{l} \ )\right|^2=
    &J^W_1+J^W_2 \cos^2{\theta}\cos^2{\Delta}\\ \nonumber 
    &+J^W_3 \cos^2{\theta}+J^Z_4 \cos^2{\Delta}+J^Z_5\cos{\theta}\cos{\Delta} \\ \nonumber 
    &+(J^W_6 \sin{\theta}\sin{\Delta}+J^W_7 \sin{(2\theta)}\sin{(2\Delta)})\cos{\phi}\\  \nonumber
    &+J^W_8\sin^2{\theta}\sin^2{\Delta}\cos{(2\phi)} \\  \nonumber
    &+J^W_{9} \cos{\theta}+J^W_{10}\cos{\Delta}+J^W_{11} \cos^3{\theta}+J^W_{12}\cos^3{\Delta}\\ \nonumber 
    &+(J^W_{13} \cos{\theta}+J^W_{14}\cos{\Delta})\cos^2{\theta}\cos^2{\Delta}\\ \nonumber 
    &+(J^W_{15} \cos{\theta}+J^W_{16}\cos{\Delta})\cos{\theta}\cos{\Delta}\\ \nonumber 
    &+(J^W_{17} \cos{\theta}+J^W_{18}\cos{\Delta})\sin^2{\theta}\sin^2{\Delta}\\ \nonumber 
    &+(J^W_{19} \cos{\theta}+J^W_{20}\cos{\Delta})\sin^2{\theta}\sin^2{\Delta}\cos{(2\phi)}\\ \nonumber 
    &+(J^W_{21}\sin{(2\theta)}\sin{\Delta}+J^W_{22}\sin{(2\Delta)}\sin{\theta})\cos{\phi}\\ \nonumber 
    &+(J^W_{23}\sin{(2\theta)}\sin{\Delta}+J^W_{24}\sin{(2\Delta)}\sin{\theta})\sin{\phi}\\ \nonumber 
    &+(J^W_{25}\sin{(2\theta)}\sin{\Delta}\cos^2{\theta}+J^W_{26}\sin{(2\Delta)}\sin{\theta}\cos^2{\Delta})\cos{\phi}\\ \nonumber
\end{align}

As advertised, the angular distribution of $h \to \ell\bar{\nu}_\ell\nu_{\ell'}\bar{\ell'}$ is richer than the one for \linebreak $h\to Z(\xrightarrow{}\ell'\bar{\ell'})\ell\bar{\ell}$. To illustrate the operators responsible for these differences, we have organized the operators contributing to each of the $J$-functions in Table \ref{hww operators}. The main takeaway from Table \ref{hww operators} is that the \textit{non-SM angular distribution} is purely $\mathcal{O}(1/\Lambda^4)$, either coming from the interference of the dimension eight operators $Q_{l^2WH^2D}^{(5)}$ and $ Q_{l^2H^2D^3}^{(4)}$ with the SM, or from the dipole operator $Q_{eW}^{(6)}$ squared terms. 

\begin{table}[t!]
\centering
\begin{tabular}{|l|l|l|l|l|}
\hline
& \multirow[c]{2}{*}{\raisebox{-3pt}{\textbf{\quad $J$-Function}}}& \multicolumn{3}{c|}{\textbf{Contribution SMEFT Operators}} \\  \cline{3-5}
&  & \makecell{Dimension 6} & Dipole & \makecell{Dimension 8} \\ \hline
\multirow{7}{*}{\raisebox{-28pt}[0pt][0pt]{\rotatebox{90}{SM like Distribution}}}
& \makecell{$J^W_1$, $J^W_2$, $J^W_3$, $J^W_4$, \\  $J^W_7$, $J^W_8$} & \makecell{$Q_{H\Box}^{(6)}$, $Q_{HD}^{(6)}$ \\$Q_{HW}^{(6)}$, $Q_{Hl}^{3,(6)}$} & \makecell{$Q_{eW}^{(6)}$} & \makecell{$Q_{HD}^{(8)}$, $Q_{HD,2}^{(8)}$, $Q_{HW}^{(8)}$ \\ $Q_{HDHW}^{(8)}$, $Q_{Hl}^{3,(8)}$,  \\$Q_{l^2WH^2D}^{(3)}$, $Q_{l^2H^2D^3}^{(3)}$} \\ \cline{2-5}
&   \makecell{$J^W_5$, $J^W_6$} & \makecell{$Q_{H\Box}^{(6)}$, $Q_{HD}^{(6)}$ \\$Q_{HW}^{(6)}$, $Q_{Hl}^{3,(6)}$} & \makecell{$Q_{eW}^{(6)}$} & \makecell{$Q_{HD}^{(8)}$, $Q_{HD,2}^{(8)}$, $Q_{HW}^{(8)}$ \\ $Q_{HDHW}^{(8)}$, $Q_{Hl}^{3,(8)}$,  \\$Q_{l^2WH^2D}^{(3)}$, $Q_{l^2H^2D^3}^{(3)}$}  \\  \hline
\multirow{7}{*}{\raisebox{-50pt}[0pt][0pt]{\rotatebox{90}{Non-SM like Distribution}}}
& \makecell{$J^W_{9}$, $J^W_{10}$,\\ $J^W_{15}$, $J^W_{16}$, \\  $J^W_{21}$, $J^W_{22}$} &  & \makecell{$Q_{eW}^{(6)}$} & \makecell{$Q_{l^2H^2D^3}^{(4)}$} \\ \cline{2-5} 
& \makecell{$J^W_{11}$, $J^W_{12}$, $J^W_{13}$, $J^W_{14}$,\\ $J^W_{17}$, $J^W_{18}$, $J^W_{19}$, $J^W_{20}$\\ $J^W_{25}$, $J^W_{26}$} &  &  & \makecell{$Q_{l^2H^2D^3}^{(4)}$} \\ \cline{2-5} 
& \makecell{$J^W_{23}$, $J^W_{24}$} &  &  & \makecell{$Q_{l^2WH^2D}^{(5)}$}  \\ \hline
\end{tabular}
\caption{SMEFT operators contribution to J-functions. This table has been divided on SM like and non-SM like distributions. The functions in SM-like are present without the need of SMEFT operators. The Non-SM require the presence of at least one of the SMEFT operators displayed on the table. }
\label{hww operators}
\end{table}
The operator that contributes the most to the \textit{non-SM angular distribution} is  $Q_{l^2H^2D^3}^{(4)}$. It is present in almost every non-SM $J$-function, and the new angular dependency can be traced back to its explicit dependence on the lepton momentum. The $J_{23}^W$ and $J_{24}^W$ terms also exhibit non-SM angular behavior, though the origin is more subtle, coming from the imaginary part of $Q_{l^2WH^2D}^{(5)}$ (effective coupling $g^{(5)}_{hW^\pm l}$)

The dipole operator $Q_{eW}^{(6)}$ deserves particular attention. In the four charged lepton case, dipole operators contributed with a \textit{SM like angular distribution}. In this case, however, the dipole operator contributes with a \textit{non-SM like angular distribution}, despite the dipole kinematic structures being the same. The explanation lies in the reduced number of helicities present. Specifically, the dipole operator contributes with the same, non-SM angular distribution to both $h\to Z(\to l'\bar{l'})l\bar{l}$ and $h \to \ell \bar{\nu}_\ell \nu_{\ell'} \bar{\ell'}$, but for $h\to Z(\to l'\bar{l'})l\bar{l}$ the non-SM pieces exactly cancel after summing over all the helicity configurations. For $h \to \ell \bar{\nu}_\ell \nu_{\ell'} \bar{\ell'}$, there aren't as many helicity configurations to sum over and the \textit{non-SM angular distribution} survives. 

With the squared amplitude decomposed in the form (\ref{J^W functions}), we can follow the same path as the four charged lepton case and extract the $J$-functions with angular asymmetries. Since we can't reconstruct the Higgs rest frame in $h\xrightarrow{}l\bar{\nu}_l\nu_l\bar{l}$ at a hadron collider because of the neutrinos and the fact that the initial longitudinal momentum of the partonic collision is unknown, this is somewhat an academic exercise. Nonetheless, it is useful to see where non-SM $J$-functions appear in an idealized, `best case' scenario. At a future lepton collider, it is possible to determine the Higgs rest frame, e.g. in $e^+e^- \to Z(\ell^+\ell^-) h$, by measuring the momenta of the recoiling $Z$, but the presence of two neutrinos means we can't pin down all of the angles in Eq.~\eqref{J^W functions}. 

The single-angle asymmetries \cite{Altmannshofer:2008dz, Kruger:1999xa}, which vanish in the four charged lepton case, isolate the $\mathcal{O}(1/\Lambda^4)$ SMEFT effects proportional to $Q_{eW}^{(6)}$ and $Q_{l^2WH^2D}^{(5)}$.

\begin{align}
  \mathcal{A}^{(1)}_{\theta}= & \left(\frac{d\Gamma}{dsdr}\right)^{-1} \int_{-1}^{1} \sgn{(\cos{\theta})}\frac{d\Gamma}{dsdrd\cos{\theta}} \, d\cos{\theta} \\ \nonumber =&\frac{3 \left(6J^W_9 + 3 J^W_{11} + J^W_{13} + 2 \left(J^W_{16} + J^W_{17}\right)\right)}{4 \left(9 J^W_1 + J^W_2 + 3 \left(J^W_3 + J^W_4\right)\right)}\\
  \mathcal{A}^{(1)}_{\Delta}= & \left(\frac{d\Gamma}{dsdr}\right)^{-1} \int_{-1}^{1} \sgn{(\cos{\Delta})}\frac{d\Gamma}{dsdrd\cos{\Delta}} \, d\cos{\Delta} \\ \nonumber =&\frac{3 \left(6J^W_{10} + 3 J^W_{12} + J^W_{14} + 2 \left(J^W_{17} + J^W_{18}\right)\right)}{4 \left(9 J^W_1 + J^W_2 + 3 \left(J^W_3 + J^W_4\right)\right)}
\end{align}
To study the effects of $Q_{l^2H^2D^3}^{(4)}$ we can define the following new asymmetries:

\begin{align}
  \mathcal{A}^{(1)}_{\phi,\theta}= & \left(\frac{d\Gamma}{dsdr}\right)^{-1} \int_{0}^{2\pi}\sgn{(\cos{\phi})}\left(\int_{-1}^{1} \sgn{(\cos{\theta})}\frac{d \Gamma}{dsdrd(\cos{\theta})d\phi}d\cos{\theta}\right)d\phi \\ \nonumber
  =&\frac{3 \left(4 J^W_{21} + J^W_{26}\right)}{4 \left(9 J^W_1 + J^W_2 + 3 \left(J^W_3 + J^W_4\right)\right)}\\
  \mathcal{A}^{(1)}_{\phi,\Delta}= & \left(\frac{d\Gamma}{dsdr}\right)^{-1} \int_{0}^{2\pi}\sgn{(\cos{\phi})}\left(\int_{-1}^{1} \sgn{(\cos{\Delta})}\frac{d \Gamma}{dsdrd(\cos{\Delta})d\phi}d\cos{\Delta}\right)d\phi \\ \nonumber
  =&\frac{3 \left(4 J^W_{22} + J^W_{25}\right)}{4 \left(9 J^W_1 + J^W_2 + 3 \left(J^W_3 + J^W_4\right)\right)}\\
  \mathcal{A}^{(2)}_{\phi,\theta}= & \left(\frac{d\Gamma}{dsdr}\right)^{-1} \int_{0}^{2\pi}\sgn{(\cos{2\phi})}\left(\int_{-1}^{1} \sgn{(\cos{\theta})}\frac{d \Gamma}{dsdrd(\cos{\theta})d\phi}d\cos{\theta}\right)d\phi \\ \nonumber
  =&\frac{3 J^W_{19}}{\pi \left(9 J^W_1 + J^W_2 + 3 \left(J^W_3 + J^W_4\right)\right)}\\
  \mathcal{A}^{(2)}_{\phi,\Delta}= & \left(\frac{d\Gamma}{dsdr}\right)^{-1} \int_{0}^{2\pi}\sgn{(\cos{2\phi})}\left(\int_{-1}^{1} \sgn{(\cos{\Delta})}\frac{d \Gamma}{dsdrd(\cos{\Delta})d\phi}d\cos{\Delta}\right)d\phi \\ \nonumber
  =&\frac{3 J^W_{20}}{\pi \left(9 J^W_1 + J^W_2 + 3 \left(J^W_3 + J^W_4\right)\right)}
\end{align}
Furthermore, to isolate the two $J$-functions generated by $Q_{l^2WH^2D}^{(5)}$, we define:
\begin{align}
  \mathcal{A}^{(3)}_{\phi,\theta}= & \left(\frac{d\Gamma}{dsdr}\right)^{-1} \int_{0}^{2\pi}\sgn{(\sin{\phi})}\left(\int_{-1}^{1} \sgn{(\cos{\theta})}\frac{d \Gamma}{dsdrd(\cos{\theta})d\phi}d\cos{\theta}\right)d\phi \\ \nonumber
  =&\frac{3 J^W_{23}}{9 J^W_1 + J^W_2 + 3 \left(J^W_3 + J^W_4\right)}\\
  \mathcal{A}^{(3)}_{\phi,\Delta}= & \left(\frac{d\Gamma}{dsdr}\right)^{-1} \int_{0}^{2\pi}\sgn{(\sin{\phi})}\left(\int_{-1}^{1} \sgn{(\cos{\Delta})}\frac{d \Gamma}{dsdrd(\cos{\Delta})d\phi}d\cos{\Delta}\right)d\phi \\ \nonumber
  =&\frac{3 J^W_{24}}{9 J^W_1 + J^W_2 + 3 \left(J^W_3 + J^W_4\right)}.
\end{align}

In addition to showing an idealized case for $h\xrightarrow{}\ell\bar{\nu}_\ell\nu_\ell\bar{\ell}$, the asymmetries above motivate studying fully reconstructible $h \to WW^*$ modes such as $h\to W W\to \nu \ell j j$. This mode suffers from an enormous $W+\text{jets}$ background at the LHC, but it is worth revisiting in the HL-LHC era, perhaps augmented by modern~\cite{Fabbri:2023ncz} or machine learning analysis techniques. At a future lepton collider, such as the FCC-ee, the $W+\text{jets}$ background is far less and this mode (along with the fully hadronic $W^*(jj)W(jj)$ mode~\cite{Selvaggi:2025kmd}) become more viable. A detailed and careful study of this or others fully reconstructible processes, either at the LHC or a future collider, is beyond the scope of this paper.\footnote{$h\to W W^*\to \nu l j j$ or $h \to WW^* \to 4j$ would feature slightly different operators (those involving quarks instead of leptons), but the same asymmetries will be present.Perhaps the biggest difference is that SMEFT generates couplings of $W$ with right-handed quarks. This introduces additional helicity configurations that could lead to additional $J$-functions or possible cancellations. The fully leptonic mode at the FCC-ee was recently studied in~\cite{Kemp:2026mnh}.}.

\subsubsection{Distributions}
Because our analytic calculation assumes we can reconstruct the Higgs rest frame and all final states are visible, it doesn't apply to a realistic experimental setup. Given the two neutrinos in the final state and the fact that the initial longitudinal momentum is not fixed at a hadron collider, this process can only be studied upon the inclusion of the Higgs production, so $pp \to h\to \ell \bar{\nu}_\ell\nu_{\ell'}\bar{\ell'}$. For this reason, we deviate from our analytic analysis and turn to Monte Carlo, generating Higgs production plus decay events with MadGraph5\_aMC@NLO \cite{Alwall:2014hca}\footnote{We have only considered the Higgs produced by gluon-gluon fusion, since this is the dominant production mode.}.

The aforementioned factors also limit the accessible observables for this process. Among these observables, we investigate the $\mathcal{O}(1/\Lambda^4)$ SMEFT effects on the invariant mass of the charged leptons, $m_{\ell \ell'}$, and on the transverse angle, $\Phi_T$, defined as
\begin{equation}
    \cos{\Phi_T}=\frac{\vec{p}_{1,T}\cdot \vec{p}_{4,T}}{|\vec{p}_{1,T}| |\vec{p}_{4,T}|}
\end{equation}
where $\vec{p_{i,T}}$ are the transverse components of the charged leptons. A final consequence of the non-reconstructible final state is that we need to include analysis cuts meant to remove background (meaning, in this case, processes other than $h\to \ell \bar{\nu}_\ell\nu_{\ell'}\bar{\ell'}$) events. We impose the cuts, summarized in Table \ref{cuts}, used by the ATLAS collaboration in their $h\to W^+W^-\to e\nu\mu\nu$ measurement~\cite{ATLAS:2021upe}.

\begin{table}[t!]\centering
\begin{tabular}{c}
\hline
\hline
\makecell{Kinematic Cuts}         \\ 
\hline
\hline
$p_T^{lead}>22  \text{ GeV}$                              \\
$p_T^{sublead}>15  \text{ GeV}$                           \\
$p_T^{miss}>20$\text{ GeV}                                \\
$p^{\ell\ell'}_{T}>30\text{ GeV}$                         \\
$55 \text{ GeV} >m_{\mathcal{\ell \ell'}}>10 \text{ GeV}$ \\
$\Delta\phi_{\ell\ell',E_T^{miss}}>\pi/2$                 \\
$\Phi_T<1.8$                                              \\
\hline
\hline
\end{tabular}
\caption{Kinematic cuts used by the Atlas collaboration, where $p^{\ell\ell'}_{T}$ is the magnitude of vector sum of the charged leptons' transverse momenta, $E_T^{miss}$ is the magnitude of missing transverse momentum, and $\Delta\phi_{\ell\ell',E_T^{miss}}$ is the angle between $\vec{E}_T^{miss}$ and $\vec{p}^{\ \ell\ell'}_{T}$ }
\label{cuts}
\end{table}

While our calculation in Section \ref{Section 4.2.1} doesn't directly apply to $pp \to h\to \ell \bar{\nu}_\ell\nu_{\ell'}\bar{\ell'}$, it is still helpful for building intuition about the impact SMEFT operators have on the transverse angle distribution. In analogy to the \textit{SM-like angular distribution} and the \textit{non-SM like angular distribution} explained in Section \ref{Section 3.2.1}, we can think of something similar for the differential distribution of $\Phi_T$ . Here the differential distribution of $\Phi_T$ in the SM limit has a particular shape, so SMEFT effects can manifest in two ways: with a similar shape to the SM $\Phi_T$ distribution, a \textit{SM-like $\Phi_T$ distribution}, or with a different shape to the SM $\Phi_T$ distribution, a  
\textit{non SM-like $\Phi_T$ distribution}.

Naturally, the inclusion of the Higgs production makes the kinematics of the process more complex. Additionally, since $\Phi_T$ is a transverse variable, it inevitably loses some information relative to $(\phi,\theta,\Delta)$ in the Higgs rest frame. These factors lead to mixing of the rest frame angles, resulting in a nontrivial mapping from $(\phi,\theta,\Delta)$ to $\Phi_T$. Nevertheless, we argue that the transformation of SM-like and non SM-like angular distributions is the same (or, the same for SM and SMEFT). This means that if an operator in Table \ref{hww operators} contributes with a  \textit{SM-like angular distribution}  in the Higgs' rest frame, it will contribute to the $\Phi_T$ distribution with a \textit{SM-like $\Phi_T$ distribution}. Likewise, an operator with a \textit{non SM-like angular distribution}, will contribute with a  \textit{non SM-like $\Phi_T$ distribution}.

\begin{figure}[t]
  \begin{subfigure}[b]{0.5\textwidth}
    \includegraphics[width=\textwidth]{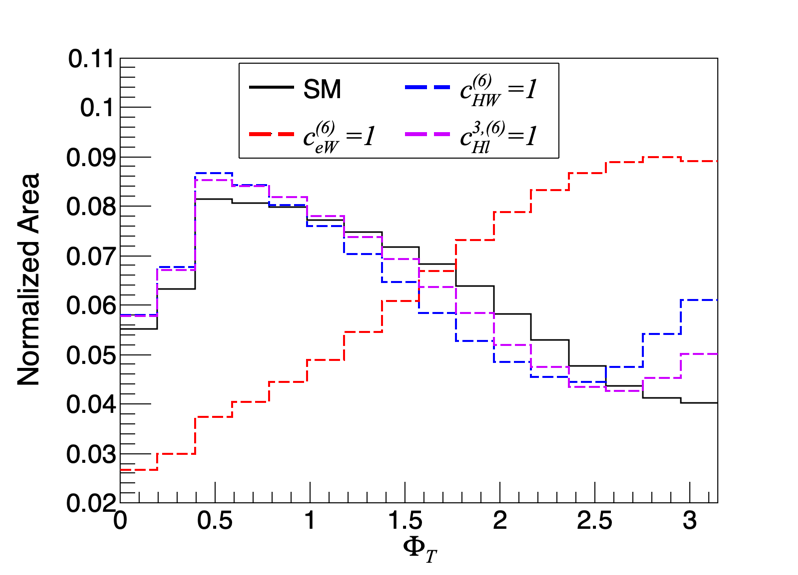}
    \subcaption{}
    \label{isolated dim6 distribution}
  \end{subfigure}
  \begin{subfigure}[b]{0.5\textwidth}       
    \includegraphics[width=\textwidth]{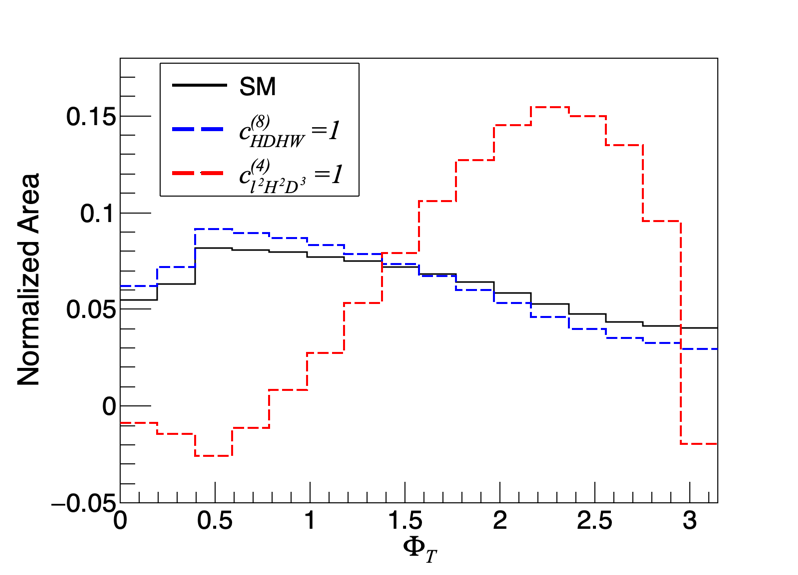}
    \subcaption{}
    \label{isolated dim8 distribution}
  \end{subfigure}
  \caption{Normalized $\Phi_T$ distribution of SM interference terms with SMEFT. Figure (a) shows the interference of dimension six operators with the SM, plus the squared contribution of the respective operator. Figure (b) shows the interference with the SM of the respective dimension 8 operator for. In both cases, the SM is represented in black for comparison. We set $\Lambda=1\, \text{TeV}$ and the nonzero coefficient Wilson Coefficient is set to 1.}
  \label{isolated distribution}
\end{figure}

As a proof of concept, in Figure \ref{isolated distribution} we show the area-normalized differential $\Phi_T$   distribution, formed with the events generated with MadGraph5. In these distributions, we only consider the interference terms of the SM with SMEFT, turning on one operator at a time, \textit{i.e.},  $2\text{Re}(A^*_{SM}A_{6}(c_i))+|A_{6}(c_i)|^2$ for dimension six operators, and $2\text{Re}(A^*_{SM}A_{8}(c_i))$ for dimension eight operators. This way, by focusing on the shape of these distributions, one can discern the type of $\Phi_T$ distribution arising solely from higher dimensional operators. We picked the dimension six operators $Q_{HW}^{(6)}$, $Q_{Hl}^{3,(6)}$, and $Q_{eW}^{(6)}$ in Figure \ref{isolated distribution}\subref{isolated dim6 distribution}, and the dimension 8 operators $Q_{HDHW}^{(8)}$ , and $Q_{l^2H^2D^3}^{(4)}$ Figure \ref{isolated distribution}\subref{isolated dim8 distribution}. We set $\Lambda=1 \text{ TeV}$ and, for illustrative purposes, the nonzero coefficient is set to 1.

In agreement with Table \ref{hww operators}, the $\Phi_T$ distributions of $Q_{HW}^{(6)}$ and $Q_{Hl}^{3,(6)}$ in Figure \ref{isolated distribution}\subref{isolated dim6 distribution} have a similar shape as the SM distribution. In contrast, $Q_{eW}^{(6)}$ has a very different shape than the SM.  Similarly, in Figure \ref{isolated distribution}\subref{isolated dim8 distribution} we see that $Q_{HDHW}^{(8)}$ has a similar shape to the SM distribution, while $Q_{l^2H^2D^3}^{(4)}$ has a radically different shape. Therefore, while we can't predict the exact $\Phi_T$ distribution from our analytic calculation, it can give us some intuition for how SMEFT operators modify it. 
Of course, once we stack the SMEFT distributions on top of the SM contribution, (rather than showing each piece area normalized), the deviations will be suppressed by the relative magnitude of the SM vs. SMEFT terms.

\begin{figure}[t]
  \begin{subfigure}[b]{0.5\textwidth}
  \includegraphics[width=\textwidth]{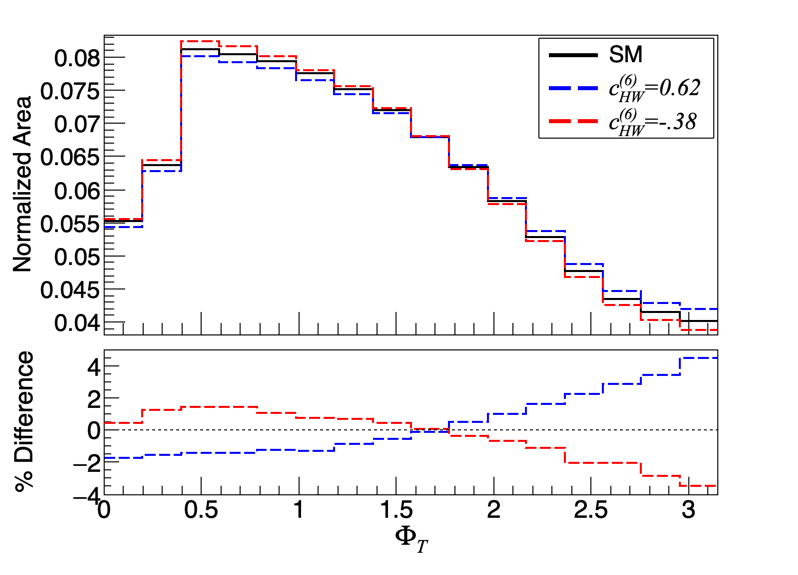}
  \end{subfigure}
    \begin{subfigure}[b]{0.5\textwidth}    \includegraphics[width=\textwidth]{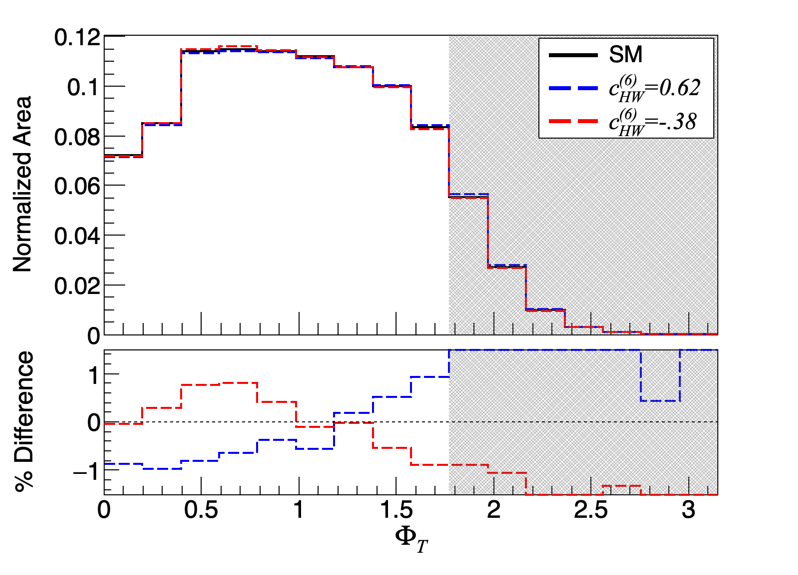}
  \end{subfigure}
  \begin{subfigure}[b]{0.5\textwidth}
  \includegraphics[width=\textwidth]{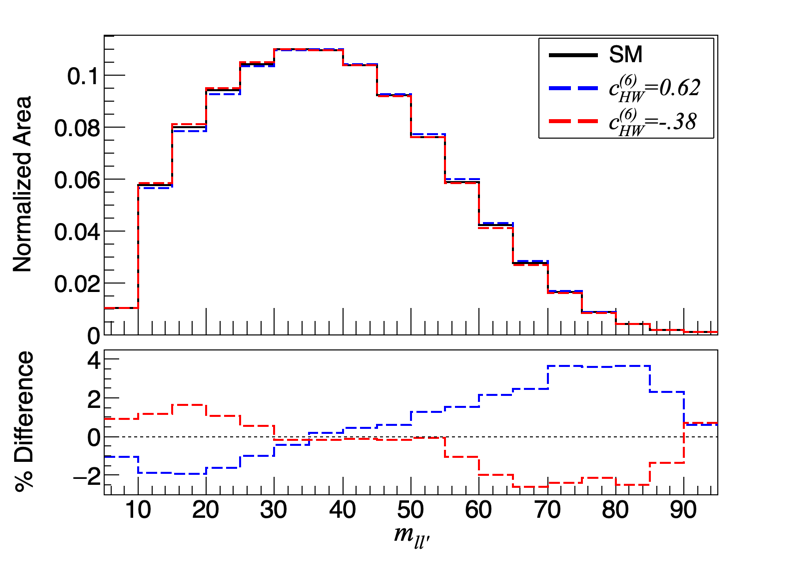}
  \end{subfigure}
    \begin{subfigure}[b]{0.5\textwidth}    \includegraphics[width=\textwidth]{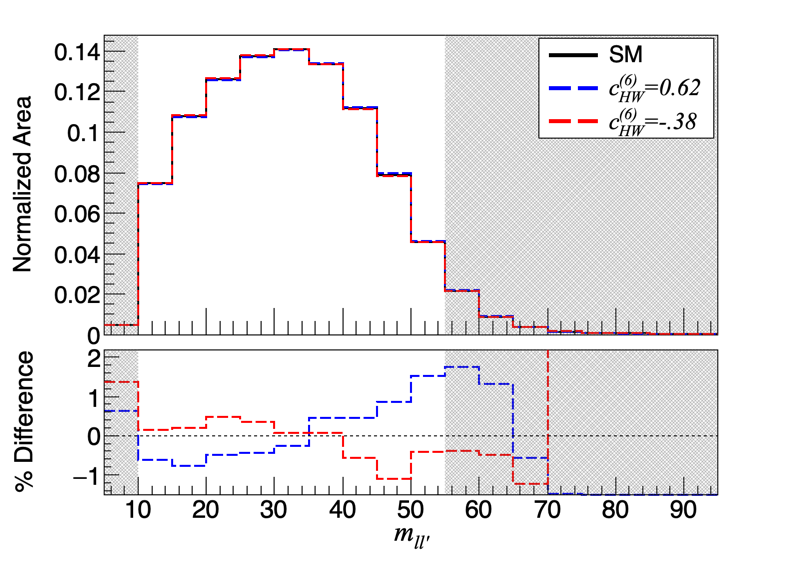}
  \end{subfigure}
  \caption{Contribution of $Q_{HW}^{(6)}$ to the normalized differential distributions of $\Phi_T$ (top two panels) and $m_{\ell\ell'}$ (bottom two panels).  In the two left panels, no cuts have been implemented, while in the two right panels, the cuts from Table \ref{cuts} have been implemented. The cuts on the respective observable discard the shaded regions. For the SMEFT distributions, we set $\Lambda=1\, \text{TeV}$ and all other Wilson coefficients to zero.}
  \label{histograms cHW}
\end{figure}

The SM plus SMEFT differential distributions for several different Wilson coefficient choices at dimension six and eight are shown below in Figs.~\ref{histograms cHW}-\ref{histograms dim8}. Based on our analytic intuition, we have chosen to study the operators that introduce a \textit{non SM-like $\Phi_T$ distribution}. Specifically, these operators are $Q_{eW}^{(6)}$ and $Q_{l^2H^2D^3}^{(4)}$. However, as we saw in the $h\xrightarrow{}Z(\xrightarrow{}\ell'\bar{\ell'})\ell\bar{\ell}$ analysis, operators contributing with a \textit{SM-like $\Phi_T$ distributions} can still have a noticeable effect. To show these, we have chosen $Q_{HW}^{(6)}$, $Q_{Hl}^{3,(6)}$, $Q_{HDHW}^{(8)}$, $Q_{l^2H^2D^3}^{(3)}$ and $Q_{l^2WH^2D}^{(3)}$, which span all the effective couplings on Table \ref{coupling factors hww}.~\footnote{With the exception of $h_{IJ}$. We have not considered operators from this connection because these operators change the normalization of the Higgs, and shift the $hWW$ vertex. Since we are focusing on normalized distributions, the operators in this connection won't affect the normalized distribution.} As in the four charged lepton case, we take $\Lambda=1\, \text{TeV}$, and set the non-dipole dimension six operators to the upper and lower bounds set by ~\cite{Ellis:2020unq}. For the dipole operator we take an $\mathcal{O}(1)$ Wilson coefficient. In all cases, we present distributions with and without the implemented cuts to show how they affect SMEFT contributions.

\begin{figure}[t]
  \begin{subfigure}[b]{0.5\textwidth}
  \includegraphics[width=\textwidth]{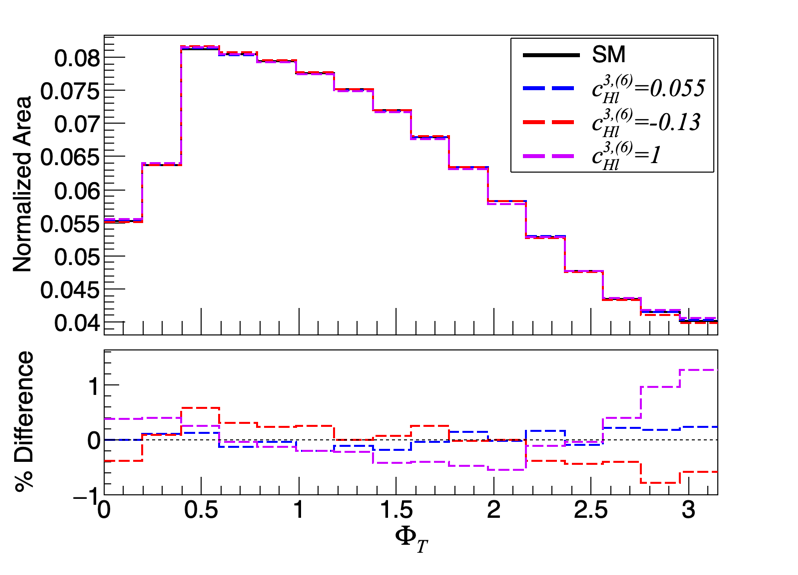}
  \end{subfigure}
    \begin{subfigure}[b]{0.5\textwidth}    \includegraphics[width=\textwidth]{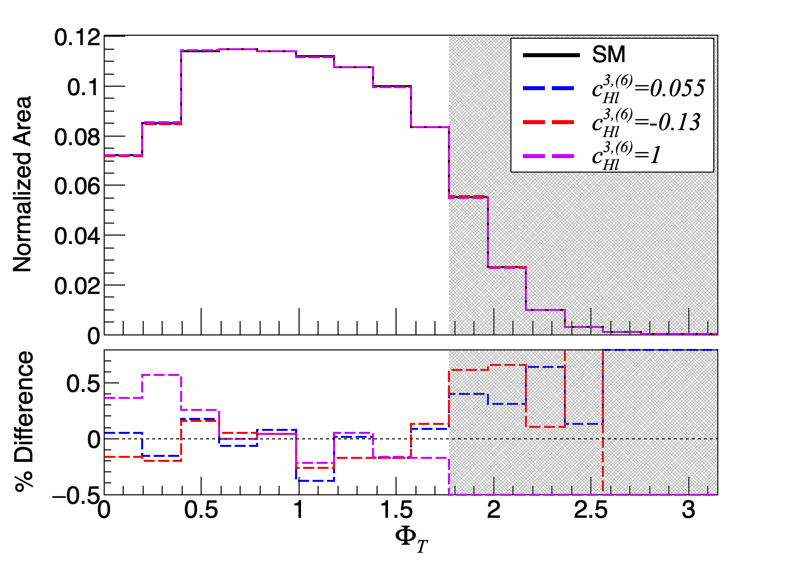}
  \end{subfigure}
  \begin{subfigure}[b]{0.5\textwidth}
  \includegraphics[width=\textwidth]{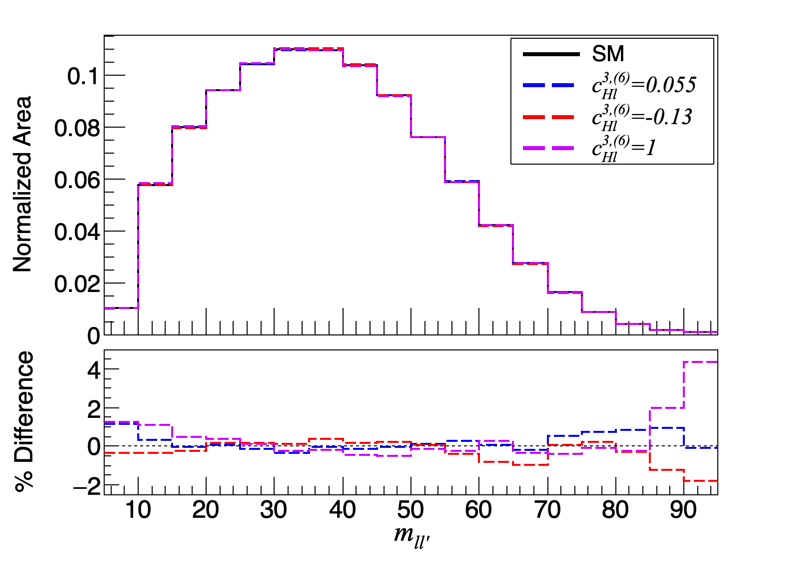}
  \end{subfigure}
    \begin{subfigure}[b]{0.5\textwidth}    \includegraphics[width=\textwidth]{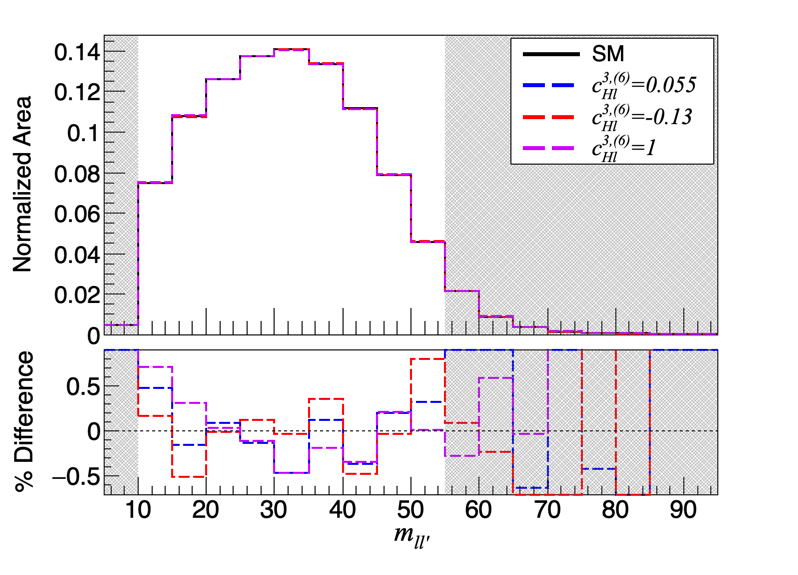}
  \end{subfigure}
  \caption{Contribution of $Q_{Hl}^{3,(6)}$ to the normalized differential distributions of $\Phi_T$ (top two panels) and $m_{\ell\ell'}$ (bottom two panels). In the two left panels, no cuts have been implemented, while in the two right panels, the cuts from Table \ref{cuts} have been implemented. The cuts on the respective observable discard the shaded regions. For the SMEFT distributions, we set $\Lambda=1\, \text{TeV}$ and all other Wilson coefficients to zero.}
  \label{histograms cHl3}
\end{figure}

\begin{figure}[t]
  \begin{subfigure}[b]{0.5\textwidth}
  \includegraphics[width=\textwidth]{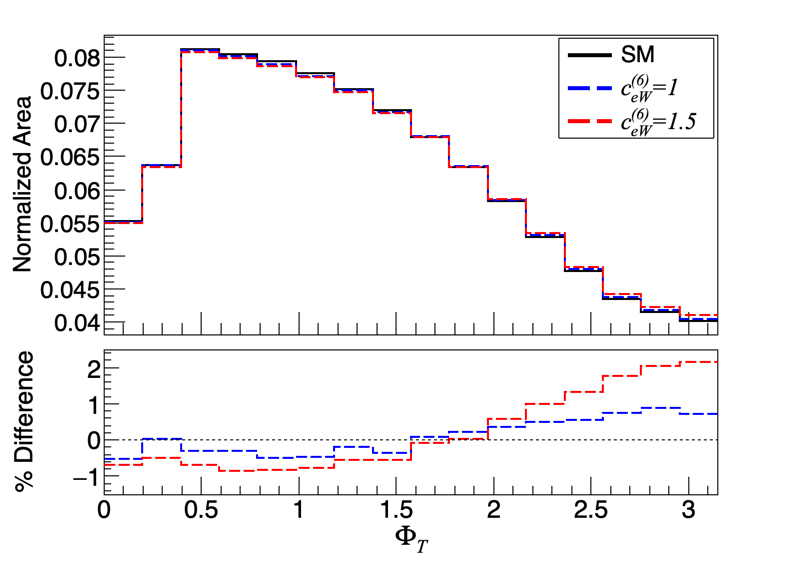}
  \end{subfigure}
    \begin{subfigure}[b]{0.5\textwidth}    \includegraphics[width=\textwidth]{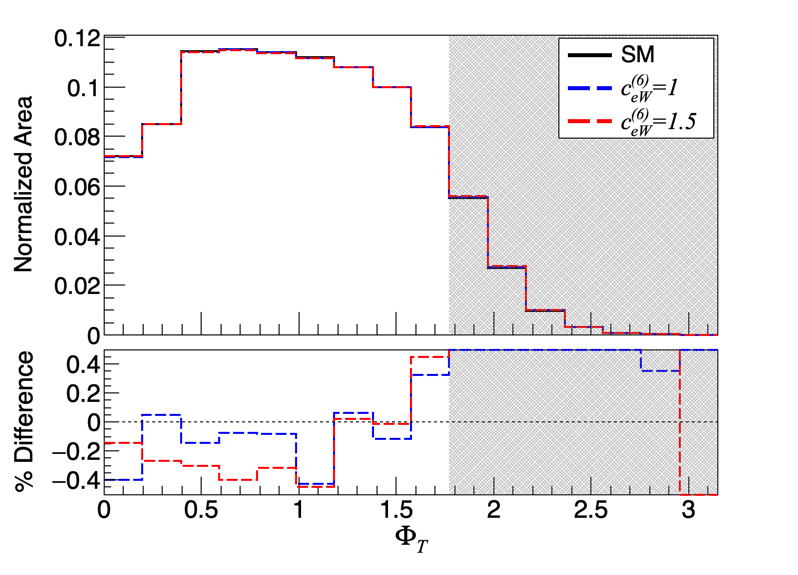}
  \end{subfigure}
  \begin{subfigure}[b]{0.5\textwidth}
  \includegraphics[width=\textwidth]{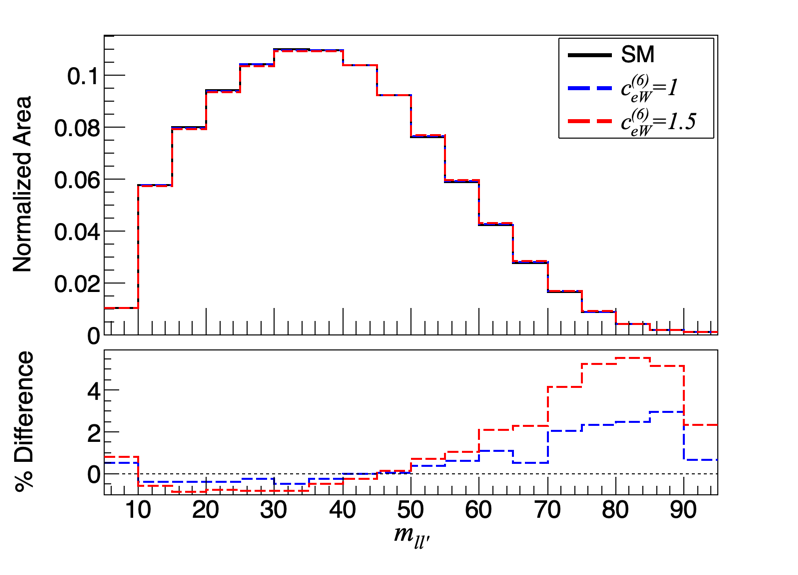}
  \end{subfigure}
    \begin{subfigure}[b]{0.5\textwidth}    \includegraphics[width=\textwidth]{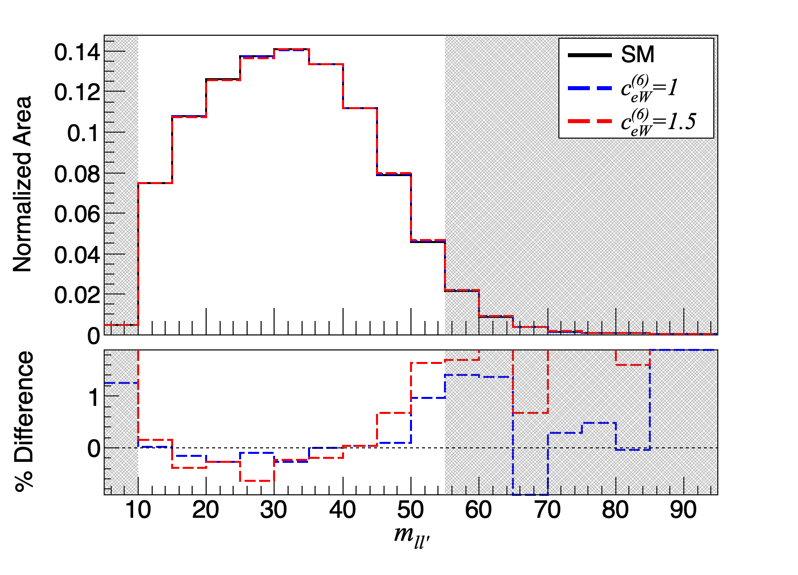}
  \end{subfigure}
  \caption{Contribution of $Q_{eW}^{(6)}$ to the normalized differential distributions of $\Phi_T$ (top two panels) and $m_{\ell\ell'}$ (bottom two panels). In the two left panels, no cuts have been implemented, while in the two right panels, the cuts from Table \ref{cuts} have been implemented. The cuts on the respective observable discard the shaded regions. For the SMEFT distributions, we set $\Lambda=1\, \text{TeV}$ and all other Wilson coefficients to zero.}
  \label{histograms dipole}
\end{figure}

We don't consider the operator $Q_{l^2WH^2D}^{(5)}$, despite the fact that it contributes with a \textit{non SM-like $\Phi_T$ distribution}, as it's contribution to the $\Phi_T$ distribution is three orders of magnitude smaller than any other operator. We can understand this from our analytic calculation. Specifically, we see that this operator contributes to the $J$-functions $J_{24}^W$ and $J_{25}^W$, and from the explicit expressions for the $J$-functions in Appendix \ref{J^W explicit expressions}, we find that the contribution of $Q_{l^2WH^2D}^{(5)}$ is proportional to $(s-r)\sim (k_1^2-k_2^2)$. Therefore, depending on which $W$ boson's momentum is larger, the interference of $Q_{l^2WH^2D}^{(5)}$ with the SM can be negative or positive. Integrated over phase space, this results in a cancellation and a small  -- but non-zero -- contribution.~\footnote{While $Q_{l^2WH^2D}^{(5)}$ is irrelevant for our study, it could be relevant in a fully reconstructible process, like the mentioned semi-leptonic decay $h\to WW^*\to \ell\nu jj$, were $J_{23}^W$ and $J_{24}^W$ can be studied with the use of asymmetries.}

\begin{figure}[t]
  \begin{subfigure}[b]{0.5\textwidth}
    \includegraphics[width=\textwidth]{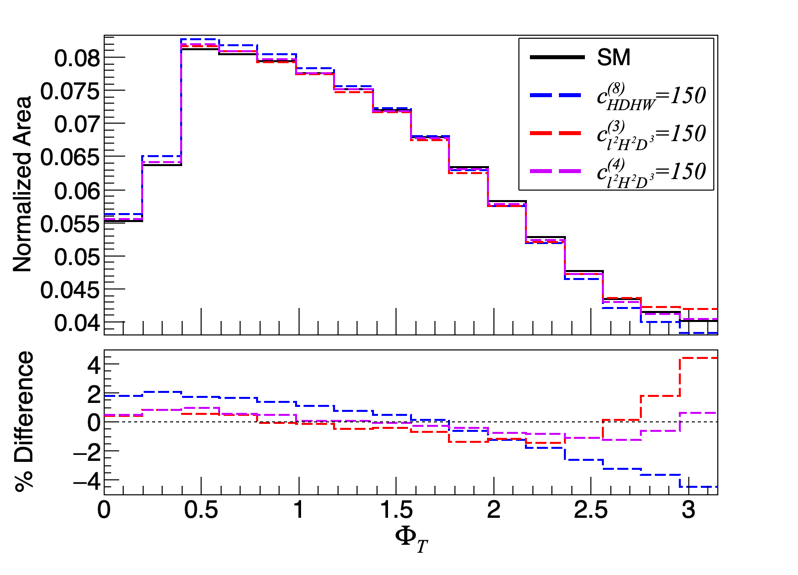}
  \end{subfigure}
  \begin{subfigure}[b]{0.5\textwidth}    
    \includegraphics[width=\textwidth]{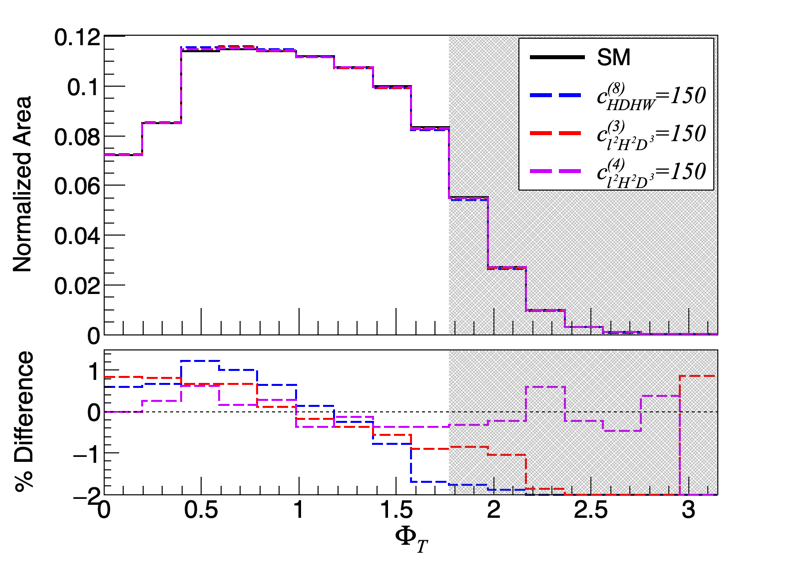}
  \end{subfigure}
  \begin{subfigure}[b]{0.5\textwidth}
  \includegraphics[width=\textwidth]{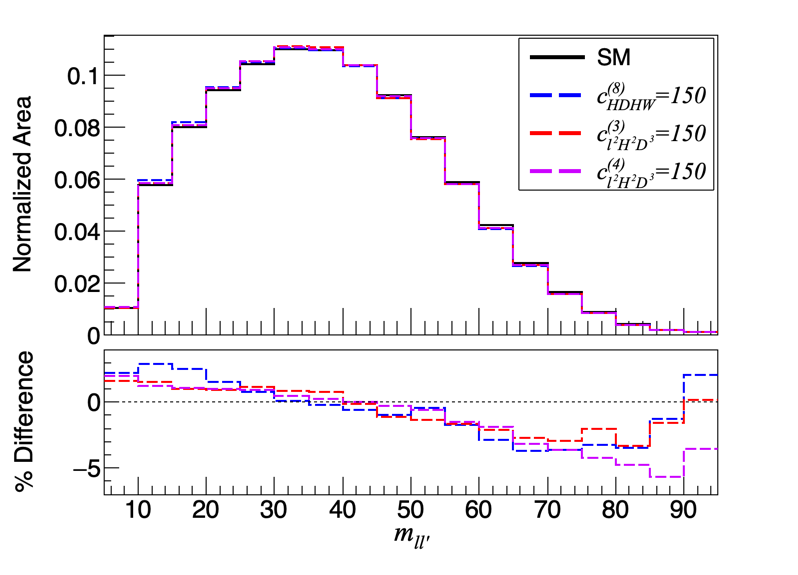}
  \end{subfigure}
    \begin{subfigure}[b]{0.5\textwidth}    \includegraphics[width=\textwidth]{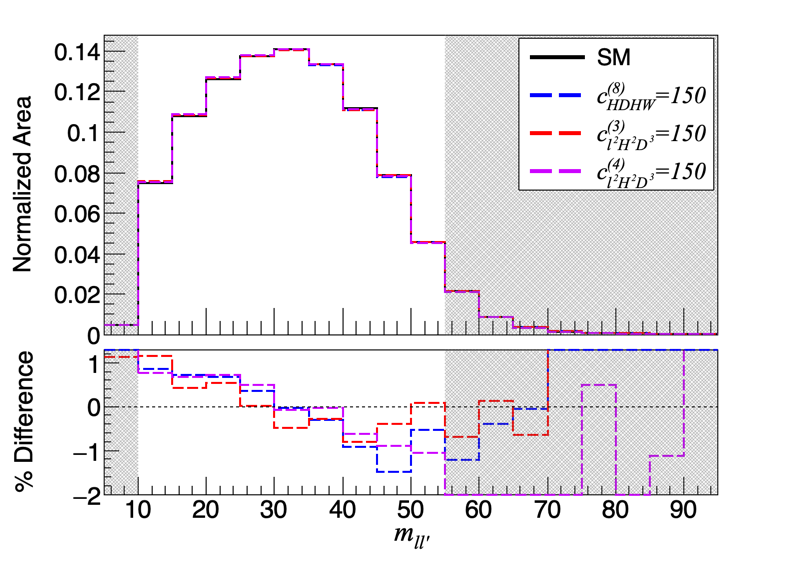}
  \end{subfigure}
  \caption{Dimension 8 operators' contribution to  $\Phi_T$ (top two panels) and $m_{\ell\ell'}$ (bottom two panels).  In the two left panels, no cuts have been implemented, while in the two right panels, the cuts from Table \ref{cuts} have been implemented. The cuts on the respective observable discard the shaded regions. For the SMEFT distributions, we set $\Lambda=1\, \text{TeV}$ and all other Wilson coefficients to zero.}
  \label{histograms dim8}
\end{figure}
In Fig.~\ref{histograms cHW}, we find that $Q_{HW}^{(6)}$ has the largest contribution to both observables. Before cuts, using values consistent with the bounds in Ref~\cite{Ellis:2020unq}, the deviation from the SM is visible at the 4\% level. This is clearly driven by the $\mathcal O(1/\Lambda^2)$ interference piece, as the interference changes from destructive to constructive as we vary the sign of $c_{HW}^{(6)}$. After cuts, we find it's contribution fades to sub-percent level. This is expected as these cuts assume and exploit SM like interactions, particularly the cuts imposed to $\Phi_T$ and $m_{\ell\ell'}$ \cite{Dittmar:1996ss}, forcing the SMEFT distributions to a SM distribution. While most of these cuts are unavoidable by experimental limitations and to exclude the background, relaxing the cuts on $\Phi_T$ and $m_{\ell\ell'}$ may help to study $Q_{HW}^{(6)}$ effects.

Figure~\ref{histograms cHl3} shows the contribution from $Q_{Hl}^{3,(6)}$. We find it is small compared to $Q_{HW}^{(6)}$. This can partially be attributed to the small Wilson coefficient for $Q_{Hl}^{3,(6)}$ (set by the constrains in Ref. \cite{Ellis:2020unq}), however there is also some cancellation present along the lines of what we observed in Sec.~\ref{sec:hzzasymm}. To show this, we artificially increased the  Wilson coefficient to $\mathcal{O}(1)$, shown in the purple in Fig.~\ref{histograms cHl3}. The effect is larger with the enhanced coefficient, however, it is still smaller than what we observed for $Q_{HW}^{(6)}$.

The dipole operator $Q_{eW}^{(6)}$, shown in the top left panel of Fig.~\ref{histograms dipole} generates a $\sim$ few percent contribution to the $\Phi_T$ distribution difference despite being suppressed by $1/\Lambda^4$. The deviations are largest at large values of $\Phi_T$. The fact that we see a deviation despite the $1/\Lambda^4$ is due -- just as in the four charged lepton case -- to the fact that the dipole term is a `self square' and is not accompanied by any powers of SM couplings. However, as with $Q_{HW}^{(6)}$, the current analysis cuts significantly reduce the differences between the dipole terms and the SM distributions.

Continuing with the dimension eight operators, we find that the contribution is minimal if we consider $\mathcal{O}(1)$ Wilson coefficients. The same reasons why dimension eight operators' contribution are suppressed in $h\to\ell\bar{\ell}(Z\to \ell' \bar{\ell'})$ apply here, namely the lack of any energy enhancement to compensate for extra powers of $v^2/\Lambda^2$ and the fact that all interference terms carry powers of (small) SM couplings. In this case there is no $Z$ boson propagator enhancement, since here we don't take the narrow-width approximation. The absence of energy enhancement may not be obvious as the Higgs production has been included. However, the Higgs is approximately on shell because $\Gamma_h\ll m_h$, therefore the energy scale of the process is $\sqrt{s}\sim m_H$. To appreciate the effects dimension eight operators have, in Figure \ref{histograms dim8} we have taken large enough Wilson coefficients to observe a noticeable effect on the $\Phi_T$ and $m_{\ell\ell'}$ normalized distributions. As before, we present distributions with and without cuts implemented.

In the top-left panel of Figure \ref{histograms dim8} we observe that the three operators under consideration contribute similarly to the $\Phi_T$ distribution, with very subtle differences around $\Phi_T\approx3$. Among these operators, $Q_{HDHW}^{(8)}$ show the largest deviation from the SM, whereas $Q_{l^2H^2D^3}^{(3)}$ and $Q_{l^2H^2D^3}^{(4)}$ show comparable contributions. Likewise, in the bottom-left panel all operators contribute similarly to the $m_{\ell\ell'}$ distribution,with $Q_{HDHW}^{(8)}$ showing the largest deviation, although the contribution is smaller compared with that for $\Phi_T$.

Inspecting the $\Phi_T$ distributions for $Q_{eW}^{(6)}$ and $Q_{l^2H^2D^3}^{(4)}$ we note that the \textit{non SM-like $\Phi_T$ distribution} we observed in Figure \ref{isolated distribution} gets diluted by the dominant SM contribution. Furthermore, all the SMEFT contributions coming from different operators aren't particularly distinguishable between them. For $m_{\ell\ell'}$ the situation is not better, as in general we find it to be less sensitive to SMEFT effects than $\Phi_T$. Therefore, we conclude that the limited observables available for $h \to \ell \bar{\nu}_\ell \nu_{\ell'} \bar{\ell'}$ obscure the novel kinematics SMEFT induces to this process.

 \section{Conclusions} \label{section 5}
In this paper, we have performed the first analysis of Higgs decay to four leptons up to $\mathcal{O}(1/\Lambda^4)$ in the SMEFT framework, extending previous $\mathcal{O}(1/\Lambda^2)$ studies by including both quadratic dimension six and interference dimension eight operator contributions. Using the geoSMEFT formulation, we analytically calculated the spinor-helicity amplitudes and differential decay rates for both the four charged lepton case, $h \to \ell \bar{\ell} \left(Z \to \ell' \bar{\ell'}\right)$, and the two charged lepton case $h \to \ell \bar{\nu}_\ell \nu_{\ell'} \bar{\ell'}$,  focusing on the kinematic structures induced by higher-dimensional operators.

For the four charged-lepton final state, we find that, despite the presence of new kinematic structures, this process exhibits an angular dependence similar to the SM. Since this process is fully reconstructible, we studied the $\mathcal{O}(1/\Lambda^4)$ SMEFT effects on the dilepton invariant mass and angular asymmetries, with the former proving generally more sensitive. Within the current bounds on dimension six operators, we identified $Q_{HW}^{(6)}$ as the operator with the largest contribution due a photon pole enhancement, which arises at  $\mathcal{O}(1/\Lambda^4)$ as a dimension six squared contribution.

In contrast, for the two charged lepton case we find that the new kinematic structures lead to novel angular dependence. We identify this as purely $\mathcal{O}(1/\Lambda^4)$ effect, generated by the operators $Q_{l^2H^2D^3}^{(4)}$, $Q_{l^2WH^2D}^{(5)}$, and  $Q_{eW}^{(6)}$ (as a dimension six squared effect). 
However, the non reconstructible nature of this process obscures these BSM effects. At a hadron collider, we cannot determine the Higgs rest frame and must consider the combined production plus decay, $pp \to h \to \ell \bar{\nu}_\ell \nu_{\ell'} \bar{\ell}'$. Not only does $pp \to h \to \ell \bar{\nu}_\ell \nu_{\ell'} \bar{\ell}'$ posses fewer useful observables, these observables are less sensitive to SMEFT effects than the asymmetries and invariant masses in $h \to \ell\bar\ell(Z\to \ell'\bar{\ell}')$, especially after analysis cuts are applied. The cuts are designed to pick out SM-like interactions, and relaxing them could help regain sensitivity to SMEFT.
`
In both decay channels, we find that dimension eight operators have a small contribution compared with the dimension six contributions, requiring  Wilson coefficient of $\mathcal{O}(50-250)$ to be observable at the LHC. This is mainly due to the additional SM couplings suppression accompanying dimension eight contributions, and the fact that $\sqrt{s}=m_h\sim v$ so no `energy enhancement' is possible. Additionally, dipole operators, though tightly constrained, could have a significant impact if these bounds are relaxed, as their non-interfering nature and non-SM helicity structures can lead to sizable effects at $\mathcal{O}(1/\Lambda^4)$.

Our analytic, rest frame calculations can also be applied to a future lepton collider such as the FCC-ee. There, the reduced SM background from process like $W + \text{jets}$ and the fact that the Higgs rest frame can be reconstructed -- regardless of its decay -- by measuring all other final momenta mean the semi-leptonic and fully hadronic $h \to WW^*$ decay modes are viable channels. This opens the possibility to study asymmetries that are sensitive to novel $\mathcal O(1/\Lambda^4)$ kinematics with no pollution from the SM, rather than looking for effects on top of a SM distribution.

\section*{Acknowledgements}
The authors thank Diogo Boito and Olivier Mattelaer for helpful correspondence. The work of AM is partially supported by the National Science Foundation under Grant Number PHY-2412701.

\appendix
\section{Kinematics}\label{kinematics appendix}
\begin{figure}[t] 
    \centering
    \includegraphics[width=\textwidth]{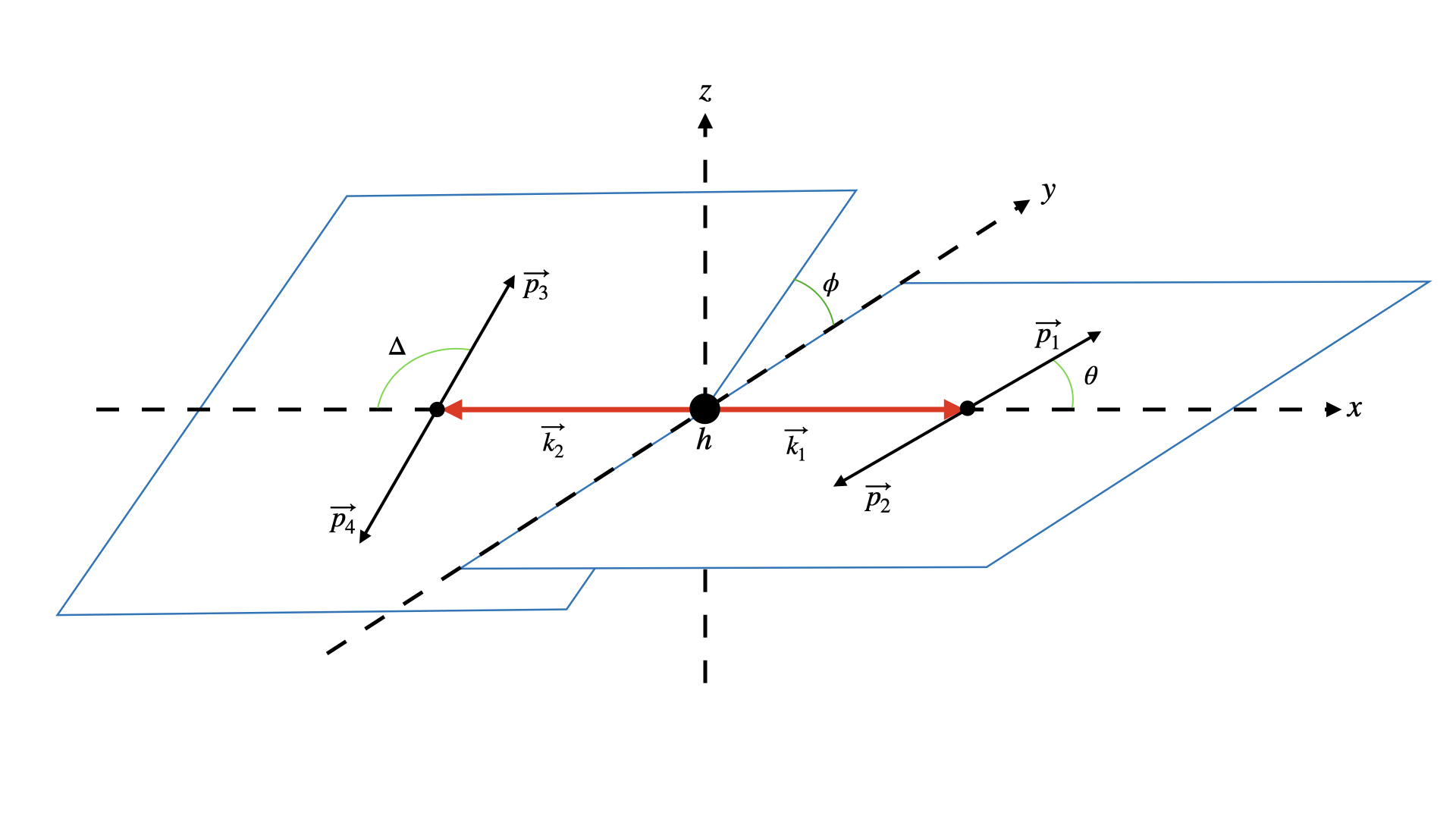}
  \caption{Kinematics of the four-body decay $h(p)\to\psi_1(p_1)\psi_2(p_2)\psi_3(p_3)\psi_4(p_4)$.}
  \label{higgs rest frame}
\end{figure}

In this Appendix we present the kinematics used in our analytic results following \cite{Buchalla:2013mpa, Beneke:2014sba}, which is based on \cite{Cabibbo:1965zzb,Pais:1968zza}. We take the momenta label as $h(p)\to\psi_1(p_1)\psi_2(p_2)\psi_3(p_3)\psi_4(p_4)$ to generically refer to both the four charged lepton and the two charged lepton case, furthermore we label the momenta of the two dilepton pairs $\psi_1\psi_2$ and $\psi_3\psi_4$ as $k_1=p_1+p_2$ and $k_2=p_3+p_4$, respectively. We set the $xyz$ coordinate frame in Higgs rest frame, as depicted in Figure \ref{higgs rest frame}, with the $x$-axis aligned with $\vec{k}_1$, the spatial component of $k_1$. In this frame, a convenient set of kinematic variables is $(s,r,\phi, \theta,\Delta)$, defined as:
\begin{itemize}
    \item $r\equiv k_2^2/m_h^2$, defined on the range $0\leq r\leq1$, is the invariant mass of the dilepton system $\psi_3\psi_4$ normalized by the Higgs' mass. In the narrow-width approximation, this variable takes the value $r\equiv m_Z^2/m_h^2\approx.53$
    \item $s\equiv k_1^2/m_h^2$, defined on the range $0\leq s\leq (1-\sqrt{r})^2$, is the invariant mass of the dilepton system $\psi_1\psi_2$ normalized by the Higgs' mass. In the narrow-width approximation, the range of this variable becomes $0\leq s \lesssim 0.075$.
    \item $\theta$, defined on the range $0\leq\theta\leq\pi$, is the angle between spatial momentum  $\vec{p}_1$ and the $x$-axis measured in the $\psi_1\psi_2$ rest frame.
    \item $\Delta$, defined on the range $0\leq\Delta\leq\pi$, is the angle between spatial momentum  $\vec{p}_3$ and the $x$-axis measured in the $\psi_3\psi_4$ rest frame.
    \item $\phi$, defined on the range $0\leq\theta<2\pi$, is the angle between the decay planes formed by the dilepton systems $\psi_1\psi_2$ and $\psi_3\psi_4$. It is measured positively from the $\psi_1\psi_2$ plane to the $\psi_3\psi_4$ plane.
\end{itemize}

In terms of these kinematic variables and in the massless lepton limit, the leptons' momenta in their respective di-lepton center of mass frame are given by:
\begin{align}
    p_1^\mu&=\frac{m_H\sqrt{s}}{2}(1,\cos{\theta},\sin{\theta},0), \\  \nonumber
    p_2^\mu&=\frac{m_H\sqrt{s}}{2}(1,-\cos{\theta},-\sin{\theta},0), \\ \nonumber
    p_3^\mu&=\frac{m_H\sqrt{r}}{2}(1,-\cos{\Delta},\sin{\Delta}\cos{\phi},\sin{\Delta}\sin{\phi}), \\ \nonumber
    p_4^\mu&=\frac{m_H\sqrt{r}}{2}(1,\cos{\Delta},-\sin{\Delta}\cos{\phi},-\sin{\Delta}\sin{\phi}).
\end{align}
Boosting these momenta to the Higgs rest frame, and taking the appropriate inner products, the relevant Lorentz invariants in terms of $(s,r,\phi, \theta,\Delta)$ are:
\begin{align}
    s_{12}&=m_h^2s, \quad  s_{34}=m_h^2r, \quad \epsilon_{\mu\nu\lambda\rho}p_1^\mu p_2^\nu p_3^\lambda p_1^\rho=\frac{m_h^2}{8}\lambda \sqrt{rs}s_\theta s_\Delta s_\phi,\\ \nonumber 
    s_{13}&=\frac{m_h^2}{4}\left[\kappa(1+c_\theta c_\Delta)+\lambda(c_\theta+c_\Delta)-2\sqrt{r s}\ s_\theta s_\Delta c_\phi\right], \\ \nonumber
    s_{14}&=\frac{m_h^2}{4}\left[\kappa(1-c_\theta c_\Delta)+\lambda(c_\theta-c_\Delta )+2\sqrt{r s}\ s_\theta s_\Delta c_\phi\right], \\ \nonumber
    s_{23}&=\frac{m_h^2}{4}\left[\kappa(1-c_\theta c_\Delta)-\lambda(c_\theta-c_\Delta )+2\sqrt{r s}\ s_\theta s_\Delta c_\phi\right], \\ \nonumber
    s_{24}&=\frac{m_h^2}{4}\left[\kappa(1+c_\theta c_\Delta)-\lambda(c_\theta+c_\Delta)-2\sqrt{r s}\ s_\theta s_\Delta c_\phi\right].
\end{align}
Where we used the shorthand $c_\alpha \equiv \cos{\alpha},$ $s_\alpha \equiv \sin{\alpha}$ $( \alpha=\theta,\Delta,\phi)$, and introduced the kinematic functions
\begin{equation}
    \kappa\equiv1-r-s, \quad \lambda\equiv\sqrt{1+s^2+r^2-2s-2r-2rs}
\end{equation}

\section{Example helicity combination and effective couplings for  $h \to \ell \bar{\nu}_\ell \nu_{\ell'} \bar{\ell'}$ }\label{example Appendix}
In this appendix we define the effective couplings and present the explicit expression for the kinematic structures introduced in Section \ref{section 4.1} . The effective couplings introduced in Table \ref{coupling factors hww} and Table \ref{helicity combinations hww} are defined in terms of the Feynman rules as the coefficients multiplying the different Lorentz structures. Their expansion in terms of Wilson coefficients can be found in Appendix \ref{effective couplings appendix}
\begin{itemize}
    \item $g^{W^\pm}$ is the coupling of $W^\pm$ with left handed leptons. This coupling is complex and under complex conjugation satisfies $(g^{W^-})^*=g^{W^+}$. This vertex has the same Lorentz structure as the SM, so SMEFT effects only enter in the coupling strength. 
    \item $c^{(i)}_{hWW}$ are the couplings of the different type of Lorentz structures involved in the $hWW$ vertex:
    \begin{align}
            h(p)W^\mu(k_1) W^\nu(k_2): \quad & ic^{(1)}_{hWW}g^{\mu\nu}+ic^{(2)}_{hWW}4(k_1^\nu k_2^\mu -k_1\cdot k_2 g^{\mu\nu})\\ \nonumber  &+ic^{(3)}_{hWW}(p^\mu k_1^\nu-p\cdot k_1g^{\mu\nu})+ic^{(4)}_{hWW}(p^\mu k_2^\nu-p\cdot k_2g^{\mu\nu}). 
    \end{align}
    Here, $c^{(4)}_{hWW}=\left(c^{(3)}_{hWW}\right)^*$
    \item $g^{(i)}_{hW^\pm\psi}$ are the couplings of the Lorentz structures of the 4-point contact vertex $hW\nu l$. Taking all momentum incoming, the Lorentz structures for this vertex are:
    \begin{align}
            \label{couplings hWlnu}
            h(p)W^{-\mu}(k_2)l^+(p_1)\nu(p_2): \quad & ig^{(1)}_{hW^- l}\gamma^\mu-ig^{(2)}_{hW^-l}p \cdot p_1\gamma^\mu-ig^{(3)}_{hW^- l}p \cdot p_2 \gamma^\mu \\ 
            \nonumber &-ig^{(4)}_{hW^- l}p \cdot k_2 \gamma^\mu-ig^{(5)}_{hW^- l}p^\mu \slashed{k_2}-ig^{(6)}_{hW^- l}p_2^\mu \slashed{p} \\
            &\nonumber -ig^{(7)}_{hW^- l}p^\mu \slashed{p}-ig^{(8)}_{hW^- l}p_1^\mu \slashed{p}. \\ 
           h(p)W^{+\mu}(k_1)\nu(p_3)l^-(p_4): \quad & ig^{(1)}_{hW^+l}\gamma^\mu-ig^{(2)}_{hW^+l}p \cdot p_3\gamma^\mu-ig^{(3)}_{hW^+l}p \cdot p_4 \gamma^\mu
           \\& \nonumber-ig^{(4)}_{hW^+l}p \cdot k_2 \gamma^\mu -ig^{(5)}_{hW^+l}p^\mu \slashed{k_2}-ig^{(6)}_{hW^+l}p_4^\mu \slashed{p}
           \\& \nonumber-ig^{(7)}_{hW^+l}p^\mu \slashed{p}-ig^{(8)}_{hW^+l}p_3^\mu \slashed{p}
   \end{align}
    From these couplings,  $g^{(1)}_{hW^\pm\psi}$, $g^{(4)}_{hW^\pm\psi}$ and $g^{(5)}_{hW^\pm\psi}$ are complex, and under complex conjugation satisfy $(g^{(i)}_{hW^+\psi})^*=g^{(i)}_{hW^-\psi}$. The rest of the couplings are real, and in general $g^{(i)}_{hW^+\psi}\neq g^{(i)}_{hW^-\psi} $, with the exception of $g^{(7)}_{hW^+\psi}= g^{(7)}_{hW^-\psi}$. Moreover, at dimension eight the effective couplings satisfy $g^{(2)}_{hW^\pm l}=g^{(8)}_{hW^\pm l}=-g^{(3)}_{hW^\pm l}=-g^{(6)}_{hW^\pm l}$
    \item $d_{W^\pm}$ is coupling of the Lorentz structure induced by dipole operators to $W\nu \ell$. Taking all momenta incoming, the Lorentz structure is
    \begin{align*}
    W^{\pm,\mu}(k)\nu(p_1)\ell^\mp(p_2) \  :  \quad  -d_{W^\pm}k_{2,\nu}\sigma^{\mu\nu} P_R 
    \end{align*}
    were $(d_{W^+})^*=d_{W^-}$. Notice that $d_{W^+}$ couples right-handed leptons with neutrinos.
    \item $d_{hZ}$ is coupling of the Lorentz structure induced by dipole operators to $
    hW \nu \ell$
    \begin{align*}
    h(p)W^{\pm,\mu}(k_2)\nu(p_1)\ell^\mp(p_2) \  :  \ \  -d_{hW^\pm}k_{2,\nu}\sigma^{\mu\nu} P_R 
    \end{align*}
    Similarly, $d_{hW^+}$ couples right-handed leptons with neutrinos and $(d_{hW^+})^*=d_{hW^-}$. At dimension six, $d_{W^\pm}$ and $d_{hW^\pm}$ are related by $d_{hW^\pm}=vd_{W^\pm}$, which further simplifies the analytic expression of the differential decay rate.
\end{itemize}

Now we present the kinematic structures for the \textit{interfering contributions} in Table \ref{coupling factors hww} and the \textit{non-interfering contributions} in Table \ref{helicity combinations hww}. For the \textit{interfering contributions}, the only helicity configuration is $(h_1,h_2,h_3,h_4)=(-,+,-,+)$, with the three topologies in Figure \ref{htoWW diagrams} receiving a contribution. Starting with $\mathcal{A}_{hWW}$ for topology (d), the four kinematic structures are given by:
\begin{align}
\label{A1hWW}
    A^{(1)}_{hWW}(p,p_1^{-},p_2^{+},p_3^{-},p_4^{+})=& \ \frac{\hat{p}^{W}_{12}}{s_{12}}\frac{\hat{p}^{W}_{34}}{s_{34}}\langle 3 \gamma_\mu 4] \langle 1 \gamma^\mu2] \\
\label{A2hWW}
    A^{(2)}_{hWW}(p,p_1^{-},p_2^{+},p_3^{-},p_4^{+})=& \ 2\frac{\hat{p}^{W}_{12}}{s_{12}}\frac{\hat{p}^{W}_{34}}{s_{34}}\bigl( \langle 3 \slashed{k_1}4] \langle 1 \slashed{k_2}2]-(k_1\cdot k_2) \langle 3 \gamma_\mu 4] \langle 1 \gamma^\mu2]\bigr) \\
\label{A3hWW}    
    A^{(3)}_{hWW}(p,p_1^{-},p_2^{+},p_3^{-},p_4^{+})=& \ -\frac{\hat{p}^{W}_{12}}{s_{12}}\frac{\hat{p}^{W}_{34}}{s_{34}}\bigl(\langle 3 \slashed{k_1}4] \langle 1 \slashed{k_2}2]+(k_1\cdot p) \langle 3 \gamma_\mu 4] \langle 1 \gamma^\mu2]\bigr) \\ 
\label{A4hWW}
    A^{(4)}_{hWW}(p,p_1^{-},p_2^{+},p_3^{-},p_4^{+})=& \ -\frac{\hat{p}^{W}_{12}}{s_{12}}\frac{\hat{p}^{W}_{34}}{s_{34}}\bigl(\langle 3 \slashed{k_1}4] \langle 1 \slashed{k_2}2]+(k_2\cdot p) \langle 3 \gamma_\mu 4] \langle 1 \gamma^\mu2]\bigr)
\end{align}
Notice $A^{(1)}_{hWW}$ and $A^{(2)}_{hWW}$ are analogous to the kinematic structures in eq. (\ref{A1hZZ}) and (\ref{A2hZZ}), with $A^{(1)}_{hWW}$ containing the SM part of $\mathcal{A}_{hWW}$.

Continuing with $\mathcal{A}_{hW^-\psi}$ for topology (e). There are eight different kinematic structures, corresponding to the eight Lorentz structures in eq. (\ref{couplings hWlnu}).\\
\begin{align}
    \label{A1hWpsipsi}
    A^{(1)}_{hW^+\psi}(p,p_1^{-},p_2^{+},p_3^{-},p_4^{+})=& \ \frac{\hat{p}_{34}^W}{s_{34}}\langle 3 \gamma_\mu 4] \langle 1 \gamma^\mu2], \\
    \label{A2hWpsipsi}
    A^{(2)}_{hW^+\psi}(p,p_1^{-},p_2^{+},p_3^{-},p_4^{+})=& \ -\frac{\hat{p}_{34}^W}{s_{34}}(p_1\cdot p)\langle 3 \gamma_\mu 4] \langle 1 \gamma^\mu2], \\
        \label{A3hWpsipsi}
    A^{(3)}_{hW^+\psi}(p,p_1^{-},p_2^{+},p_3^{-},p_4^{+})=& \ -\frac{\hat{p}_{34}^W}{s_{34}}(p_2\cdot p)\langle 3 \gamma_\mu 4] \langle 1 \gamma^\mu2],  \\
        \label{A4hWpsipsi}
    A^{(4)}_{hW^+\psi}(p,p_1^{-},p_2^{+},p_3^{-},p_4^{+})=& \ -\frac{\hat{p}_{34}^W}{s_{34}}(k_2\cdot p)\langle 3 \gamma_\mu 4] \langle 1 \gamma^\mu2],  
\end{align}
\begin{align}
    \label{A5hWpsipsi}
    A^{(5)}_{hW^+\psi}(p,p_1^{-},p_2^{+},p_3^{-},p_4^{+})=& \ \frac{\hat{p}_{34}^W}{s_{34}}\langle 3 \slashed{k_1}4] \langle 1 \slashed{k_2}2], \\
    \label{A6hWpsipsi}
    A^{(6)}_{hW^+\psi}(p,p_1^{-},p_2^{+},p_3^{-},p_4^{+})=& \ \frac{\hat{p}_{34}^W}{s_{34}}\langle 3 \slashed{p_2}4]\langle 1 \slashed{k_2}2], \\
    \label{A7hWpsipsi}
    A^{(7)}_{hW^+\psi}(p,p_1^{-},p_2^{+},p_3^{-},p_4^{+})=& \ -\frac{\hat{p}_{34}^W}{s_{34}}\langle 3 \slashed{k_1}4] \langle 1 \slashed{k_2}2], \\
    \label{A8hWpsipsi}
    A^{(8)}_{hW^+\psi}(p,p_1^{-},p_2^{+},p_3^{-},p_4^{+})=& \ \frac{\hat{p}_{34}^W}{s_{34}}\langle 3 \slashed{p_1}4]\langle 1 \slashed{k_2}2].
\end{align}
Note the explicit dependence on the leptons' momenta in the kinematic structures $A^{(2)}_{hW^+\psi}$, $A^{(3)}_{hW^+\psi}$, $A^{(6)}_{hW^+\psi}$ , and $A^{(8)}_{hW^+\psi}$, which have no analogous in the four charged lepton case and are generated by $Q_{l^2H^2D^3}^{(4)}$. The kineamatic structures $A^{(1)}_{hW^+\psi}$, $A^{(4)}_{hW^+\psi}$, $A^{(5)}_{hW^+\psi}$, and $A^{(7)}_{hW^+\psi}$ are analogous to those in eqs. (\ref{A1hZpsi}), (\ref{A2hZpsi}), (\ref{A3hZpsi}) and (\ref{A4hZpsi}) respectively.

Finally, the kinematic structures contributing to $\mathcal{A}_{hW^-\psi}$ for topology (f) are 
\begin{align}
    A^{(1)}_{hW^-\psi}(p,p_1^{-},p_2^{+},p_3^{-},p_4^{+})=& \ \frac{\hat{p}_{12}^W}{s_{12}}\langle 3 \gamma_\mu 4] \langle 1 \gamma^\mu2], \\
    A^{(2)}_{hW^-\psi}(p,p_1^{-},p_2^{+},p_3^{-},p_4^{+})=& \ -\frac{\hat{p}_{12}^W}{s_{12}}(p_3\cdot p)\langle 3 \gamma_\mu 4] \langle 1 \gamma^\mu2],  \\
    A^{(3)}_{hW^-\psi}(p,p_1^{-},p_2^{+},p_3^{-},p_4^{+})=& \ -\frac{\hat{p}_{12}^W}{s_{12}}(p_4\cdot p)\langle 3 \gamma_\mu 4] \langle 1 \gamma^\mu2],  \\
    A^{(4)}_{hW^-\psi}(p,p_1^{-},p_2^{+},p_3^{-},p_4^{+})=& \ -\frac{\hat{p}_{12}^W}{s_{12}}(k_1\cdot p)\langle 3 \gamma_\mu 4] \langle 1 \gamma^\mu2],  \\
    A^{(5)}_{hW^-\psi}(p,p_1^{-},p_2^{+},p_3^{-},p_4^{+})=& \ \frac{\hat{p}_{12}^W}{s_{12}}\langle 3 \slashed{k_1}4] \langle 1 \slashed{k_2}2], \\
    A^{(6)}_{hW^-\psi}(p,p_1^{-},p_2^{+},p_3^{-},p_4^{+})=& \ \frac{\hat{p}_{12}^W}{s_{12}}\langle 1 \slashed{p_4}2] \langle 3 \slashed{k_1}4], \\
    A^{(7)}_{hW^-\psi}(p,p_1^-,p_2^{+},p_3^{-},p_4^{+})=& \ -\frac{\hat{p}_{12}^W}{s_{12}}\langle 3 \slashed{k_1}4] \langle 1 \slashed{k_2}2], \\
    A^{(8)}_{hW^-\psi}(p,p_1^{-},p_2^{+},p_3^{-},p_4^{+})=& \ \frac{\hat{p}_{12}^W}{s_{12}}\langle 1 \slashed{p_3}2] \langle 3 \slashed{k_1}4].
\end{align}
Immediately, one can see the similarities between these kinematic structures and those of $\mathcal{A}_{hW^+\psi}$. In fact, the kinematic structures for $\mathcal{A}_{hW^-\psi}$ can be obtained from those  for  $\mathcal{A}_{hW^+\psi \psi}$ as follows:
\begin{equation}
    A^{(i)}_{hW^-\psi}(p,p_1^{-},p_2^{+},p_3^{-},p_4^{+})=A^{(i)}_{hW^+\psi}(p,p_3^{-},p_4^{+},p_1^{-},p_2^{+})
\end{equation}
Continuing with the kinematic structures for the \textit{non-interfering contributions} in Table \ref{helicity combinations hww}, there are two possible helicity combinations. Considering the helicity configuration \linebreak $(h_1,h_2,h_3,h_4)=(+,+,-,+)$, diagrams (d) and (e) get a $\mathcal{O}(1/\Lambda^2)$ contribution, the corresponding dipole subamplitudes are given by
\begin{align}
    \label{AhWWdipole1}
    A^{dipole}_{hWW}(p,p_1^+,p_2^+,p_3^-,p_4^+)&=\hat{p}^W_{12}\hat{p}^W_{34}\big([1\gamma_{\mu}\slashed{k}_1 2]-[1\slashed{k}_1\gamma_{\mu} 2]\big)\langle3\gamma^{\mu}4],\\ 
    \label{AhWpsidipole1}
    A^{dipole}_{hW\psi}(p,p_1^+,p_2^+,p_3^-,p_4^+)&=-\frac{1}{2}\hat{p}^W_{34}\big([1\gamma_{\mu}\slashed{k}_2 2]-[1\slashed{k}_2\gamma_{\mu} 2]\big)\langle3\gamma^{\mu}4]. 
\end{align}
These are analogous to the kinematic structures in eqs. \eqref{AhZZdipole} and \eqref{AhZpsidipole}. The other helicity combination is $(h_1,h_2,h_3,h_4)=(-,+,+,+)$, for which diagram (d) and (f) get a $\mathcal{O}(1/\Lambda^2)$ contribution given by
\begin{align}
    \label{AhWWdipole2}
    A^{dipole}_{hWW}(p,p_1^-,p_2^+,p_3^+,p_4^+)&=\hat{p}^W_{12}\hat{p}^W_{34}\big([3\gamma_{\mu}\slashed{k}_2 4]-[3\slashed{k}_2\gamma_{\mu} 4]\big)\langle1\gamma^{\mu}2],\\ 
    \label{AhWpsidipole2}
    A^{dipole}_{hW\psi}(p,p_1^-,p_2^+,p_3^+,p_4^+)&=-\frac{1}{2}\hat{p}^W_{12}\big([3\gamma_{\mu}\slashed{k}_1 4]-[3\slashed{k}_1\gamma_{\mu} 4]\big)\langle3\gamma^{\mu}4]. 
\end{align}
$A^{dipole}_{hWW}$ is analogous to eq. \eqref{AhZZdipole2}, one the other hand $A^{dipole}_{hW^-\psi}$ has no analogous kinematic structure in the four charged lepton case. This is because this kinematic structure comes from topology (f) in Figure \ref{htoWW diagrams}, and in the four charged lepton case there is no analogous topology.

\section{Mapping of effective couplings and Wilson coefficients}\label{effective couplings appendix}
The effective couplings can be expressed in terms of geoSMEFT connections, these connections can then be written in terms of Wilson coefficients to obtain the effective couplings in terms of Wilson coefficients. Both of these expressions have been worked out in \cite{Helset:2020yio, Corbett:2023yhk}, here we directly present the effective couplings in terms of Wilson coefficients. In the following, the parameters with bar are the canonically normalized geoSMEFT Lagrangian parameters, whose expansion to $\mathcal{O}(1/\Lambda^4)$ can be found in \cite{Helset:2020yio}.

Starting with the effective couplings relevant for $h\to\ell\bar{\ell}\left(Z\to\ell'\bar{\ell'}\right)$. For the $hZZ$ vertex defined in eq. (\ref{hZZ vertex}), the effective couplings in terms of Wilson coefficients are given by:
\begin{flalign}
    c^{(1)}_{hZZ}=&\frac{\bar{g}_Z^2}{4}v+\frac{\bar{g}_Z^2}{16}v^3\left(4 c_{H\Box}^{(6)}+3c^{(6)}_{HD} \right) && \\ \nonumber
    & +\frac{\bar{g}_Z^2}{128}v^5\left(20c^{(8)}_{HD}+20c^{(8)}_{HD,2}+3\left[4 c_{H\Box}^{(6)}-c^{(6)}_{HD} \right]\left[4 c_{H\Box}^{(6)}+3c^{(6)}_{HD} \right]\right), && \\ \nonumber
    c^{(2)}_{hZZ}=& \bar{g}_Z^2 v \left(c_{HW}^{(6)}\frac{c_Z^4}{g_2^2}+c_{HWB}^{(6)}\frac{c_Z^2s^2_Z}{g_2g_1}+c_{HB}^{(6)}\frac{s_Z^4}{g_1^2} \right)&& \\
    & +\frac{\bar{g}_Z^2}{4}v^3\left(\frac{c_Z^4}{g_2^2}\left[4c_{HW}^{(8)}+4c_{HW,2}^{(8)}+c_{HW}^{(6)}\left(4c_{H\Box}^{(6)}-c_{HD}^{(6)}\right)\right]\right.&& \\ \nonumber
    & \hspace{50pt} +\frac{c_Z^2s^2_Z}{g_2g_1}\left[4c_{HWB}^{(8)}+c_{HWB}^{(6)}\left(4c_{H\Box}^{(6)}-c_{HD}^{(6)}\right)\right]&& \\ \nonumber 
    & \hspace{90pt} \left.+\frac{s_Z^4}{g_1^2}\left[4c_{HB}^{(8)}+c_{HB}^{(6)}\left(4c_{H\Box}^{(6)}-c_{HD}^{(6)}\right)\right] \right),&& \\ 
    c^{(3)}_{hZZ}=& \frac{\bar{g}_Z^2}{4} v^3 \left(c_{HDHW}^{(8)}\frac{c_Z^2}{g_2}+c_{HDHB}^{(8)}\frac{s_Z^2}{g_1} \right). &&
\end{flalign}
The effective couplings for the effective vertex $hAZ$ defined in eq. (\ref{hZA vertex}) are given by:
\begin{flalign} 
    \nonumber
    c^{(1)}_{hZA}=& \bar{g}_Z \bar{e} v \left(2 c_{HW}^{(6)}\frac{c_Z^4}{g_2^2} -c_{HWB}^{(6)}\frac{c_Z^2-s^2_Z}{g_2g_1}-2 c_{HB}^{(6)}\frac{s_Z^4}{g_1^2} \right)\\ 
    & +\frac{\bar{g}_Z \bar{e}}{4}v^3\bigg(2\frac{c_Z^4}{g_2^2}\left[4c_{HW}^{(8)}+4c_{HW,2}^{(8)}+c_{HW}^{(6)}\left(4c_{H\Box}^{(6)}-c_{HD}^{(6)}\right)\right]&& \\ \nonumber
    & \hspace{50pt} -\frac{c_Z^2-s^2_Z}{g_1g_2}\left[[4c_{HWB}^{(8)}+c_{HWB}^{(6)}\left(4c_{H\Box}^{(6)}-c_{HD}^{(6)}\right)\right]\\ \nonumber 
    & \hspace{90pt} \left.-2\frac{s_Z^4}{g_1^2}\left[[4c_{HB}^{(8)}+c_{HB}^{(6)}\left(4c_{H\Box}^{(6)}-c_{HD}^{(6)}\right)\right] \right), && \\
    c^{(2)}_{hZA}=& - \frac{\bar{g}_Z\bar{e}}{4} v^3 \left(c_{HDHW}^{(8)}\frac{1}{g_2}+c_{HDHB}^{(8)}\frac{1}{g_1} \right).
\end{flalign}
The effective coupling for the $Z\psi\psi$ vertex are given by:
\begin{flalign}
    g^{eZ}=&-\bar{g}_Zs_Z^2 + \frac{\bar{g}_Z}{2}v^2c^{(6)}_{He}+\frac{\bar{g}_Z}{4}v^4c^{(8)}_{He}, && \\
    g^{lZ}=&-\bar{g}_Z(2s_Z^2-1) + \frac{\bar{g}_Z}{2}v^2\left(c^{1,(6)}_{Hl}+c^{3,(6)}_{Hl}\right)+\frac{\bar{g}_Z}{4}v^4\left(c^{1,(8)}_{Hl}+^{2,(8)}_{Hl}+c^{3,(6)}_{Hl}\right). &&
\end{flalign}
The effective couplings for the 4-point contact vertex $hZ\psi\psi$ defined in eq. (\ref{hZpsipsi vertex}) are given by:
\begin{flalign}
    g^{(1)}_{hZe}=&\bar{g}_Z^2vc^{(6)}_{He}+\frac{\bar{g}_Z^2}{4}v^3\left(4c^{(8)}_{He}+c^{(6)}_{He}\left[4c^{(6)}_{H\Box}-c^{(6)}_{HD}\right]\right),&& \\
    g^{(1)}_{hZl}=&\bar{g}_Z^2v\left(c^{1,(6)}_{Hl}+c^{3,(6)}_{Hl}\right)&& \\ \nonumber
    &+\frac{\bar{g}_Z^2}{4}v^3\left(4\left[c^{1,(8)}_{Hl}+^{2,(8)}_{Hl}+c^{3,(8)}_{Hl}\right]+\left[c^{1,(6)}_{Hl}+c^{3,(6)}_{Hl}\right]\left[4c^{(6)}_{H\Box}-c^{(6)}_{HD}\right]\right),&& \\
    g^{(2)}_{hZ\psi}=&-\bar{s}_Wv\left(c^{(1)}_{\psi^2BH^2D}+(\sigma_3)_\psi c^{(3)}_{l^2BH^2D}\right)-\bar{c}_Wv\left(c^{(1)}_{\psi^2WH^2D}-(\sigma_3)_\psi c^{(3)}_{l^2WH^2D}\right)&& \\ \nonumber
    &-\frac{\bar{e}v}{2\bar{c}_W\bar{s}_W}\left(c^{(1)}_{\psi^2H^2D^3}-(\sigma_3)_\psi c^{(3)}_{l^2H^2D^3}\right),&& \\
    g^{(3)}_{hZ\psi}=&\bar{s}_Wv\left(c^{(1)}_{\psi^2BH^2D}+(\sigma_3)_\psi c^{(3)}_{l^2BH^2D}\right)+\bar{c}_Wv\left(c^{(1)}_{\psi^2WH^2D}-(\sigma_3)_\psi c^{(3)}_{l^2WH^2D}\right)&& \\ \nonumber
    &-\frac{\bar{e}v}{2\bar{c}_W\bar{s}_W}\left(c^{(1)}_{\psi^2H^2D^3}-(\sigma_3)_\psi c^{(3)}_{l^2H^2D^3}\right),&& \\
    g^{(4)}_{hZ\psi}=&\frac{\bar{e}v}{2\bar{c}_W\bar{s}_W}\left(c^{(1)}_{\psi^2H^2D^3}-(\sigma_3)_\psi c^{(3)}_{l^2H^2D^3}\right),
\end{flalign}
were $(\sigma_3)_e=0$ and $(\sigma_3)_l=-1$. The dipole contribution to $Z\psi\psi$ and $hZ\psi\psi$ defined in eqs. (\ref{Zpsipsi dipole vertex}) and (\ref{hZpsipsi dipole vertex}) are given by:
\begin{flalign}
    d_Z = & \sqrt{2}\bar{g}_Zv\left(\frac{\bar{c}^2_Z}{g_2}c^{(6)}_{eW}+\frac{\bar{s}^2_Z}{g_1}c^{(6)}_{eB}\right), && \\
    d_{hZ} = & \sqrt{2}\bar{g}_Z\left(\frac{\bar{c}^2_Z}{g_2}c^{(6)}_{eW}+\frac{\bar{s}^2_Z}{g_1}c^{(6)}_{eB}\right). &&
\end{flalign}
Continuing with the effective couplings relevant for $h \to \ell \bar{\nu}_\ell \nu_{\ell'} \bar{\ell'}$. The effective couplings for the $hWW$ vertex are given by:
\begin{flalign}
    c^{(1)}_{hWW}=&\frac{\bar{g}_2^2}{4}v+\frac{\bar{g}_2^2}{8}v^3\left(4c_{H\Box}^{(6)}-c_{HD}^{(6)}\right), && \\ \nonumber 
    &+\frac{\bar{g}_2^2}{64}v^5\left(20c_{HD}^{(8)}-28c^{(6)}_{HD,2}+3\left[4c_{H\Box}^{(6)}-c_{HD}^{(6)}\right]^2 \right) && \\ \nonumber
    c^{(2)}_{hWW}=&4vc_{HW}^{(6)}+v^3\left(4c_{HW}^{(8)}+c_{HW}^{(6)}\left[4c_{H\Box}^{(6)}-c_{HD}^{(6)}+8c_{HW}^{(6)}\right]\right), && \\ 
    c^{(3)}_{hWW}=& -\frac{\bar{g}_2}{4} v^3 \left(c_{HDHW}^{(8)}-ic_{HDHB}^{(8)} \right), && \\
    c^{(4)}_{hWW}=& -\frac{\bar{g}_2}{4} v^3 \left(c_{HDHW}^{(8)}+ic_{HDHB}^{(8)} \right). &&
\end{flalign}
The effective coupling for the $Wl\nu$ vertex is given by:
\begin{flalign}
    g^{W^\pm}=-\frac{\bar{g}_2}{\sqrt{2}}-\frac{\bar{g}_2}{\sqrt{2}}v^2c^{3,(6)}_{Hl}-\frac{\bar{g}_2}{\sqrt{2}}v^4\left(c^{3,(8)}_{Hl}\pm ic^{\epsilon,(8)}_{Hl}\right). &&
\end{flalign}
The effective couplings for the 4-point contact term $hWl\nu$ are:
\begin{flalign}
    g^{(1)}_{hW^\pm l}&=-\frac{2\bar{2g}_2}{\sqrt{2}}vc^{3,(6)}_{Hl}-\frac{\bar{g}_2}{2\sqrt{2}}v^3\left(c^{3,(8)}_{Hl}\pm ic^{\epsilon,(8)}_{Hl}+c^{3,(6)}_{Hl}\left[4c_{H\Box}^{(6)}-c_{HD}^{(6)}\right]\right), && \\
    g^{(2)}_{hW^\pm l}&=\mp\frac{\bar{e}v}{\sqrt{2}\bar{s}_W}c^{(4)}_{l^2H^2D^3}, && \\
    g^{(3)}_{hW^\pm l}&=\pm\frac{\bar{e}v}{\sqrt{2}\bar{s}_W}c^{(4)}_{l^2H^2D^3}, && \\
    g^{(4)}_{hW^\pm l}&=\sqrt{2}v\left(c^{(3)}_{l^2WH^2D}\pm i c^{(5)}_{l^2WH^2D}\right)+\frac{\bar{e}v}{\sqrt{2}\bar{s}_W}c^{(3)}_{l^2H^2D^3}, && \\
    g^{(5)}_{hW^\pm l}&=-\sqrt{2}v\left(c^{(3)}_{l^2WH^2D}\pm i c^{(5)}_{l^2WH^2D}\right)+\frac{\bar{e}v}{\sqrt{2}\bar{s}_W}c^{(3)}_{l^2H^2D^3}, && \\
    g^{(6)}_{hW^\pm l}&=\pm\frac{\bar{e}v}{\sqrt{2}\bar{s}_W}c^{(4)}_{l^2H^2D^3}, && \\
    g^{(7)}_{hW^\pm l}&=-\frac{\sqrt{2}\bar{e}v}{\bar{s}_W}c^{(3)}_{l^2H^2D^3}, && \\
    g^{(8)}_{hW^\pm l}&=\mp\frac{\bar{e}v}{\sqrt{2}\bar{s}_W}c^{(4)}_{l^2H^2D^3}. && 
\end{flalign}
Finally, the dipole contributions to $Wl\nu$ and $hWl\nu$ are given by:
\begin{flalign}
    d_W & =-2v\frac{\bar{g_2}}{g_2}c^{(6)}_{eW}, && \\ 
    d_{hW} & =-2\frac{\bar{g_2}}{g_2}c^{(6)}_{eW}. && 
\end{flalign}

\section{Explicit expressions for the $J$-functions}
In this appendix we present the explicit expressions for the $J$-functions introduced in Section \ref{Section 3.2.1} and \ref{Section 4.2.1} in terms of the effective couplings.
\subsection{$J^Z_i$ functions} \label{J^Z explicit expressions}
The $J^Z_i$ functions are defined as the coefficient multiplying the angular functions in eq. (\ref{J^Z functions}). To simplify the expressions for $J^Z_i$ we have introduced the following effective couplings combinations
$$
\begin{array}{lll}
    G_0^2=\big(g^{Zl}\big )^2+\big(g^{Ze}\big )^2, & G_V=g^{Zl}+g^{Z e}, & G_A=g^{Zl}-g^{Z e}, \\ \nonumber
    H_0^2=\big(g^{(1)}_{h Z l}\big )^2+\big(g^{(1)}_{h Z e}\big )^2, & H_V=g^{(1)}_{h Z l}+g^{(1)}_{h Z e}, & H_A=g^{(1)}_{h Z l}-g^{(1)}_{h Z e}, \\
\multicolumn{3}{c}{
  \begin{array}{ll}
    H_1=g^{Z l}g^{(1)}_{h Z l}+g^{Ze}g^{(1)}_{h Z e}, & K_1=g^{Z l}g^{(1)}_{h Z l}-g^{Ze}g^{(1)}_{h Z e}, \\ 
    H_2=g^{Z l}g^{(2)}_{h Z l}+g^{Ze}g^{(2)}_{h Z e} &     K_2=g^{Z l}g^{(2)}_{h Z l}-g^{Ze}g^{(2)}_{h Z e}, \\
    H_3=g^{Z l}g^{(3)}_{h Z l}+g^{Ze}g^{(3)}_{h Z e}, &K_3=g^{Z l}g^{(3)}_{h Z l}-g^{Ze}g^{(3)}_{h Z e} \\
    H_4=g^{Z l}g^{(4)}_{h Z l}+g^{Ze}g^{(4)}_{h Z e}, & K_4=g^{Z l}g^{(4)}_{h Z l}-g^{Ze}g^{(4)}_{h Z e}
  \end{array}
}
\end{array}$$
Note that in the following expressions, $J_1^Z$ and $J_2^Z$ differ only by their dipole contribution, which has been subtracted to simplify the expressions. The same applies to $J_3^Z$ and $J_4^Z$.

\begin{align}
    J^Z_1&-\frac{1}{2}  r m_H^2G_0^2\left|d_Z\right|^2\left(\left(\kappa^2-2 r s\right)m_H^4+\frac{4 \kappa s  v m_H^2c_{h Z Z}^{(1)}}{r-s}+\frac{8 s (r+s)v^2\left(c_{h Z Z}^{(1)}\right)^2}{(r-s)^2}\right) \\
    =&J^Z_2+\frac{1}{2} m_H^2 G_0^2\left|d_Z\right|^2\left(r \left(\kappa^2+2 r s\right)m_H^4-\frac{12 \kappa r s  v m_H^2c_{h Z Z}^{(1)}}{r-s}-\frac{4 (r+s)\left(\kappa^2+2 r s\right)v^2\left(c_{h Z Z}^{(1)}\right)^2}{(r-s)^2}\right) \\ \nonumber    
    =& -\frac{G_0^4}{(r-s)^2}\left(m_H^2c_{h Z Z}^{(1)} \left(12 \kappa r s c_{h Z Z}^{(2)}+\left(\kappa^2-\kappa \lambda^2+2 r s\right)c_{h Z Z}^{(3)}\right)\right. \\ \nonumber & \hspace{140pt} \left.-\left(\kappa^2+2 r s\right)\left(c_{h Z Z}^{(1)}\right)^2-2 r s\left(\kappa^2+8 r s\right) m_H^4\left(c_{h Z Z}^{(2)}\right)^2\right) \\ \nonumber
     &-\frac{ m_H^2G_0^2}{2(r-s)}\left( m_H^2c_{h Z Z}^{(1)}\left((\kappa+2 r)\left(\kappa^2+2 r s\right)H_2+ \kappa \lambda^2\left(H_3-H_4\right)\right)\right. \\ \nonumber & \hspace{190pt} \left.+2 H_1\left(\left(\kappa^2+2 r s\right)c_{h Z Z}^{(1)}-6 \kappa r s c_{h Z Z}^{(2)} m_H^2\right)\right) \\ \nonumber
     &-\frac{1}{4} m_H^4G_0^2 \left(12 \kappa r \bar{e} c_{h Z A}^{(1)} H_V+\left(\kappa^2+2 r s\right)H_0^2+\frac{4 r\left(\kappa^2+8 r s\right) \bar{e}^2 \left(c_{h Z A}^{(1)}\right)^2}{s}\right) \\ \nonumber
    &-\frac{\bar{e} m_H^2G_0^2 G_V }{r-s}\left(c_{h Z Z}^{(1)}\left(6 \kappa r c_{h Z A}^{(1)}+\left(\kappa^2+3 \kappa r+2 r s\right)c_{h Z A}^{(2)}\right)-2 r\left(\kappa^2+8 r s\right)m_H^2 c_{h Z A}^{(1)} c_{h Z Z}^{(2)}\right)
\end{align}
\begin{align} 
    \nonumber
    J^Z_3&-\frac{ r m_H^2 G_0^2\left|d_Z\right|^2}{2}\left(\left(\lambda^2+2 r s\right)m_H^4+\frac{4 \kappa s v m_H^2c_{h Z Z}^{(1)} }{r-s}+\frac{4\left(\lambda^2+2 r s+2 s^2\right)v^2\left(c_{h Z Z}^{(1)}\right)^2 }{(r-s)^2}\right) \\ \nonumber
    =& J_4^Z - \frac{1}{2}m_H^2 G_0^2 \left|d_Z\right|^2\left(r \left(2 r s-\lambda^2\right)m_H^4+\frac{4  r s\kappa  v m_H^2c_{h Z Z}^{(1)}}{r-s}+\frac{4 s\left(\lambda^2+2 r^2+2 r s\right)v^2\left(c_{h Z Z}^{(1)}\right)^2 }{(r-s)^2}\right) \\
    =& \frac{G_0^4}{(r-s)^2}\left(m_H^2c_{h Z Z}^{(1)} \left(2 r s\left(2 \kappa c_{h Z Z}^{(2)}+c_{h Z Z}^{(3)}\right)-(\kappa-1) \lambda^2 c_{h Z Z}^{(3)}\right)\right. \\ \nonumber & \hspace{140pt}\left.-\left(\lambda^2+2 r s\right)\left(c_{h Z Z}^{(1)}\right)^2-2 r s \left(\kappa^2-2 \lambda^2\right)m_H^4\left(c_{h Z Z}^{(2)}\right)^2\right) \\ \nonumber
    & +\frac{ m_H^2G_0^2}{2(r-s)}\left(2\left(\lambda^2+2 r s\right)c_{h Z Z}^{(1)} H_1 \right. \\ \nonumber & \hspace{45pt}\left.+m_H^2\left((\kappa+2 r)\left(\lambda^2+2 r s\right)c_{h Z Z}^{(1)}H_2+ \kappa \lambda^2 c_{h Z Z}^{(1)}\left(H_3-H_4\right) -4  r s\kappa c_{h Z Z}^{(2)}H_1 \right)\right) \\ \nonumber
    & -\frac{1}{4} m_H^4 G_0^2\left(4r\kappa \bar{e}c_{h Z A}^{(1)} H_V+\left(\lambda^2+2 r s\right)H_0^2+\frac{4 r\bar{e}^2\left(\kappa^2-2 \lambda^2\right)  \left(c_{h Z A}^{(1)}\right)^2}{s}\right) \\ \nonumber
    & +\frac{ \bar{e}m_H^2G_0^2 G_V }{r-s}\left(2 \kappa r c_{h Z A}^{(1)} c_{h Z Z}^{(1)}+\left(\lambda^2+r(\kappa+2 s)\right)c_{h Z A}^{(2)} c_{h Z Z}^{(1)}-2 r \left(\kappa^2-2 \lambda^2\right)m_H^2c_{h Z A}^{(1)} c_{h Z Z}^{(2)} \right)
\end{align}
\begin{flalign}
    J^Z_5=&-\frac{8 r s G_A^2 G_v^2\left(m_H^2c_{h Z Z}^{(1)} \left(2 \kappa c_{h Z Z}^{(2)}+c_{h Z Z}^{(3)}\right)-\kappa^2m_H^4\left(c_{h Z Z}^{(2)}\right)^2 -\left(c_{h Z Z}^{(1)}\right)^2\right)}{(r-s)^2} && \\ \nonumber 
    & -\frac{4 r s  m_H^2 G_A G_V}{r-s}\left(2  c_{h Z Z}^{(1)}K_1+m_H^2\left( (\kappa+2 r)c_{h Z Z}^{(1)}K_2-2 \kappa c_{h Z Z}^{(2)} K_1\right)\right)&& \\ \nonumber
    & +2 r m_H^4G_A G_V H_A \left(s H_V+2 \kappa \bar{e} c_{h Z A}^{(1)}\right) && \\ \nonumber 
    & -\frac{4 r\bar{e} m_H^2 G_A^2 G_V}{r-s}\left(2 \kappa c_{h Z A}^{(1)} c_{h Z Z}^{(1)}+(\kappa+2 s)c_{h Z A}^{(2)} c_{h Z Z}^{(1)}-2 \kappa^2 m_H^2 c_{h Z A}^{(1)} c_{h Z Z}^{(2)}\right) &&
\end{flalign}
\begin{align}
    J^Z_6=&\frac{2 \sqrt{r s} G_A^2 G_V^2}{(r-s)^2}\left( m_H^2c_{h Z Z}^{(1)}\left(2 \left(\kappa^2+4 r s\right)c_{h Z Z}^{(2)}+4 r s c_{h Z Z}^{(3)}-(\kappa-2) \kappa c_{h Z Z}^{(3)}\right) \right.\\ \nonumber & \hspace{230pt} \left. -8 \kappa r s m_H^4\left(c_{h Z Z}^{(2)}\right)^2-2 \kappa\left(c_{h Z Z}^{(1)}\right)^2\right) \\ \nonumber
    &+\frac{\sqrt{r s}G_A G_V }{r-s}\left(m_H^2c_{h Z Z}^{(1)} \left(4 \kappa H_1 +2 \kappa  (\kappa+2 r)m_H^2K_2+\lambda^2 m_H^2\left(K_3-K_4\right)\right)\right. \\ \nonumber & \hspace{250pt}\left.-2 \left(\kappa^2+4 r s\right) m_H^4c_{h Z Z}^{(2)} K_1\right) \\ \nonumber
    & -\frac{r m_H^4G_V G_A H_A \left(\kappa s H_V+\left(\kappa^2+4 r s\right)\bar{e} c_{h Z A}^{(1)}\right)}{\sqrt{r s}} \\ \nonumber
    &+\frac{\bar{e} r G_A^2 G_V m_H^2\left(-16  r s \kappa m_H^2c_{h Z A}^{(1)} c_{h Z Z}^{(2)} +2 \left(\kappa^2+4 r s\right)c_{h Z A}^{(12)} c_{h Z Z}^{(1)}+(4 r s+\kappa(\kappa+4 s))c_{h Z A}^{(2)} c_{h Z Z}^{(1)}\right)}{(r-s) \sqrt{r s}} \\
    J^Z_7=& -\frac{\sqrt{r s} G_0^4 }{2(r-s)^2}\bigg(m_H^2c_{h Z Z}^{(1)} \left(2 c_{h Z Z}^{(2)}\left(\lambda^2-2 \kappa^2\right)+c_{h Z Z}^{(3)}\left(\lambda^2-2 \kappa\right)\right) \\ \nonumber   & \hspace{200pt} +2 \kappa\left(c_{h Z Z}^{(1)}\right)^2 +8 \kappa r s m_H^4\left(c_{h Z Z}^{(2)}\right)^2\bigg)\\ \nonumber
    & +\frac{\sqrt{r s}G_0^2 }{4(r-s)}\left( m_H^2c_{h Z Z}^{(1)}\left(4 \kappa \mathrm{H}_1+m_H^2\left(2 \kappa(\kappa+2 r) H_2+\lambda^2\left(H_3-H_4\right) \right)\right)\right.\\ \nonumber  & \hspace{235pt} \left.-2\left(2 \kappa^2-\lambda^2\right) m_H^4 c_{h Z Z}^{(2)}H_1 \right) \\ \nonumber
    &-\frac{ r m_H^4G_0^2}{4 \sqrt{r s}}\left(e\left(2 \kappa^2-\lambda^2\right) c_{h Z A}^{(1)} H_V+8 \kappa r  \bar{e}^2\left( c_{h Z A}^{(1)}\right)^2+ \kappa s H_0^2\right) \\  \nonumber
    & +\frac{r \bar{e} m_H^2 G_0^2  G_V}{4(r-s) \sqrt{r s}} \left(2 \left(2 \kappa^2-\lambda^2\right)c_{h Z A}^{(1)} c_{h Z Z}^{(1)}+\left(2 \kappa^2-\lambda^2+4 \kappa s\right)c_{h Z A}^{(2)} c_{h Z Z}^{(1)}-16 \kappa r s m_H^2c_{h Z A}^{(1)} c_{h Z Z}^{(2)} \right)\\  \nonumber
    &+\frac{1}{2} \sqrt{r s} m_H^2  G_0^2\left|d_Z\right|^2\left(\kappa r m_H^4+\frac{ \left(2 \kappa^2-\lambda^2\right)v m_H^2c_{h Z Z}^{(1)}}{r-s}+\frac{4 \kappa(r+s) v^2\left(c_{h Z Z}^{(1)}\right)^2}{(r-s)^2}\right) \\ 
     J^Z_8=& -\frac{2 r s G_0^4}{(r-s)^2}\left(c_{h Z Z}^{(1)} m_H^2\left(2 \kappa c_{h Z Z}^{(2)}+c_{h Z Z}^{(3)}\right)-\left(c_{h Z Z}^{(1)}\right)^2-\kappa^2 m_H^4\left(c_{h Z Z}^{(2)}\right)^2\right) && \\ \nonumber
     & -\frac{r s m_H^2G_0^2 }{r-s}\left(2 c_{h Z Z}^{(1)} H_1+ (\kappa+2 r)m_H^2c_{h Z Z}^{(1)}H_2-2 \kappa m_H^2 c_{h Z Z}^{(2)}  H_1\right) && \\ \nonumber
     & +\frac{1}{2} r m_H^4G_0^2\left(2\kappa ec_{h Z A}^{(1)} H_V+sH_0^2+\frac{2\kappa^2 \bar{e}^2  \left(c_{h Z A}^{(1)}\right)^2}{s}\right) && \\ \nonumber
     &-\frac{r m_H^2 \bar{e} G_0^2 G_V}{r-s}\left(c_{h Z Z}^{(1)}\left(2 \kappa c_{h Z A}^{(1)}-(\kappa+2 s)c_{h Z A}^{(2)}\right)+2 \kappa^2  m_H^2c_{h Z A}^{(1)} c_{h Z Z}^{(2)}\right) && \\ \nonumber
     & -r s m_H^2G_0^2\left|d_Z\right|^2\left(r m_H^4+\frac{2 \kappa v m_H^2c_{h Z Z}^{(1)}}{r-s}+\frac{4(r+s)v^2\left(c_{h Z Z}^{(1)}\right)^2}{(r-s)^2}\right)     &&   
\end{align}
\subsection{$J^W_i$ functions} \label{J^W explicit expressions}
The $J^W_i$ functions are defined as the coefficient multiplying the angular functions in eq. (\ref{J^W functions}). To simplify the expressions for $J^W_i$ we have used that at dimension eight, the effective couplings satisfy $g^{(2)}_{hW^\pm l}=g^{(8)}_{hW^\pm l}=-g^{(3)}_{hW^\pm l}=-g^{(6)}_{hW^\pm l}$. The imaginary parts of the effective couplings are generated at $\mathcal{O}(1/\Lambda^4)$, which by interfering with the SM lead to terms proportional to $\Gamma_W$. This contributions are suppressed by $\Gamma_W \ll m_W$, so we take $\Gamma_W \sim 0$. Additionally we introduced the dimensionless variable $\mu_W\equiv m_W/m_h$ and the replacement rule $W^-\leftrightarrow W^+$, which stands for the following simultaneous replacements
$$W^-\leftrightarrow W^+ \equiv \left\{g^{W^-}\leftrightarrow g^{W^+}, g^{(i)}_{hW^-l}\leftrightarrow g^{(i)}_{hW^+l}\right\}$$
Like in the precious $J$-Functions, $J_1^W$ and $J_2^W$ differ by their dipole contribution, so we subtract it to simply their expressions.
\begin{flalign}
    &J_1^W -  \frac{r s v c_{h W W}^{(1)} \left|d_W\right|^2\left|g^W\right|^2\left(2c_{h W W}^{(1)} v(r+s)+\kappa  m_h^2\left(2\mu_W-s-r\right)\right)}{m_H^2\left(\mu_W-r\right)^2\left(\mu_W-s\right)^2} && \\ \nonumber
    & \qquad +\frac{\left|d_W\right|^2\left|g^W\right|^2 m_H^2\left(2 r s-\kappa^2\right)\left( r\left(\mu_W-s\right)^2+s\left(\mu_W-r\right)^2\right)}{4\left(\mu_W-s\right)^2\left(\mu_W-r\right)^2} &&  \\
    &=J_2^W +  \frac{ v c_{h W W}^{(1)} \left|d_W\right|^2\left|g^W\right|^2\left(2c_{h W W}^{(1)} v(r+s)\left(2 r s+\kappa^2\right)+3\kappa r s m_h^2\left(2\mu_W-s-r\right)\right)}{m_H^2\left(\mu_W-r\right)^2\left(\mu_W-s\right)^2} && \\ \nonumber
    & \qquad +\frac{\left|d_W\right|^2\left|g^W\right|^2 m_H^2\left(2 r s+\kappa^2\right)\left( r\left(\mu_W-s\right)^2+s\left(\mu_W-r\right)^2\right)}{4\left(\mu_W-s\right)^2\left(\mu_W-r\right)^2} \\ \nonumber
    & = \frac{\kappa^2 +2 r s}{4}\left|\frac{g^{(1)}_{hW^-l}g^{W^+}}{\mu_W-r}+\frac{g^{(1)}_{hW^+l}g^{W^-}}{\mu_W-s} \right|^2&& \\ \nonumber
    &\qquad +\frac{\left|g^W\right|^4\left(c_{h W W}^{(1)} m_H^2\left(2 \operatorname{Re}\left(c_{h W W}^{(3)}\right)\left((r+s)\left(\kappa^2+2 r s\right)+6 \kappa r s\right)-12 \kappa r s c_{h W W}^{(2)}\right)\right)}{4 m_H^4\left(\mu_W-r\right)^2\left(\mu_W-s\right)^2} \\ \nonumber
    &\qquad +\frac{\left|g^W\right|^4\left(2 r s\left(c_{h W W}^{(2)}\right)^2 m_H^4\left(\kappa^2+8 r s\right)+\left(c_{h W W}^{(1)}\right)^2\left(\kappa^2+2 r s\right)\right)}{4 m_H^4\left(\mu_W-r\right)^2\left(\mu_W-s\right)^2}\\ \nonumber  
    &\qquad +\left[\frac{\operatorname{Re}\left(g_{h W^-l}^{(1)} g^{W^{+}}\right)\left|g^W\right|^2\left(6 \kappa r s c_{h W W}^{(2)} m_H^2-c_{h W W}^{(1)}\left(\kappa^2+2 r s\right)\right)}{2 m_H^2\left(\mu_W-r\right)^2\left(\mu_W-s\right)}\right.  \\ \nonumber
    &\qquad +\frac{c_{h W W}^{(1)}\left|g^W\right|^2\left(\kappa \lambda^2\left( g_{h W^-l}^{(7)} \operatorname{Re}\left(g^{W^{+}}\right)- \operatorname{Re}\left(g_{h W^{-}}^{(5)} g^{W^{+}}\right)\right)\right)}{4\left(\mu_W-r\right)^2\left(\mu_W-s\right)} \\ \nonumber
    &\qquad \left. -\frac{c_{h W W}^{(1)}\left|g^W\right|^2  \operatorname{Re}\left(g_{h W^{-}}^{(4)} g^{W^{+}}\right)(\kappa+2 r)\left(\kappa^2+2 r s\right)}{4\left(\mu_W-r\right)^2\left(\mu_W-s\right)}+\left(\begin{array}{cc}
       s\leftrightarrow r  \\ W^+\leftrightarrow W^-
    \end{array}\right) \right] 
\end{flalign}
\begin{flalign}
    J^W_3= & -\frac{\kappa^2 +2 r s}{4}\left|\frac{g^{(1)}_{hW^-l}g^{W^+}}{\mu_W-r}+\frac{g^{(1)}_{hW^+l}g^{W^-}}{\mu_W-s} \right|^2&& \\ \nonumber
    & -\frac{ 2 c_{h W W}^{(1)} m_H^2\left|g^W\right|^4\left(\lambda^2 \operatorname{Re}\left(c_{h W W}^{(3)}\right)(r+s)+2 r s\left(\operatorname{Re}\left(c_{h W W}^{(3)}\right)-\kappa c_{h W W}^{(2)}\right)\right)}{4 m_H^4\left(r-\mu_W\right)^2\left(s-\mu_W\right)^2}\\ \nonumber
    & -\frac{\left|g^W\right|^4\left(2 r s\left(c_{h W W}^{(2)}\right)^2 m_H^4\left(\kappa^2-2 \lambda^2\right)+\left(c_{h W W}^{(1)}\right)^2\left(\lambda^2+2 r s\right)\right)}{4 m_H^4\left(r-\mu_W\right)^2\left(s-\mu_W\right)^2}\\ \nonumber
    & \left[+\frac{\operatorname{Re}\left(g_{h W^-l}^{(1)} g^{W^{+}}\right)\left|g^W\right|^2\left(c_{h W W}^{(1)}\left(\lambda^2+2 r s\right)-2 \kappa r s c_{h W W}^{(2)} m_H^2\right)}{2 m_H^2\left(\mu_W-r\right)^2\left(\mu_W-s\right)}\right. \\ \nonumber 
    &+\frac{c_{h W W}^{(1)}\left|g^W\right|^2\left(2 \operatorname{Re}\left(g_{h W^-l}^{(4)} g^{W^{+}}\right)(\kappa+2 r)\left(\lambda^2+2 r s\right)\right)}{8\left(\mu_W-r\right)^2\left(\mu_W-s\right)} \\ \nonumber
    & \left.+\frac{\kappa \lambda^2c_{h W W}^{(1)}\left|g^W\right|^2\left(2 \operatorname{Re}\left(g_{h W^{-} l}^{(5)} g^{W^{+}}\right)-2 g_{h W^{-} l}^{(7)} \operatorname{Re}\left(g^{W^{+}}\right)\right)}{8\left(\mu_W-r\right)^2\left(\mu_W-s\right)}+\left(\begin{array}{cc}
       s\leftrightarrow r  \\ W^- \leftrightarrow W^+
    \end{array}\right)\right] \\ \nonumber
    &+\frac{r v\left|d_W\right|^2\left|g^W\right|^2c_{h WW}^{(1)}\left(c_{h W}^{(1)} v\left(\lambda^2+2 r s+2 s^2\right)+\kappa s  m_H^2\left(2 \mu_W-r-s\right)\right)}{m_H^2\left(r-\mu_W\right)^2\left(s-\mu_W\right)^2}\\ \nonumber
    & -\frac{1}{4}\left|d_W\right|^2\left|g^W\right|^2 m_H^2\left(\frac{s\left(2 r s-\lambda^2\right)}{\left(s-\mu_W\right)^2}+\frac{r\left(\lambda^2+2 r s\right)}{\left(r-\mu_W\right)^2}\right) \\ 
    J_4^W=&J_3^W\Big|_{\begin{subarray}{l}
    s \leftrightarrow r \\
    W^+ \leftrightarrow W^-\end{subarray}} && \\ 
    J_5^W=& 2rs\left|\frac{g^{(1)}_{hW^-l}g^{W^+}}{\mu_W-r}+\frac{g^{(1)}_{hW^+l}g^{W^-}}{\mu_W-s} \right|^2&& \\ \nonumber &+ \frac{r s\left|g^W\right|^4\left(c_{h W W}^{(1)} m_H^2\left(2 \operatorname{Re}\left(c_{h W W}^{(3)}\right)-2 \kappa c_{h W W}^{(2)}\right)+\kappa^2\left(c_{h W W}^{(2)}\right)^2 m_H^4+\left(c_{h W W}^{(1)}\right)^2\right)}{2 m_H^4\left(\mu_W-r\right)^2\left(\mu_W-s\right)^2}&& \\ \nonumber   
    & + \left[\frac{4 r s\left|g^W\right|^2\operatorname{Re}\left(g_{h W^-l}^{(1)} g^{W^{+}}\right)\left(\kappa c_{h W W}^{(2)} m_H^2-c_{h W W}^{(1)}\right)}{m_H^2\left(\mu_W-r\right)^2\left(\mu_W-s\right)}\right. \\ \nonumber 
    & \left. \hspace{20pt}-\frac{2 r  s c_{h W W}^{(1)}\left|g^W\right|^2\operatorname{Re}\left(g_{h W^{-}l}^{(4)} g^{W^{+}}\right)(\kappa+2 r)}{\left(\mu_W-r\right)^2\left(\mu_W-s\right)} +\left(\begin{array}{cc}
       s\leftrightarrow r  \\ W^+\leftrightarrow W^-
    \end{array}\right) \right] \nonumber
\end{flalign}

\begin{flalign}
    \nonumber
    J_6^W= & \frac{\kappa\sqrt{rs}}{4}\left|\frac{g^{(1)}_{hW^-l}g^{W^+}}{\mu_W-r}+\frac{g^{(1)}_{hW^+l}g^{W^-}}{\mu_W-s} \right|^2 -\frac{ \sqrt{r s}\kappa\left|g^W\right|^4\left( \left(c_{h W W}^{(1)}\right)^2+ r s\left(c_{h W W}^{(2)}\right)^2 m_H^4\right)}{ m_H^4\left(\mu_W-r\right)^2\left(\mu_W-s\right)^2}&& \\   
    & -\frac{\sqrt{r s}c_{h W W}^{(1)} \left|g^W\right|^4 \left( \operatorname{Re}\left(c_{h W W}^{(3)}\right)\left(2 \kappa-\lambda^2\right)- c_{h W W}^{(2)}\left(2 \kappa^2-\lambda^2\right)\right)}{ m_H^2\left(\mu_W-r\right)^2\left(\mu_W-s\right)^2} && \\ \nonumber 
    & +\left[\frac{\sqrt{r s}\left|g^W\right|^2 \operatorname{Re}\left(g_{h W^-l}^{(1)} g^{W^{+}}\right)\left(c_{h W W}^{(2)} m_H^2\left(\lambda^2-2 \kappa^2\right)+2 \kappa c_{h W W}^{(1)}\right)}{m_H^2\left(\mu_W-r\right)^2\left(\mu_W-s\right)}\right. && \\ \nonumber
    & \qquad +\frac{\sqrt{r s}\lambda^2c_{h W W}^{(1)}\left|g^W\right|^2\left(2 \operatorname{Re}\left(g_{h W^{-}l}^{(5)} g^{W^{+}}\right)-2 g_{h W^{-}}^{(7)} \operatorname{Re}\left(g^{W^{-}l}\right)\right)}{2\left(\mu_W-r\right)^2\left(\mu_W-s\right)}&& \\ \nonumber
    & \qquad \left. +\frac{\sqrt{r s}\kappa (\kappa+2 r)c_{h W W}^{(1)}\left|g^W\right|^2  \operatorname{Re}\left(g_{h W^{-}l}^{(4)} g^{W^{+}}\right)}{\left(\mu_W-r\right)^2\left(\mu_W-s\right)} +\left(\begin{array}{cc}
       s\leftrightarrow r  \\ W^+\leftrightarrow W^-
    \end{array}\right) \right] && \\
    J_7^W= & J_6^W + \frac{\kappa \left|d_W\right|^2\left|g^W\right|^2 m_H^2 \sqrt{r s}}{4}\left(\frac{s}{\left(\mu_W-s\right)^2}+\frac{r}{\left(\mu_W-r\right)^2} \right)&& \\  \nonumber 
    &+\frac{\sqrt{r s}vc_{h W W}^{(1)}\left|d_W\right|^2\left|g^W\right|^2\left(4 (r+s)\kappa   vc_{h W W}^{(1)}+ m_H^2\left(2 \kappa^2-\lambda^2\right)\left(2 \mu_W-r-s\right)\right)}{4 m_H^2\left(r-\mu_W\right)^2\left(s-\mu_W\right)^2} && \\
    J_8^W= &  \frac{J_5^W}{4} - \frac{r s m_H^2\left|d_W\right|^2\left|g^W\right|^2 }{2}\left(\frac{s}{\left(\mu_W-s\right)^2}+\frac{r}{\left(\mu_W-r\right)^2} \right)&& \\  \nonumber 
    &-\frac{vc_{h W W}^{(1)} \left|d_W\right|^2\left|g^W\right|^2\left(8 r s c_{h W W}^{(1)} v(r+s)-1m_H^2\left(2 \kappa^2-\lambda^2\right) \sqrt{r s}\left(2 \mu_W-r-s\right)\right)}{4 m_H^2\left(r-\mu_W\right)^2\left(s-\mu_W\right)^2} && \\ 
    J_9^W=& \frac{\lambda c_{h W W}^{(1)}\left|g^W\right|^2 }{2\left(\mu_W-r\right)\left(\mu_W-s\right)}\left( \frac{\left(2 r s-\kappa^2\right)g_{h W^{+} l}^{(2)}\operatorname{Re}\left(g^{W^{-}}\right)}{\left(\mu_W-s\right)}+\frac{2 r s g_{h W^{-} l}^{(2)} \operatorname{Re}\left(g^{W^{+}}\right) }{\left(\mu_W-r\right)}\right) && \\ \nonumber & -\frac{r \lambda \left|d_W\right|^2\left|g^W\right|^2}{2\left(\mu_W-r\right)^2}\left(\kappa m_H^2+\frac{4 s v c_{h W W}^{(1)}}{\left(\mu_W-s\right)}\right) && \\ 
    J_{10}^W=& \frac{\lambda c_{h W W}^{(1)}\left|g^W\right|^2 }{2\left(\mu_W-r\right)\left(\mu_W-s\right)}\left( \frac{\left(2 r s-\kappa^2\right)g_{h W^{-} l}^{(2)}\operatorname{Re}\left(g^{W^{+}}\right)}{\left(\mu_W-r\right)}+\frac{2 r s g_{h W^{+} l}^{(2)} \operatorname{Re}\left(g^{W^{-}}\right) }{\left(\mu_W-s\right)}\right) && \\ \nonumber & -\frac{s \lambda \left|d_W\right|^2\left|g^W\right|^2}{2\left(\mu_W-s\right)^2}\left(\kappa m_H^2+\frac{4 r v c_{h W W}^{(1)}}{\left(\mu_W-r\right)}\right) && \\ 
    J_{11}^W=&\frac{\lambda^3 c_{h W W}^{(1)}\left|g^W\right|^2 g_{h W^{+} l}^{(2)} \operatorname{Re}\left(g^{W^{-}}\right)}{2\left(\mu_W-r\right)\left(\mu_W-s\right)^2} && 
\end{flalign}

\begin{flalign}
    J_{12}^W=&\frac{\lambda^3 c_{h W W}^{(1)}\left|g^W\right|^2 g_{h W^{-} l}^{(2)} \operatorname{Re}\left(g^{W^{+}}\right)}{2\left(\mu_W-r\right)^2\left(\mu_W-s\right)} && \\
    J_{13}^W=& -\frac{ \lambda \kappa^2 c_{h W W}^{(1)}\left|g^W\right|^2  g_{h W^+l}^{(2)} \operatorname{Re}\left(g^{W^{-}}\right)}{2\left(\mu_W-r\right)\left(\mu_W-s\right)^2} && \\ 
    J_{14}^W=& -\frac{ \lambda \kappa^2 c_{h W W}^{(1)}\left|g^W\right|^2  g_{h W^-l}^{(2)} \operatorname{Re}\left(g^{W^{+}}\right)}{2\left(\mu_W-r\right)^2\left(\mu_W-s\right)} && \\ 
    J_{15}^W=&\frac{\lambda c_{h W W}^{(1)}\left|g^W\right|^2 }{4\left(\mu_W-r\right)\left(\mu_W-s\right)}\left( \frac{\left(\kappa^2+\lambda^2\right)g_{h W^{-}l}^{(2)}\operatorname{Re}\left(g^{W^{+}}\right)}{\left(\mu_W-r\right)}-\frac{12 r s g_{h W^{+}l}^{(2)}\operatorname{Re}\left(g^{W^{-}}\right)}{\left(\mu_W-s\right)}\right)&& \\ 
    & + \frac{\lambda s\left|d_W\right|^2\left|g^W\right|^2}{2\left(\mu_W-s\right)^2}\left(\kappa m_H^2+\frac{4 r c_{h W W}^{(1)} v}{\left(\mu_W-r\right)}\right) && \\
    J_{16}^W=&\frac{\lambda c_{h W W}^{(1)}\left|g^W\right|^2 }{4\left(\mu_W-r\right)\left(\mu_W-s\right)}\left( \frac{\left(\kappa^2+\lambda^2\right)g_{h W^{+}l}^{(2)}\operatorname{Re}\left(g^{W^{-}}\right)}{\left(\mu_W-s\right)}-\frac{12 r s g_{h W^{-}l}^{(2)}\operatorname{Re}\left(g^{W^{+}}\right)}{\left(\mu_W-r\right)}\right)&& \\ 
    & + \frac{\lambda r\left|d_W\right|^2\left|g^W\right|^2}{2\left(\mu_W-r\right)^2}\left(\kappa m_H^2+\frac{4 s c_{h W W}^{(1)} v}{\left(\mu_W-s\right)}\right) && \\
    J^W_{17}=&J^W_{19}=-\frac{\lambda r s\left|g^W\right|^2 c_{h W W}^{(1)} g_{h W^{+} l}^{(2)} \operatorname{Re}\left(g^{W^{-}}\right)}{\left(\mu_W-r\right)\left(\mu_W-s\right)^2} \\
    J^W_{18}=&J^W_{20}=-\frac{\lambda r s\left|g^W\right|^2 c_{h W W}^{(1)} g_{h W^{-} l}^{(2)} \operatorname{Re}\left(g^{W^{+}}\right)}{\left(\mu_W-r\right)^2\left(\mu_W-s\right)} \\
    J^W_{21}=&\frac{\kappa \lambda\sqrt{r s} \left|g^W\right|^2c_{h W W}^{(1)}  }{4\left(\mu_W-r\right)\left(\mu_W-s\right)}\left(\frac{3 g_{h W^{+} l}^{(2)}\operatorname{Re}\left(g^{W^{+}}\right)}{\mu_W-s}-\frac{g_{h W^{-}l}^{(2)}\operatorname{Re}\left(g^{W^{-}}\right)}{\mu_W-r}\right)&& \\ \nonumber
    & -\frac{\lambda\left|g^W\right|^2\left|d_W\right|^2 \sqrt{r s}}{2\left(\mu_W-s\right)^2}\left(s m_H^2+\frac{\kappa c_{h W W}^{(1)} v}{\mu_W-r}\right)&& \\
    J^W_{22}=&\frac{\kappa \lambda\sqrt{r s}\left|g^W\right|^2 c_{h W W}^{(1)}  }{4\left(\mu_W-r\right)\left(\mu_W-s\right)}\left(\frac{3 g_{h W^{-} l}^{(2)}\operatorname{Re}\left(g^{W^{+}}\right)}{\mu_W-r}-\frac{g_{h W^{+}l}^{(2)}\operatorname{Re}\left(g^{W^{-}}\right)}{\mu_W-s}\right)&& \\ \nonumber
    & -\frac{\lambda\left|g^W\right|^2 \left|d_W\right|^2 \sqrt{r s}}{2\left(\mu_W-r\right)^2}\left(r m_H^2+\frac{\kappa c_{h W W}^{(1)} v}{\mu_W-s}\right)&& \\
    J^W_{23}=&J^W_{24}=-\frac{\lambda(r-s) \sqrt{r s} \left|g^W\right|^2 c_{h W W}^{(1)} \operatorname{Im}\left(g^{W^{+}} g_{h W^{-} l}^{(5)}\right)}{2\left(\mu_W-r\right)^2\left(\mu_W-s\right)^2}&& 
\end{flalign}
\begin{flalign}
    J^W_{25}=&\frac{\kappa \lambda \sqrt{r s} \left|g^W\right|^2 c_{h W W}^{(1)} g_{h W^{+} l}^{(2)} \operatorname{Re}\left(g^{W^-}\right)}{\left(\mu_W-r\right)\left(\mu_W-s\right)^2}&& \\
    J^W_{26}=&\frac{\kappa \lambda \sqrt{r s} c_{h W W}^{(1)} \left|g^W\right|^2g_{h W^{-} l}^{(2)}  \operatorname{Re}\left(g^{W^+}\right)}{\left(\mu_W-r\right)^2\left(\mu_W-s\right)}
\end{flalign}

\bibliographystyle{JHEP}
\bibliography{references}

\end{document}